\documentclass[aps,pre,twocolumn,longbibliography,showpacs,amsmath,amsmath,amssymb,amsfonts,superscriptaddress]{revtex4-1}

\usepackage[utf8]{inputenc}
\usepackage[T1]{fontenc}
\usepackage{amsmath}
\usepackage{amssymb}
\usepackage{mathtools}
\usepackage{bbold}
\usepackage{float}
\usepackage[toc,page,title]{appendix}
\usepackage[section]{placeins}
\usepackage[margin=1in]{geometry}
\usepackage[textsize=scriptsize]{todonotes}
\makeatletter \def\@captype{figure} \makeatother 
\usepackage{color}
\definecolor{extra}{RGB}{34,180,104}

\usepackage[
        colorlinks=true,urlcolor=blue,linkcolor=blue,citecolor=blue]%
        {hyperref}

\renewcommand{\Re}{\mathrm{Re}}
\renewcommand{\Im}{\mathrm{Im}}
\newcommand{\transposesymb}{{\sf T}}
\newcommand{\commutator}[2]{[#1, #2]}
\newcommand{\pdv}[2]{\partial_{#2}#1}
\newcommand{\bra}[1]{\langle #1 |}
\newcommand{\ket}[1]{| #1 \rangle}
\newcommand{\braket}[2]{\langle #1| #2 \rangle}
\newcommand{\ketbra}[1]{| #1 \rangle \langle #1|}
\newcommand{\bbra}[1]{\langle\langle #1 ||}
\newcommand{\kket}[1]{|| #1 \rangle\rangle}
\newcommand{\bbraket}[2]{\langle\langle #1|| #2 \rangle\rangle}

\newcommand{\avg}[1]{\langle #1 \rangle}
\newcommand{\avgT}[1]{\avg{#1}_{\currenttilt}}
\newcommand{\abs}[1]{\lvert #1 \rvert}
\newcommand{\identityop}{\hat{\mathbb{1}}}
\newcommand{\be}{\begin{equation}}
\newcommand{\ee}{\end{equation}}
\DeclarePairedDelimiter\floor{\lfloor}{\rfloor}
\newcommand{\observable}{O}
\newcommand{\trajobservable}{A}
\newcommand{\Doobletter}{\mathrm{D}}
\newcommand{\field}{E}
\newcommand{\crit}[1]{#1_{\mathrm{c}}}
\newcommand{\nprtcls}{N}
\newcommand{\jumprate}{p}
\newcommand{\jumprateR}{\jumprate_{+}}
\newcommand{\jumprateL}{\jumprate_{-}}
\newcommand{\jumprateRL}{\jumprate_{\pm}}
\newcommand{\rateinL}{\alpha}
\newcommand{\rateoutR}{\beta}
\newcommand{\rateoutL}{\gamma}
\newcommand{\rateinR}{\delta}
\newcommand{\dens}{\rho}
\newcommand{\diffusivity}{D}
\newcommand{\mobility}{\sigma}

\newcommand{\timev}{t}
\newcommand{\intcurrent}{Q}
\newcommand{\current}{q}

\newcommand{\conf}{C}
\newcommand{\prob}{P}

\newcommand{\occup}{n}
\newcommand{\currenttilt}{\lambda}
\newcommand{\transitionrate}[2]{W_{#1 \to #2}}
\newcommand{\escaperate}[1]{R_{#1}}

\newcommand{\genmat}{\hat{\mathbb{W}}}
\newcommand{\genmatT}{\genmat^{\currenttilt}}

\newcommand{\genmatD}{\genmatT_{\Doobletter}}
\newcommand{\genmatDrate}{\genmatDrate_{\Doobletter}}
\newcommand{\densL}{\dens_{L}}
\newcommand{\densR}{\dens_{R}}
\newcommand{\mass}{\mu}
\newcommand{\lattsize}{L}

\newcommand{\cm}{\mathbf{z}}
\newcommand{\Zsymm}[1]{\mathbb{Z}_{#1}}
\newcommand{\symmop}{\hat{S}}
\newcommand{\symmeigval}[1]{\phi_{#1}}

\newcommand{\symmopR}{\symmop_{\frac{2\pi}{r}}}
\newcommand{\nstatesP}{r}
\newcommand{\hamilt}{H}
\newcommand{\invtemp}{\beta}
\newcommand{\spinangle}{\varphi}
\newcommand{\meanmag}{\mathbf{m}}
\newcommand{\energytilt}{\lambda}
\newcommand{\DPF}{Z}
\newcommand{\DFE}{\theta}

\newcommand{\trajduration}{\tau}

\newcommand{\traj}{\omega_{\trajduration}}
\newcommand{\currentcrit}{\crit{\current}}
\newcommand{\currenttiltcrit}{\crit{\lambda}}
\newcommand{\currenttiltcritmax}{\currenttiltcrit^{+}}
\newcommand{\currenttiltcritmin}{\currenttiltcrit^{-}}
\newcommand{\currenttiltcritminmax}{\currenttiltcrit^{\pm}}

\newcommand{\occupop}{\hat{\occup}}
\newcommand{\creationop}{\hat{\sigma}^{+}}
\newcommand{\destructop}{\hat{\sigma}^{-}}

\newcommand{\eigvalT}[1]{\theta^{\currenttilt}_{#1}}

\newcommand{\eigvalD}[1]{\eigvalT{#1, \Doobletter}}

\newcommand{\eigvectTRnk}[1]{R^{\currenttilt}_{#1}}
\newcommand{\eigvectTLnk}[1]{L^{\currenttilt}_{#1}}
\newcommand{\eigvectDRnk}[1]{R^{\currenttilt}_{#1, \Doobletter}}
\newcommand{\eigvectDLnk}[1]{L^{\currenttilt}_{#1, \Doobletter}}
\newcommand{\eigvectTR}[1]{\ket{\eigvectTRnk{#1}}}
\newcommand{\eigvectTL}[1]{\bra{\eigvectTLnk{#1}}}
\newcommand{\eigvectDR}[1]{\ket{\eigvectDRnk{#1}}}
\newcommand{\eigvectDL}[1]{\bra{\eigvectDLnk{#1}}}
\newcommand{\flatvectnk}{-}
\newcommand{\flatvect}{\bra{\flatvectnk}}
\newcommand{\eigvectTLop}[1]{\hat{L}^{\currenttilt}_{#1}}

\newcommand{\probvecttimenk}[1]{\prob_{#1}}
\newcommand{\probvecttime}[1]{\ket{\probvecttimenk{#1}}}

\newcommand{\probT}{\prob^{\currenttilt}}

\newcommand{\probTvecttimenk}[1]{\probT_{#1, P_0}}
\newcommand{\probTvecttime}[1]{\ket{\probTvecttimenk{#1}}}
\newcommand{\probTvectstnk}{\probTvecttimenk{\text{ss}}}
\newcommand{\probTvectst}{\ket{\probTvectstnk}}

\newcommand{\probDphasevectnk}[1]{\Pi^{\currenttilt}_{#1}}
\newcommand{\probDphasevect}[1]{\ket{\probDphasevectnk{#1}}}
\newcommand{\probDphasevecttime}[1]{\ket{\probDphasevectnk{#1}(\timev)}}

\newcommand{\isep}{\mathrel{{.}\,{.}}\nobreak}

\begin{document}
\title{Spectral signatures of symmetry-breaking dynamical phase transitions}
\author{R. Hurtado-Guti\'errez}
\affiliation{Institute Carlos I for Theoretical and Computational Physics, and Departamento de Electromagnetismo y F\'{\i}sica de la Materia, Universidad de Granada, Granada 18071, Spain.}
\author{P.I. Hurtado}
\affiliation{Institute Carlos I for Theoretical and Computational Physics, and Departamento de Electromagnetismo y F\'{\i}sica de la Materia, Universidad de Granada, Granada 18071, Spain.}
\author{C. P\'erez-Espigares}
\affiliation{Institute Carlos I for Theoretical and Computational Physics, and Departamento de Electromagnetismo y F\'{\i}sica de la Materia, Universidad de Granada, Granada 18071, Spain.}
\date{\today}

\begin{abstract}
Large deviation theory provides the framework to study the probability of rare fluctuations of time-averaged observables, opening new avenues of research in nonequilibrium physics. One of the most appealing results within this context are dynamical phase transitions (DPTs), which might occur at the level of trajectories in order to maximize the probability of sustaining a rare event. While the Macroscopic Fluctuation Theory has underpinned much recent progress on the understanding of symmetry-breaking DPTs in driven diffusive systems, their microscopic characterization is still challenging. In this work we shed light on the general spectral mechanism giving rise to continuous DPTs not only for driven diffusive systems, but for any jump process in which a discrete $\mathbb{Z}_n$ symmetry is broken. By means of a symmetry-aided spectral analysis of the Doob-transformed dynamics, we provide the conditions whereby symmetry-breaking DPTs might emerge and how the different dynamical phases arise from the specific structure of the degenerate eigenvectors. In particular, we show explicitly how all symmetry-breaking features are encoded in the subleading eigenvectors of the degenerate subspace. Moreover, by partitioning configuration space into equivalence classes according to a proper order parameter, we achieve a substantial dimensional reduction which allows for the quantitative characterization of the spectral fingerprints of DPTs. We illustrate our predictions in several paradigmatic many-body systems, including (i) the one-dimensional boundary-driven weakly asymmetric exclusion process (WASEP), which exhibits a particle-hole symmetry-breaking DPT for current fluctuations, (ii) the $3$- and $4$-state Potts model for spin dynamics, which displays discrete rotational symmetry-breaking DPTs for energy fluctuations, and (iii) the closed WASEP which presents a continuous symmetry-breaking DPT into a time-crystal phase characterized by a rotating condensate.
\end{abstract}

\maketitle

\section{Introduction}
    
The study of dynamical large deviations in classical and quantum nonequilibrium systems has allowed for a better understanding of the emerging patterns both in the steady states and their fluctuations \cite{bertini01a,derrida01a,bertini02a,derrida02a,derrida03a,bodineau04a,bertini05a,bertini05b,bodineau05a,bertini06a,bodineau06a,bertini07a,derrida07a,bodineau07a,garrahan07a,lecomte07c,touchette09a,garrahan09a,garrahan10a,bertini15a,garrahan18a}. Particularly relevant has been the discovery of fluctuation theorems \cite{evans93a,gallavotti95a,kurchan98a,lebowitz99a,maes99a,harris07a,jarzynski97a,jarzynski97b,crooks00a,hatano01a}, concerning the symmetries in the probabilities of fluctuations of dynamical observables --such as the current or the entropy production-- \cite{maes06a,hurtado11b,maes14a,marcantoni20a,marcantoni21a}, and the thermodynamic uncertainty relations \cite{barato15a,gingrich16a}, which yield bounds on dissipation in terms of current fluctuations.
   
One of the most intriguing phenomena which have gained attention in the last two decades are the so-called dynamical phase transitions (DPTs) \cite{bertini05a,bodineau05a,bertini06a,jack15a,perez-espigares18b,perez-espigares19a,banuls19a}.
Unlike standard phase transitions, which occur when modifying a physical parameter, these might occur when a system sustains an atypical value, i.e~a rare fluctuation, of a trajectory-dependent observable. DPTs are accompanied by a drastic change in the structure of those trajectories responsible for such fluctuation, and they are revealed as non-analyticities in the associated large deviation functions, which play the role of thermodynamic potentials for nonequilibrium settings \cite{touchette09a}.

From a macroscopic perspective in driven diffusive systems, the existence of DPTs is governed by the \emph{action} functional of a certain fluctuation provided by the Macroscopic Fluctuation Theory \cite{bertini15a}. Within this framework, the occurrence of a DPT is determined according to the functional form of the transport coefficients characterizing the system, namely the diffusivity and the mobility \cite{bertini05a,bodineau05a,bertini06a,baek17a}. In this context, a myriad of emerging structures associated with DPTs have been discovered, including symmetry-breaking density profiles \cite{baek17a,baek18a,perez-espigares18a}, localization effects \cite{gutierrez21a}, condensation phenomena \cite{chleboun17a} or traveling waves \cite{hurtado11a,perez-espigares13a,tizon-escamilla17b} displaying time-crystalline order \cite{hurtado-gutierrez20a}. Moreover, DPTs have been also predicted and observed in active media \cite{cagnetta17a,whitelam18b,tociu19a,gradenigo19a,nemoto19a,cagnetta20a,chiarantoni20a,fodor20a,grandpre21a,keta21a,yan22a}, where individual particles can consume free energy to produce directed motion, as well as in many different open quantum systems \cite{flindt09a,garrahan10a,garrahan11a,ates12a,hickey12a,genway12a,flindt13a,lesanovsky13a,maisi14a,manzano14a,manzano18a,manzano21a}. Interestingly, many of these DPTs involve the spontaneous breaking of a $\mathbb{Z}_n$ symmetry (or invariance under discrete rotations of angles $2\pi/m$ with $m=1,2,...,n$ in the order parameter space).
     
A different, complementary path to investigate the physics of DPTs consists in analyzing them in terms of the microscopic dynamics, governed by the corresponding stochastic generator. For long times, such generator may be \emph{tilted}  \cite{lecomte07c,garrahan10a} so as to obtain the scaled cumulant generating function of the relevant observable from the eigenvalue with the largest real part, which is the Legendre transform of the associated large deviation function, just as the free energy with respect to the entropy in equilibrium statistical mechanics. However, the tilted generator is not a proper stochastic generator, as it does not conserve probability, but it can be turned into a physical stochastic generator by means of the Doob transform \cite{simon2009,popkov10a,jack2010,chetrite13a,chetrite15b}. The Doob dynamics reweights the statistics of trajectories to focus on those responsible for a fluctuation and can be interpreted as the original dynamics supplemented with a the appropriate driving field which makes typical the rare fluctuations of the original problem \cite{hurtado-gutierrez20a}. In this way the Doob steady state contains all the information on the most likely path to a fluctuation.

From a spectral perspective, the hallmark of a symmetry-breaking DPT is the emergence of a collection of Doob eigenvectors with a vanishing spectral gap \cite{gaveau98a,hanggi82a,ledermann50a,gaveau06a,minganti18a}. Such degenerate subspace, which can be further classified using the underlying symmetry operator, defines the stationary subspace of the Doob stochastic generator, so that the typical states responsible for a given fluctuation in the original system can be retrieved from these degenerate Doob eigenvectors. Similar ideas have been put forward for standard phase transitions, where the equivalence between emergent degeneracy of the leading eigenspace of the stochastic generator and the appearance of a phase transition has been demonstrated \cite{gaveau98a,gaveau06a}. Moreover, recent results leveraging on this idea have been derived to construct metastable states in open quantum systems \cite{macieszczak16a,macieszczak21a} and to obtain optimal trajectories of symmetry-breaking DPTs in driven diffusive systems \cite{perez-espigares18a,hurtado-gutierrez20a,gutierrez21a}. Yet, the general structure of the Doob eigenvectors, their relation to the underlying symmetry of the stochastic generator, and the particular mechanism of symmetry breaking have been elusive so far. Previous works have focused on particular models, but the common underlying spectral mechanism giving rise to the different dynamical phases when a discrete $\mathbb{Z}_n$ symmetry is broken is still lacking.

In this work we address this problem by shedding light on the general behavior and structure of the Doob eigenvectors involved in $\mathbb{Z}_n$ symmetry-breaking DPTs. We discuss the equivalence between an emergent degeneracy of the leading eigenspace of the Doob generator and the appearance of a DPT as characterized by different steady states (with different values of an appropriate order parameter). This motivates the introduction of a transformation in the degenerate subspace to construct the physical phase probability vectors from the gapless, degenerate Doob eigenvectors. These different phase probability vectors are connected by the symmetry operator, thus restoring the symmetry of the original generator. The Doob steady state can be then written as a weighted sum of these phase probability vectors, and the different weights are governed by the projection of the initial state on the subleading Doob eigenvectors and their eigenvalues under the symmetry operator (hereafter referred to as the symmetry eigenvalues). This clear picture explains how the system breaks the symmetry by singling out a particular dynamical phase out of the multiple possible phases present in the first Doob eigenvector, and enables to identify phase-selection mechanisms by initial state preparation somewhat similar to those already described in open quantum systems with strong symmetries \cite{manzano14a,manzano18a,manzano21a}. Moreover, by assuming that the different phases are disjoint (so that statistically-relevant configurations belong to one phase at most), we derive an explicit expression for the components of the subleading Doob eigenvectors in the degenerate subspace in terms of the leading eigenvector and the symmetry eigenvalues, which hence contain all the information on the symmetry-breaking process. This highlights the stringent spectral structure imposed by symmetry on DPTs. 

The analysis of this spectral structure in particular problems is unfeasible due to the high-dimensional character of configuration space (which typically grows exponentially with the system size). 
We overcome this issue by first introducing a partition of configuration space into equivalence classes according to a proper order parameter of the DPT under study, and then using it to perform a strong dimensional reduction of the space.
The resulting reduced vectors live in a Hilbert space with much lower dimension (which usually scales linearly with the system size), allowing the statistical confirmation of our predictions in different models.

Remarkably, the above-described symmetry-breaking spectral mechanism, demonstrated here for DPTs in the large deviation statistics of time-averaged observables, is completely general for $\mathbb{Z}_n$-invariant systems and expected to hold valid also in standard (steady-state) critical phenomena \cite{gaveau98a,gaveau06a,binney92a}.
   
We illustrate our general results by analyzing in detail three distinct DPTs in different paradigmatic many-body systems: the one dimensional boundary-driven (or open) weakly asymmetric simple exclusion process (WASEP), the $3-$ and $4-$state Potts model for spin dynamics, and the closed (or periodic) WASEP. In the open WASEP a particle-hole ($\Zsymm{2}$) symmetry is broken when the system either crowds or depletes the lattice with particles in order to sustain current fluctuations well below the average, so that the previous ideas apply in a straightforward way. On the other hand, the $3-$ and $4-$state Potts model exhibits a spontaneous breaking of a discrete rotational symmetry ($\Zsymm{3}$ and $\Zsymm{4}$, respectively) to a ferromagnetic (dynamical) phase in order to sustain time-averaged energy fluctuations well below the average. Finally, the large deviation physics of the closed WASEP is even more compelling: for currents below a critical threshold, the system self-organizes into a macroscopic jammed state in the form of a rotating particle condensate, which hinders transport thus facilitating a current fluctuation much lower than the average. This rotating condensate breaks time-translation invariance and the spatial translation symmetry of the ring ($\Zsymm{\lattsize}$, with $\lattsize$ being the number of lattice sites). In particular, we show that the different phases here correspond to the different locations of the condensate along the lattice, with motion encoded in the imaginary part of the spectrum, which shifts the selection of the phase making the condensate to travel at constant velocity.

The paper is organized as follows. In Section \S\ref{sec:model_tools} we review the quantum Hamiltonian formalism for the master equation in stochastic many-body systems, as well as its application to study
the statistics of trajectories and the large deviation theory of time-averaged observables. In this section we also introduce the Doob transform to build an auxiliary stochastic dynamics that makes typical the rare fluctuation of the original dynamics. Section \S\ref{secDPTsymm} is devoted to study the spectral fingerprints of DPTs using the machinery presented in \S\ref{sec:model_tools} and exploiting the symmetry of the dynamics. This analysis provides general predictions on the spectral signatures of symmetry-breaking DPTs which we proceed to test in concrete examples in the subsequent sections. In particular, in Section \S\ref{secoWASEP} we focus on particle current fluctuations in the boundary-driven WASEP. On the other hand, Section \S\ref{secPotts} is concerned with energy fluctuations in the $r-$state Potts model of spin dynamics (with $r=3,4$) while Section \S\ref{seccWASEP} is devoted to study the symmetry-breaking in space and time giving rise to the time crystal structure observed in the open WASEP, rendering a fresh view on this intriguing phenomenon. The general predictions of \S\ref{secDPTsymm} are confirmed in every example, offering physical insights on the different DPTs reported. We end the paper with a discussion of our results in Section \S\ref{seccWASEP}, while some technical notes are described in the appendices.

\section{Statistics of trajectories and Doob transform}
\label{sec:model_tools}

In this work we focus on many-body jump processes, namely discrete-state stochastic processes such as interacting-particle systems defined on a lattice and evolving in continuous time. We represent the states or configurations as vectors $\ket{C}$ of an orthonormal basis in a Hilbert space ${\cal H}$ satisfying  $\braket{\conf}{\conf^{\prime}} = \delta_{\conf \conf^{\prime}}$ \cite{schutz01a}. This allows us to write the state of the system at time $\timev$ as a probability vector,
\begin{equation}
    \probvecttime{\timev} = \sum_{\conf} \prob(\conf, \timev) \ket{\conf} \, ,
\end{equation}
whose real entries, $0\le \prob(\conf, \timev)\le 1$, correspond to the probability of finding the system in configuration $\conf$ at time $\timev$. The time evolution of this probability vector $\probvecttime{\timev}\in {\cal H}$ is given by a \emph{master equation} written in an operatorial form, $\pdv{\probvecttime{\timev}}{\timev} = \genmat \probvecttime{\timev}$, where $\genmat$ is the Markov generator of the dynamics \cite{kampen11a}.
In general, this generator reads
\begin{equation}
\genmat = \sum_{\conf, \conf^{\prime} \neq \conf} \transitionrate{\conf}{\conf^{\prime}} \ket{\conf^{\prime}} \bra{\conf} - \sum_{\conf} \escaperate{\conf} \ketbra{\conf} \, ,
\label{genmat}
\end{equation}
where $\transitionrate{\conf}{\conf^{\prime}}$ is the transition rate from configuration $\conf$ to $\conf^{\prime}$, and $\escaperate{\conf} \equiv \sum_{\conf^{\prime} \neq \conf} \transitionrate{\conf}{\conf^{\prime}}$ is the escape rate from configuration $\conf$.
Since this generator is stochastic (i.e.~probability conserving), we have that $\flatvect \genmat  = 0$, with $\flatvect$ the so-called ``flat'' state, $\flatvect = \sum_{\conf} \bra{\conf}$, so that the normalization of the probability vector is always conserved,
i.e. $\braket{-}{\probvecttimenk{\timev}} = \sum_\conf \prob(\conf, \timev) = 1$ $\forall \timev$.

In this work we shall focus on DPTs taking place when a time-integrated dynamical observable is conditioned to have a prescribed value. In order to study the statistics of such trajectory-dependent observable and its large deviation properties, we will consider ensembles of trajectories of duration $\tau$. Each trajectory 
$\traj\equiv\{(\conf_i,\timev_i)\}_{i=0,1,...,m}$
is completely specified by the sequence of configurations visited by the system, $\{\conf_i\}_{i=0,1,...,m}$,
and the times at which they occur, $\{\timev_i\}_{i=0,...m}$, with $m$ being the number of transitions throughout the trajectory, 
\begin{equation}
\traj :\; \conf_0 \xrightarrow{\timev_1} \conf_1 \xrightarrow{\timev_2} \conf_2 \xrightarrow{\timev_3} ( \cdots ) \xrightarrow{\timev_{m}} \conf_{m} \, ,
\label{traj}
\end{equation}
with $\timev_0=0$ setting the time origin. The probability  of a trajectory \footnote{Although this is customarily called ``probability'', it is really a probability distribution which is continuous on the transition times $\{\timev_i\}_{i=1,...,m}$ and discrete on the visited configurations $\{\conf_i\}_{i=0,1,...,m}$ and the number of transitions $m$.} is then given by 
\begin{multline}
\label{eq:traj_prob}
\prob[\traj] = \text{e}^{-(\trajduration - \timev_{m}) \escaperate{\conf_{m}}} \transitionrate{\conf_{m-1}}{\conf_{m}} \cdots \\
\cdots \text{e}^{-(\timev_2 -\timev_1) \escaperate{\conf_1}} \transitionrate{\conf_0}{\conf_1} \text{e}^{-\timev_1 \escaperate{\conf_0}} \prob(\conf_0, 0)\, .
\end{multline}
The time-extensive observables whose large deviation statistics we are interested in might depend on the state of the process and its transitions over time. For jump processes, such trajectory-dependent observables can be written in general as 
\begin{equation}
A (\omega_{\tau}) = \sum_{i=0}^{m} (t_{i+1} - t_i) g(C_i) + \sum_{i=0}^{m - 1} \eta_{C_i,C_{i+1}}\, .
\label{eq:current_def}
\end{equation}
The first sum above corresponds to the time integral of configuration-dependent observables $g(C_i)$, while the second sum stands for observables that increase by $\eta_{C_i,C_{i+1}}$ in the transitions from $C_i$ to $C_{i+1}$. In the first sum we have defined $t_0=0$ and $t_{m+1}=\tau$. If we are interested e.g. in the large-deviation statistics of the time-integrated current, we set $g(C_i)=0$ and $\eta_{C_i,C_{i+1}}=\pm 1$ depending on the direction of the particle jump, while $\eta_{C_i,C_{i+1}}=1$ for the kinetic activity. On the other hand, the statistics of the time-integrated energy can be obtained by setting $\eta_{C_i,C_{i+1}}=0$ and defining $g(C_i)$ as the energy of configuration $C_i$. Thus, the probability of having a given value of $A$ after a time $\tau$ is simply
$
\prob_\trajduration (A) = \sum_{\traj} \prob[\traj] \delta(A - A(\traj))
$.
This probability obeys in general a \emph{large deviation principle} for long times \cite{hurtado14a, bertini15a, derrida07a}
\begin{equation}
\prob_{\trajduration}\left(
\frac{A}{\trajduration} =a \right) \asymp \text{e}^{-\trajduration F(a)}\, ,
\end{equation}
where $a = A/\trajduration$ is the time-average of $A$ and ``$\asymp$'' stands for logarithmic asymptotic equality, i.e., 
\begin{equation}
\label{ldprf}
\lim_{\trajduration \to \infty} -\frac{1}{\trajduration} \ln\prob_{\trajduration}(a) = F(a)\, .
\end{equation}
This is the so-called rate function or \emph{large deviation function} (LDF), which is positive, $F(a) \ge 0$, and equal to zero only for the average or mean value, $\langle a \rangle$. The LDF $F(a)$ measures the rate at which the probability of observing a fluctuation peaks around $\langle a\rangle$ as $\trajduration$ increases \cite{touchette09a,hurtado14a}, or equivalently the exponential rate at which the likelihood of appreciable fluctuations away from $\langle a\rangle$ decays in time.

In order to obtain $F(a)$, we define an ensemble of trajectories with a fixed value $A$. In such ensemble of trajectories, the LDF plays a role akin to the entropy in the microcanonical ensemble of configurations, with the difference that the fixed quantity here is the (trajectory-dependent) $A$ instead of the (configurational) system energy. Calculations in this ensemble are usually complicated so, in the spirit of the microcanonical-canonical ensemble equivalence, it is convenient to define a \emph{biased} $\lambda$-ensemble (sometimes also called $s$-ensemble in literature) in which we allow the observable to fluctuate but fixing its average value, $\langle A \rangle_{\lambda}/\tau=a$, through the \emph{biasing field} $\currenttilt$. 
 In this new ensemble, the probability of a trajectory is \cite{lecomte07c, garrahan07a, garrahan09a, jack2010}
\begin{equation}
\label{eq:traj_probT} \prob^{\currenttilt}[\traj] = \frac{\prob[\traj] \text{e}^{\currenttilt A(\traj)}} {\DPF_{\trajduration}(\currenttilt)}\, ,
\end{equation}
where the normalizing factor $\DPF_{\trajduration}(\currenttilt)$ is the \emph{dynamical partition function} of the $\currenttilt$-ensemble, $Z_{\tau}(\lambda)= \sum_{\traj} \prob[\traj] \text{e}^{\currenttilt A(\traj)}$. The biasing field $\currenttilt$ is conjugated to the time-integrated current in a way similar to the inverse temperature and energy in equilibrium systems \cite{touchette09a}. Positive values of $\currenttilt$ bias the dynamics towards values of $A$ larger than its average (corresponding to $\lambda=0$), while negative values do the opposite. Hence, the average of a trajectory-dependent observable $\observable(\traj)$ in the biased ensemble is given in general by
\begin{equation}
\avgT{\observable} = \frac{\avg{\observable~\text{e}^{\currenttilt A}}}{\DPF_{\trajduration}(\currenttilt)}\, ,
\end{equation}
where $\avg{\cdot}$ represents an average in the unbiased trajectory ensemble ($\lambda=0$). Thus, instead of computing $F(a)$ by means of $P[\traj]$, we do it through $Z_{\tau}(\lambda)$, which is nothing but its moment generating function. 

For long times it can be readily checked that $Z_{\tau}(\lambda)$ follows a large deviation principle \cite{lecomte07c,touchette09a,hurtado14a},
\be
Z_{\tau}(\lambda)\asymp \text{e}^{\tau \theta(\lambda)}
\label{ZLDF}
\ee
with $\theta(\lambda)$ being the scaled cumulant generating function, $\DFE(\currenttilt)=\lim_{\trajduration \to \infty} \trajduration^{-1} \ln \DPF_{\trajduration}(\lambda)$, whose derivatives provide the cumulants of the time-average observable, $a$, in the biased ensemble. In particular, the first derivative gives the average, $\theta'(\lambda)=\langle A \rangle_{\lambda}/\tau$. 
As a consequence the value of $\lambda$ to be chosen is the one matching the fluctuation $a$, i.e.~such that $\theta'(\lambda)=a$, or, equivalently, $F'(a)=\lambda$.
We can see that $\theta(\lambda)$ corresponds to a dynamical free energy which is related, as in equilibrium statistical mechanics, to the rate function or dynamical entropy $F(a)$ by means of the Legendre-Frenchel transform $\theta(\lambda) =  -\min_{a} \left[F(a) - \currenttilt a\right]$ \cite{touchette09a}. Microscopically, for ergodic Markov processes and long times, $\theta (\lambda)$ is given by the eigenvalue with the largest real part of the so-called \emph{tilted} generator $\genmatT$. 
The latter modifies the original generator, Eq.~\eqref{genmat}, by introducing an exponential bias or tilt in the transition rates according to the increment of the observable in each transition and an extra term in the diagonal part of the generator \cite{garrahan09a}, namely
\begin{eqnarray}
\nonumber
\genmatT = &&
\sum_{\conf, \conf^{\prime} \neq \conf} \text{e}^{\lambda \eta_{C,C'}}\transitionrate{\conf}{\conf^{\prime}} \ket{\conf^{\prime}} \bra{\conf}
- \sum_{\conf} \escaperate{\conf} \ketbra{\conf}\\ 
&&+\lambda \sum_{\conf} g(C)\ketbra{\conf}.
\label{eq:genmatT}
\end{eqnarray}
The dynamical partition function can then be expressed in operatorial form as \cite{lecomte07c}
\begin{equation}
\DPF_{\trajduration}(\currenttilt) = \flatvect \text{e}^{\trajduration \genmatT} \probvecttime{0}\, ,
\label{DPFop}
\end{equation}
where $\probvecttime{0}\in {\cal H}$ is an arbitrary initial state. 

Next we introduce the spectrum of the (in general non-symmetric) tilted generator $\genmatT$. Let $\eigvectTR{j}$ and $\eigvectTL{j}$ be the $j$-th right and left eigenvectors of $\genmatT$, such that $\genmatT\eigvectTR{j} = \eigvalT{j} \eigvectTR{j}$ and $\eigvectTL{j}\genmatT = \eigvalT{j} \eigvectTL{j}$, with $\eigvalT{j}\in\mathbb{C}$ the corresponding eigenvalue, ordered in decreasing value of their real part. In most models of interest, the set of left and right eigenvectors form a complete biorthogonal basis of the Hilbert space 
\footnote{In most cases, the generator is either diagonalizable or diagonalizable when restricted to the invariant subspaces associated with the eigenvalues closer to zero, which are the ones contributing in the long time limit.},
such that $\braket{L^{\lambda}_i}{R^{\lambda}_j}=\delta_{ij}$. By using now a spectral decomposition, we can write $\genmatT = \sum_{j} \eigvalT{j} \eigvectTR{j} \eigvectTL{j}$. It is then straightforward to show from Eqs.~\eqref{DPFop} and \eqref{ZLDF} that for long times $\trajduration$ the dynamical free energy is given by the eigenvalue of $\genmatT$ with the largest real part, $\DFE(\currenttilt) = \eigvalT{0}$. 

Notice that, except for $\currenttilt = 0$ where the original (unbiased) dynamics is recovered, $\genmatT$ does not conserve probability, i.e., $\flatvect \genmatT \neq 0$, and therefore it is not a proper stochastic generator. This implies that it is not possible to directly retrieve from the tilted generator the physical trajectories leading to the fluctuation $a$, since $\genmatT$ does not represent an actual physical dynamics. 
To overcome this issue and obtain the trajectories generating the biased ensemble in Eq.~\eqref{eq:traj_probT} for a given $\lambda$, we introduce an auxiliary dynamics or driven process built on $\genmatT$, known as the Doob transformed generator \cite{simon2009,jack2010,popkov10a,chetrite15a}
\begin{equation}
\genmatD = \eigvectTLop{0} \genmatT (\eigvectTLop{0})^{-1} - \eigvalT{0} \identityop\, ,
\label{genmatD}
\end{equation}
where $\eigvectTLop{0}$ is a diagonal matrix with elements $(\eigvectTLop{0})_{ij} = (\eigvectTL{0})_i \delta_{ij}$, and $\identityop$ is the identity matrix.
The spectra of both generators, $\genmatT$ and $\genmatD$, are simply related by a shift in their eigenvalues, $\eigvalD{j} = \eigvalT{j} - \eigvalT{0}$, and a simple transformation of their left and right eigenvectors, $\eigvectDL{j} = \eigvectTL{j} (\eigvectTLop{0})^{-1}$ and $\eigvectDR{j} = \eigvectTLop{0} \eigvectTR{j}$. 
As a consequence, the leading eigenvalue of $\genmatD$ becomes zero and its associated leading right eigenvector, given by $\eigvectDR{0} = \eigvectTLop{0} \eigvectTR{0} $, becomes the Doob stationary state.
In addition, the leading left eigenvector is $\eigvectDL{0} = \eigvectTL{0} (\eigvectTLop{0})^{-1} = \flatvect$, confirming that the Doob generator does conserve probability, $\flatvect \genmatD = 0$.
In this way, the Doob-transformed generator provides the physical trajectories distributed according to the $\lambda$-ensemble given by Eq.~\eqref{eq:traj_probT} in the $\tau\to\infty$ limit \cite{chetrite15b}, revealing how fluctuations are created in time.
The left and right eigenvectors of the Doob generator also form a complete biorthogonal basis of the Hilbert space,
and they are further normalized such that $\max_{\conf} | \braket{\eigvectDLnk{j}}{\conf}| = 1$ and $\braket{\eigvectDLnk{i}}{\eigvectDRnk{j}} = \delta_{ij}$ \cite{gaveau98a}.
Note that this normalization specifies the eigenvectors with $j>0$ up to an arbitrary complex phase (determined in Appendix \ref{appB}).
In addition, due to conservation of probability, $\flatvect \genmatD = 0$, we have that $\braket{-}{\eigvectDRnk{j}}=0$ for all $j\neq 0$.

In the following we will make use of the above spectral tools to study the general structure of the Doob eigenvectors across a DPT, with the aim of unveiling how the different dynamical phases emerge when an underlying symmetry is broken.

\section{Dynamical phase transitions and symmetries}
\label{secDPTsymm}

Standard critical phenomena occur at the configurational level when varying a control parameter like e.g.~temperature or pressure. In contrast, as already pointed out above, DPTs appear in the trajectory space when the system is conditioned to sustain a large fluctuation of a time-averaged observable. Such DPTs often involve the emergence of distinct $\mathbb{Z}_n$ symmetry-broken patterns
\cite{baek17a,baek18a,perez-espigares18a,gutierrez21a}, which might be time dependent \cite{chleboun17a,hurtado11a,perez-espigares13a,tizon-escamilla17b,hurtado-gutierrez20a}, facilitating the corresponding fluctuation and thus making it far more probable than anticipated. 
As in standard second-order phase transitions, continuous DPTs are characterized by some type of order which continuously arises at the level of trajectories as the control (biasing) field is varied across the transition, and which is captured by an appropriate \emph{order parameter}.

In this section we study in detail the spectral fingerprints of DPTs using the mathematical tools developed in the previous section, supplemented with the microscopic symmetry of the system. For that, we first need to specify what a $\mathbb{Z}_n$ symmetry is in this context. With this information at hand, we will analyze the structure of the steady state of the system in terms of the eigenvectors of the Doob generator before and after the appearance of the DPT, discussing along the way the connection between degeneracy of the leading Doob subspace and the emergence of a DPT. We will show how the stationary state in the symmetry-broken regime is constructed as a weighted sum of different, well-defined phase probability vectors connected via the symmetry operator, and we will highlight initial-state preparation strategies to single out a given symmetry-broken phase. Partitioning configuration space into the different phases then allows us to write the most important components of the subleading Doob eigenvectors in the degenerate subspace in terms of the leading eigenvector and the symmetry eigenvalues, showing how all the information on the symmetry-breaking process is encoded in this leading eigenvector. The introduction of an order parameter vector space which inherits the spectral features associated with the DPT, and the ensuing strong dimensional reduction, opens the door to the empirical verification of our findings, that we set out to develop in the following sections.

\subsection{$\mathbb{Z}_n$ symmetry}
\label{ssecSym}

Our interest in this work is focused on DPTs involving the spontaneous breaking of $\mathbb{Z}_n$ symmetry, hence some general remarks about symmetry aspects of stochastic processes are in order \cite{hanggi82a}. In particular, we are interested in symmetry properties under transformations of state space of the original stochastic process, as defined by the generator $\genmat$, and how these properties are inherited by the Doob auxiliary process $\genmatD$ associated with the fluctuations of a time-integrated observable $A$. For discrete state space, as in our case, any such symmetries correspond to permutations in configuration space \cite{hanggi82a}. 

The set of transformations that leave invariant a stochastic process (in a sense described below) form a symmetry group. Multiple stochastic many-body systems of interest are invariant under transformations belonging to the $\mathbb{Z}_n$ group. This is a cyclic group of order $n$, so its elements are built from the repeated application of a single operator $\symmop\in \mathbb{Z}_n$, which satisfies $\symmop^n=\identityop$. This operator is then unitary, probability-conserving (i.e.~$\bra{-}\hat{S}=\bra{-}$), and invertible, with real and non-negative matrix elements, and commutes with the generator of the stochastic dynamics, $\commutator{\genmat}{\symmop} = 0$, or equivalently 
\be
\genmat = \symmop \genmat \symmop^{-1}\, .
\label{symcond}
\ee  
The action of the $\mathbb{Z}_n$ symmetry operator on configurations is described as a bijective transformation acting on state space, $\symmop\ket{C}=\ket{C_S}\in {\cal H}$. This transformation induces a map $\cal{S}$ in trajectory space
\begin{eqnarray}
\nonumber
& \omega_{\tau}: C_1 \to C_2 \to \ldots \to C_m \\ \nonumber
&\mathcal{S}\big\downarrow \\ 
&\mathcal{S} \omega_{\tau}: C_{S1} \to C_{S2} \to \ldots \to C_{Sm}\, ,
\label{Rconf}
\end{eqnarray}
that transforms the configurations visited along the path but leaves unchanged the transition times $\{t_i\}_{i=0,1,\ldots m}$ between configurations.

For the symmetry to be inherited by the 
Doob auxiliary process 
$\genmatD$ associated with the fluctuations of an observable $A$, see Eq.~\eqref{genmatD}, it is necessary that this trajectory-dependent observable remains invariant under the trajectory transformation, i.e. $A({\cal{S}}\omega_{\tau})=A(\omega_{\tau})$. 
This condition, together with the invariance of the original dynamics under $\symmop$, see Eq.~\eqref{symcond}, crucially implies that both the tilted and the Doob generators are also invariant under $\symmop$, as demonstrated in Appendix \ref{appA},
\be
\genmatT = \symmop \genmatT \symmop^{-1}\, ,~~~~\genmatD = \symmop \genmatD \symmop^{-1}.
\ee
As a consequence, both $\genmatD$ and $\symmop$ share a common eigenbasis, i.e. they can be diagonalized at the same time, so $\ket{R_{j,\text{D}}^{\lambda}}$ and $\bra{L_{j,\text{D}}^{\lambda}}$ are also eigenvectors of $\symmop$ with eigenvalues $\phi_j$, i.e. $\hat{S}\ket{R_{j,\text{D}}^{\lambda}}=\phi_j\ket{R_{j,\text{D}}^{\lambda}}$ and $\bra{L_{j,\text{D}}^{\lambda}}\hat{S}=\phi_j\bra{L_{j,\text{D}}^{\lambda}}$, designated as symmetry eigenvalues. Due to the unitarity and cyclic character of $\symmop$, the eigenvalues $\phi_j$ simply correspond to the $n$ roots of unity, i.e. $\phi_j=\text{e}^{i 2\pi k_j/n}$ with $k_j=0,1,...,n-1$.

\subsection{DPTs and degeneracy}
\label{ssecDeg}

The steady state associated with the Doob stochastic generator $\genmatD$ describes the statistics of trajectories during a large deviation event of parameter $\currenttilt$ of the original dynamics. The formal solution of the Doob master equation for any time $\timev$ and starting from an initial probability vector $\ket{P_0}$ can be written as $\probTvecttime{\timev} = \exp(+ \timev \genmatD) \ket{P_0}$. Introducing now a spectral decomposition of this formal solution, we have
\begin{equation}
\probTvecttime{t} = \eigvectDR{0} + \sum_{j>0} \text{e}^{\timev \eigvalD{j}} \eigvectDR{j} \braket{\eigvectDLnk{j}}{P_0}  \, ,
\label{specdec}
\end{equation}
where we have already used that the Doob generator is stochastic and hence has a leading zero eigenvalue, $\eigvalD{0} = 0$. Furthermore, since $\braket{-}{\eigvectDRnk{j}} = \braket{\eigvectDLnk{0}}{\eigvectDRnk{j}} = \delta_{0j}$, all the probability of $\probTvecttime{t}$ is contained in $\ket{\eigvectDRnk{0}}$, i.e. $\braket{-}{\probTvecttimenk{t}}=\braket{-}{\eigvectDRnk{0}}=1$.
Thus, each term with $j>0$
in the r.h.s of Eq.~\eqref{specdec} corresponds to a particular redistribution of the probability. 
Moreover, as the symmetry operator $\hat{S}$ conserves probability, we get $1 = \flatvect \symmop \eigvectDR{0} = \flatvect \symmeigval{0} \eigvectDR{0}$, i.e.
\begin{equation}
\phi_0=1 \, ,
\end{equation}
for the symmetry eigenvalue of the leading eigenvector. This implies that $\eigvectDR{0}$ is invariant under the symmetry operator.

To study the steady state of the Doob dynamics, $\probTvectst\equiv \lim_{\timev\to\infty} \probTvecttime{\timev}$, we now define the spectral gaps as $\Delta^{\lambda}_j=\Re(\theta_0^{\lambda}-\theta_j^{\lambda})=-\Re(\eigvalD{j})\ge 0$, which control the exponential decay of the corresponding eigenvectors, cf. Eq.~\eqref{specdec}. Note that $0\le \Delta^{\lambda}_1\le \Delta^{\lambda}_2\le \ldots$ due to the ordering of eigenvalues according to their real part, see \S\ref{sec:model_tools}. When $\Delta^{\lambda}_1$ is strictly positive, $\Delta^{\lambda}_1>0$, so that the spectrum is gapped (usually $\Delta^{\lambda}_1$ is referred to as \emph{the} spectral gap), all subleading eigenvectors decay exponentially fast for timescales $\timev \gg 1/\Delta^{\lambda}_1$ and the resulting Doob steady state is unique,
\be
\probTvectst =  \eigvectDR{0}.
\label{ssnopt}
\ee
This steady state preserves the symmetry of the generator, $\symmop \probTvectst = \probTvectst$, so no symmetry-breaking phenomenon at the fluctuating level is possible whenever the spectrum of the Doob generator $\genmatD$ is gapped. This is hence the spectral scenario before any DPT can occur. 

Conversely, any symmetry-breaking phase transition at the trajectory level will demand for an emergent degeneracy in the leading eigenspace of the associated Doob generator. This is equivalent to the spectral fingerprints of standard symmetry-breaking phase transitions in stochastic systems \cite{gaveau98a,hanggi82a,ledermann50a,gaveau06a,minganti18a}. As the Doob auxiliary process $\genmatD$ 
is indeed stochastic, these spectral fingerprints \cite{gaveau98a} will characterize also DPTs at the fluctuating level. In particular, for a many-body stochastic system undergoing a $\mathbb{Z}_n$ symmetry-breaking DPT, we expect that the difference between the real part of the first and the subsequent $n-1$ eigenvalues $\eigvalD{j}$ goes to zero in the thermodynamic limit once the DPT kicks in. In this case the Doob \emph{stationary} probability vector is determined by the first $n$ eigenvectors defining the degenerate subspace. Note that, in virtue of the Perron-Frobenius theorem, for any finite system size the steady state is non-degenerate, highlighting the relevance of the thermodynamic limit.

In general, the gap-closing eigenvalues associated with these eigenvectors may exhibit non-zero imaginary parts, $\Im(\theta_{j,\text{D}}^\lambda)\ne 0$, thus leading to a time-dependent Doob \emph{stationary} vector in the thermodynamic limit
\begin{equation}
\probTvectst (t) = \eigvectDR{0} + \sum_{j=1}^{n-1} \text{e}^{+i t \Im(\theta_{j,\text{D}}^\lambda)}\eigvectDR{j} \braket{\eigvectDLnk{j}}{\probvecttimenk{0}}\, .
\label{Psstime}
\end{equation}
Moreover, if these imaginary parts display band structure, the resulting Doob \emph{stationary} state will exhibit a periodic motion characteristic of a time crystal phase \cite{hurtado-gutierrez20a}, as we will show in the particular example of \S\ref{seccWASEP}. However, in many cases the gap-closing eigenvalues of the Doob eigenvectors in the degenerate subspace are purely real, so $\Im(\theta_{j,\text{D}}^\lambda)= 0$ and the resulting Doob steady state is truly stationary,
\begin{equation}
\probTvectst = \eigvectDR{0} + \sum^{n-1}_{j=1}  \eigvectDR{j} \braket{\eigvectDLnk{j}}{\probvecttimenk{0}}\, .
\label{eq:DPT_probvect}
\end{equation}
The number $n$ of vectors that contribute to the Doob steady state corresponds to the different number of phases that appear once the $\Zsymm{n}$ symmetry is broken. Indeed, a $n$th-order degeneracy of the leading eigenspace implies the appearance of $n$ different, linearly independent stationary distributions \cite{hanggi82a,ledermann50a}, as we shall show below. 
As in the general time-dependent solution, Eq.~\eqref{specdec}, all the probability is concentrated on the first eigenvector $\eigvectDR{0}$, which preserves the symmetry, $\symmop\eigvectDR{0}=\eigvectDR{0}$, while the subsequent eigenvectors in the degenerate subspace describe the \emph{redistribution} of this probability according to their projection on the initial state, containing at the same time all the information on the symmetry-breaking process.
Notice that, even if the degeneration of the $n$ first eigenvalues is complete, we can still single out $\eigvectDR{0}$ as the only eigenvector with eigevalue $\phi_0=1$ under $\symmop$ (all the gap-closing eigenvectors have different eigenvalues under $\symmop$, as it is shown in Appendix~\ref{appB}).
Indeed, the steady state Eq.~\eqref{eq:DPT_probvect} does not preserve in general the symmetry of the generator, i.e. $\symmop \probTvectst \ne \probTvectst$, and hence the symmetry is broken in the degenerate phase. The same happens for the time-dependent Doob asymptotic state Eq.~\eqref{Psstime}.

\subsection{Phase probability vectors}
\label{ssecPhase}

Our next task consists in finding the $n$ different and linearly independent stationary distributions $\probDphasevect{l}\in {\cal H}$, with $l=0,1,\ldots n-1$, that emerge at the DPT once the degeneracy kicks in \cite{gaveau98a,hanggi82a,ledermann50a,gaveau06a,minganti18a}. Each one of these phase probability vectors $\probDphasevect{l}$ is associated univocally with a single symmetry-broken phase $l\in[0 \isep n-1]$, and the set spanned by these vectors and their left duals defines a new basis of the degenerate subspace. In this way, a phase probability vector $\probDphasevect{l}$ can be always written as a linear combination of the Doob eigenvectors in the degenerate subspace,
\be
\probDphasevect{l} = \sum_{j=0}^{n-1} C_{l,j} \eigvectDR{j}\, ,
\label{ppvlin}
\ee
with complex coefficients $C_{l,j}\in \mathbb{C}$. Moreover, the phase probability vectors must be normalized, $\braket{-}{\Pi_l^\lambda}=1$ $\forall l\in[0\isep n-1]$, and crucially they must be related by the action of the symmetry operator, 
\be
\probDphasevect{l+1} = \symmop \probDphasevect{l} \, ,
\label{ppvsym}
\ee
which implies that $\probDphasevect{l} = \symmop^l \probDphasevect{0}$ and therefore $C_{l,j} = C_{0,j} (\phi_j)^l$,  
with $\phi_j$ the eigenvalues of the symmetry operator. Imposing that any steady state can be written as a statistical mixture (or convex combination) of the different phase probability vectors, it can be 
simply shown (see Appendix \ref{appB}) that the coefficients are $C_{0,j}=1$ $\forall j\in[0\isep n-1]$.
In this way, $C_{l,j} = (\phi_j)^l$, and the probability vectors associated with each of the symmetry-broken phases are
\begin{equation}
\probDphasevect{l} =   \sum_{j=0}^{n-1} (\phi_j)^l\eigvectDR{j} = \eigvectDR{0} +  \sum_{j=1}^{n-1} (\phi_j)^l\eigvectDR{j} \, .
\label{eq:probphasevect}
\end{equation}

It will prove useful to introduce the left duals $\bra{\pi_l^\lambda}$ of the phase probability vectors, i.e., the row vectors satisfying the biorthogonality relation $\braket{\pi_{l'}^\lambda}{\Pi_{l}^\lambda}=\delta_{l',l}$. These must be linear combinations of the left eigenvectors associated with the degenerate subspace,
\be
\bra{\pi_l^\lambda} = \sum_{j=0}^{n-1} \eigvectDL{j} D_{l,j}  \, ,
\label{ppvlin2}
\ee
with complex coefficients $D_{l,j}\in \mathbb{C}$. Imposing biorthogonality, using the spectral expansion \eqref{ppvlin}, and recalling that
the first $n$ eigenvalues $\phi_j$ correspond exactly to the $n$-th roots of unity (see end of Appendix~\ref{appB}),
we thus find $D_{l,j} = \frac{1}{n}(\phi_j)^{-l}$, so that
\be
\bra{\pi_l^\lambda} = \sum_{j=0}^{n-1} \frac{1}{n}(\phi_j)^{-l} \eigvectDL{j} \, .
\label{ppvlin3}
\ee

With these left duals, we can now easily write the right eigenvectors $\eigvectDR{j}$ in the degenerate subspace, $\forall j\in[0\isep n-1]$, in terms of the different phase probability vectors,
\be
\eigvectDR{j} = \sum_{l=0}^{n-1} \probDphasevect{l} \braket{\pi_l^\lambda}{R_{j,D}^\lambda} = \frac{1}{n} \sum_{l=0}^{n-1} (\symmeigval{j})^{-l} \probDphasevect{l}\, ,
\label{rinv}
\ee
where we have used that $\braket{\pi_l^\lambda}{R_{j,D}^\lambda} = \frac{1}{n} (\symmeigval{j})^{-l}$. Using this decomposition in Eq.~\eqref{Psstime}, we can thus reconstruct the (degenerate) Doob \emph{steady} state as a weighted sum of the phase probability vectors $\probDphasevect{l}$ associated with each of the $n$ symmetry-broken phases,
\be
\probTvectst (t)= \sum_{l=0}^{n-1} w_l(t) \probDphasevect{l} \, ,
\label{Pssdecomp}
\ee
with the weights $w_l(t)=\braket{\pi_l^\lambda}{P_{\text{ss},P_0}^\lambda}(t)$ given by
\be
w_l(t)= \frac{1}{n} + \frac{1}{n}\sum_{j=1}^{n-1}  \text{e}^{+i t \Im(\theta_{j,\text{D}}^\lambda)} (\phi_j)^{-l} \braket{\eigvectDLnk{j}}{\probvecttimenk{0}}\, .
\label{coeft}
\ee
These weights are time-dependent if the imaginary part of the gap-closing eigenvalues are non-zero, though in many applications the relevant eigenvalues are purely real. In such cases
\be
\begin{split}
&\probTvectst = \sum_{l=0}^{n-1} w_l \probDphasevect{l}\, , \\
&w_l = \frac{1}{n} + \frac{1}{n}\sum_{j=1}^{n-1}  (\phi_j)^{-l} \braket{\eigvectDLnk{j}}{\probvecttimenk{0}}\, .
\end{split}
\label{coef}
\ee 
This shows that the Doob steady state can be described as a statistical mixture of the different phases (as described by their unique phase probability vectors $\probDphasevect{l}$), i.e. $\sum_{l=0}^{n-1} w_l=1$, with $0\le w_l\le 1$ $\forall l\in[0\isep n-1]$. These statistical weights $w_l$ are determined by the projection of the initial state on the different phases, which is in turn governed by the overlaps of the degenerate left eigenvectors with the initial state and their associated symmetry eigenvalues.

Equation~\eqref{coef} shows how to prepare the system initial state $\ket{\probvecttimenk{0}}$ to single out a given symmetry-broken phase $\probDphasevect{l'}$ in the long-time limit. Indeed, by comparing Eqs.~\eqref{eq:DPT_probvect} and~\eqref{eq:probphasevect}, it becomes evident that choosing $\ket{\probvecttimenk{0}}$ such that $\braket{\eigvectDLnk{j}}{\probvecttimenk{0}}=(\phi_j)^{l'}$ $\forall j \in [1\isep n-1]$ leads to a \emph{pure} steady state $\probTvectst = \probDphasevect{l'}$, i.e. such that $w_{l}=\delta_{l,l'}$. 
This strategy provides a simple phase-selection mechanisms by initial state preparation somewhat similar to those already described in open quantum systems with strong symmetries \cite{manzano14a,manzano18a,manzano21a}.

It is important to note that, in general, degeneracy of the leading eigenspace of the stochastic generator is only possible in the thermodynamic limit. This means that for finite-size systems one should always expect small but non-zero spectral gaps $\Delta^{\lambda}_j$, $j\in[1\isep n-1]$,  and hence the long-time Doob steady state is $\probTvectst = \eigvectDR{0}$, see Eq.~\eqref{ssnopt}. This steady state, which can be written as $\eigvectDR{0} = \frac{1}{n}\sum_{l=0}^{n-1} \probDphasevect{l}$, preserves the symmetry of the generator, so that no symmetry-breaking DPT is possible for finite-size systems. 
Rather, for large but finite system sizes, one should expect an emerging quasi-degeneracy \cite{gaveau98a,gaveau06a} in the parameter regime where the DPT emerges,  i.e. with $\Delta^{\lambda}_j / \Delta^{\lambda}_{n} \ll 1$ $\forall j\in[1\isep n-1]$. In this case, and for time scales $t\ll 1/\Delta^{\lambda}_{n-1}$ but $t\gg 1/\Delta^{\lambda}_{n}$, we expect to observe a sort of metastable symmetry breaking captured by the physical phase probability vectors $\probDphasevect{l}$, with punctuated jumps between different symmetry sectors at the individual trajectory level. 
This leads in the long-time limit to an effective restitution of the original symmetry as far as the system size is finite.

It is worth mentioning that the symmetry group which is broken in the DPT may be larger than $\mathbb{Z}_n$. In such case, as long as it contains a $\mathbb{Z}_n$ group whose symmetry is also broken, the results derived in this section are still valid. Actually, this occurs for the $r$-state Potts model discussed in section \ref{secPotts}, which despite breaking the symmetry of the dihedral group $\mathbb{D}_r$, it breaks as well the symmetry of the $\mathbb{Z}_r$ subgroup.

\subsection{Structure of the degenerate subspace}
\label{ssecSpec}

A key observation is that, once a symmetry-breaking phase transition kicks in, be it either configurational (i.e. standard) or dynamical, the associated \emph{typical} configurations fall into well-defined symmetry classes, i.e. the symmetry is broken already at the individual configurational level. As an example, consider the paradigmatic 2D Ising model and its (standard) $\mathbb{Z}_2$ symmetry-breaking phase transition at the Onsager temperature $T_c$, separating a disordered paramagnetic phase for $T>T_c$ from an ordered, symmetry-broken ferromagnetic phase for $T<T_c$ \cite{binney92a}. For temperatures well below the critical one, the stationary probability of e.g. completely random (symmetry-preserving) spin configurations is extremely low, while high-probability configurations exhibit a net non-zero magnetization typical of symmetry breaking. This means that statistically-relevant configurations do belong to a specific symmetry phase, in the sense that they can be assigned to the \emph{basin of attraction} of a given symmetry sector \cite{gaveau06a}.

Something similar happens in $\mathbb{Z}_n$ symmetry-breaking DPTs. In particular, once the DPT kicks in and the symmetry is broken, statistically-relevant configurations $\ket{C}$ (i.e. such that $\braket{C}{P_{\text{ss},P_0}^\lambda}=P_{\text{ss},P_0}^\lambda(C)$ is significantly different from zero) 
belong to a well-defined symmetry class with index $\ell_C\in[0\isep n-1]$. In terms of phase probability vectors, this means that
\be
\frac{\braket{C}{\probDphasevectnk{{l}}}}{\braket{C}{\probDphasevectnk{{\ell_{\conf}}}}} \approx 0, \qquad \forall l\ne \ell_C \, .
\label{condC}
\ee
This property arises from the large deviation scaling of $\braket{C}{\probDphasevectnk{{l}}}$.
In other words, statistically-relevant configurations in the symmetry-broken Doob steady state can be partitioned into disjoint symmetry classes. This simple but crucial observation can be used now to unveil a hidden spectral structure in the degenerate subspace, associated with such configurations. In this way, if $\ket{C}$ is  one of these configurations belonging to phase $\ell_C$, from Eq.~\eqref{rinv} we deduce that
\be
\braket{C}{\eigvectDRnk{j}}= \frac{1}{n} \sum_{l=0}^{n-1} (\symmeigval{j})^{-l} \braket{C}{\probDphasevectnk{{l}}}\approx \frac{1}{n} (\symmeigval{j})^{-\ell_C} \braket{C}{\probDphasevectnk{{\ell_C}}} \, .
\ee    
In particular, for $j=0$ we have that $\braket{C}{\eigvectDRnk{0}} \approx \frac{1}{n} \braket{C}{\probDphasevectnk{{\ell_C}}}$ since $\phi_0=1$, and therefore
\be
\braket{C}{\eigvectDRnk{j}} \approx (\symmeigval{j})^{-\ell_C} \braket{C}{\eigvectDRnk{0}} 
\label{eq:eigvectsfromeigvect0}
\ee
for $j\in[1\isep n-1]$. In this way, the components $\braket{C}{\eigvectDRnk{j}}$ of the subleading eigenvectors in the degenerate subspace associated with the statistically-relevant configurations are (almost) equal to those of the leading eigenvector $\eigvectDR{0}$ except for a complex argument given by $(\phi_j)^{-\ell_C}$. This highlights how the $\mathbb{Z}_n$ symmetry-breaking phenomenon imposes a specific structure on the degenerate eigenvectors involved in a continuous DPT. Of course, this result is based on the (empirically sound) assumption that statistically-relevant configurations can be partitioned into disjoint symmetry classes. We will confirm \emph{a posteriori} 
this result in the three examples considered below.

\subsection{Order parameter space}
\label{ssecOparam}

The direct analysis of the eigenvectors in many-body stochastic systems is typically an unfeasible task, as the dimension of the configuration Hilbert space usually grows exponentially with the system size. Moreover, extracting useful information from this analysis is also difficult as configurations are not naturally cathegorized according to their symmetry properties. This suggests to introduce a partition of the configuration Hilbert space ${\cal H}$ into equivalence classes according to a proper order parameter for the DPT under study, grouping \emph{similar} configurations together (in terms of their symmetry properties) so as to reduce the effective dimension of the problem, while introducing at the same time a natural parameter to analyze the spectral properties.

We define an order parameter $\mu$ for the DPT of interest as a map $\mu:{\cal H} \to \mathbb{C}$ that gives for each configuration $\ket{C}\in {\cal H}$ a complex number $\mu(C)$ whose modulus measures the \emph{amount of order}, i.e. how deep the system is into the symmetry-broken regime, and whose argument determines in which phase it is. Of course, other types of DPTs may have their own natural order parameters, but for $\mathbb{Z}_n$ symmetry-breaking DPTs a simple complex-valued number suffices, as we shall show below. Associated with this order parameter, we now introduce a reduced Hilbert space ${\cal H}_\mu=\{\kket{\nu}\}$ representing the possible values of the order parameter as vectors $\kket{\nu}$ of a biorthogonal basis satisfying $\bbraket{\nu'}{\nu} = \delta_{\nu,\nu'}$. In general, the dimension of ${\cal H}_\mu$ will be significantly smaller than that of ${\cal H}$ since the possible values of the order parameter typically scale linearly with the system size.

In order to transform probability vectors from the original Hilbert space to the reduced one,
we define a surjective application $\widetilde{\cal T}:{\cal H} \to {\cal H}_\mu$ that maps all configurations $\ket{C}\in {\cal H}$ with order parameter $\nu$ onto a single vector $\kket{\nu}\in {\cal H}_\mu$ of the reduced Hilbert space.
Crucially, this mapping $\widetilde{\cal T}$ from configurations to order parameter equivalence classes must conserve probability, i.e. the accumulated probability of all configurations with a given value of the order parameter in the original Hilbert space ${\cal H}$ must be the same as the probability of the equivalent vector component in the reduced space. In particular, let $\ket{P}\in {\cal H}$ be a probability vector in configuration space, and $\kket{P}=\widetilde{\cal T} \ket{P}$ its corresponding reduced vector in ${\cal H}_\mu$. Conservation of probability then means that
\be
P(\nu)
= \sum_{\substack{\ket{C}\in{\cal H}: \\ \mu(C)=\nu}} \braket{C}{P} = \bbraket{\nu}{P} \, , \qquad \forall \nu \, .
\label{Pconserv}
\ee
This probability-conserving condition thus constraints the particular form of the map $\widetilde{\cal T}:{\cal H} \to {\cal H}_\mu$. In general, if $\ket{\psi}\in{\cal H}$ is a vector in the original configuration Hilbert space, the reduced vector $\kket{\psi}=\widetilde{\cal T}\ket{\psi}\in {\cal H}_\mu$ is hence defined as
\be
\kket{\psi} = \sum_\nu \bbraket{\nu}{\psi}~\kket{\nu} = \sum_\nu \Bigg[\sum_{\substack{\ket{C}\in{\cal H}: \\ \mu(C)=\nu}} \braket{C}{\psi}\Bigg] \kket{\nu} \, .
\label{Pconserv3}
\ee

But what makes a \emph{good} order parameter $\mu$? In short, a good order parameter must be sensitive to the different symmetry-broken phases and to how the symmetry operator moves one phase to another. More in detail, let $\{\ket{C}\}_\nu=\{\ket{C}\in{\cal H}: \mu(C)=\nu\}$ be the set of all configurations $\ket{C}\in{\cal H}$ with order parameter $\mu(C)=\nu$, i.e. the set of all configurations defining the equivalence class represented by the reduced vector $\kket{\nu}\in{\cal H}_\mu$. Applying the symmetry operator $\hat{S}$ to all configurations in $\{\ket{C}\}_\nu$ defines a new set $\hat{S}(\{\ket{C}\}_\nu)$. We say that $\mu$ is a good order parameter iff: (i) The new set $\hat{S}(\{\ket{C}\}_\nu)$ 
corresponds to the equivalence class
$\{\ket{C}\}_{\nu'}$
associated with another order parameter vector $\kket{\nu'}\in{\cal H}_\mu$, and (ii) it can distinguish a symmetry-broken configuration from its symmetry-transformed configuration. This introduces a bijective mapping $\hat{S}_\mu \kket{\nu}=\kket{\nu'}$ between equivalence classes that defines a reduced symmetry operator $\hat{S}_\mu$ acting on the reduced order-parameter space. Mathematically, this mapping can be defined from the relation $\widetilde{\cal T} \symmop \ket{C} = \hat{S}_\mu \widetilde{\cal T}\ket{C}$ $\forall \ket{C}\in{\cal H}$, where condition (i) ensures that $\hat{S}_\mu$ is a valid symmetry operator.

As an example, consider again the 2D Ising spin model for the paramagnetic-ferromagetic phase transition mentioned above \cite{binney92a}. Below the critical temperature, this model breaks spontaneously a $\mathbb{Z}_2$ spin up-spin down symmetry, a phase transition well captured by the total magnetization $m$, the natural order parameter. The symmetry operation consists in this case in flipping the sign of all spins in a configuration, and this operation induces a one to one, bijective mapping between opposite magnetizations. An alternative, plausible order parameter could be $m^2$. This parameter can certainly distinguish the ordered phase ($m^2\ne 0$) from the disordered one ($m^2\approx 0$), but still cannot discern between the two symmetry-broken phases, and hence it is not a good order parameter in the sense defined above.

As we shall illustrate below, the reduced eigenvectors $\kket{R_{j,\text{D}}^\lambda} = \widetilde{\cal T} \ket{R_{j,\text{D}}^\lambda}$ associated with the 
spectrum of the Doob generator in the original configuration space encode the most relevant information regarding the DPT, and can be readily analyzed. Indeed, it can be easily checked that the results obtained in the previous subsections also apply in the reduced order parameter space. In particular, before the DPT happens, the reduced Doob steady state is unique, see Eq.~\eqref{ssnopt},
\be
\kket{P_{\text{ss},P_0}^\lambda} = \kket{R_{0,\text{D}}^\lambda} \, ,
\label{redeq1}
\ee
while once the DPT kicks in and the symmetry is broken, degeneracy appears and
\be
\kket{P_{\text{ss},P_0}^\lambda} = \kket{R_{0,\text{D}}^\lambda} + \sum^{n-1}_{j=1}  \kket{R_{j,\text{D}}^\lambda}~\braket{\eigvectDLnk{j}}{\probvecttimenk{0}}\, ,
\label{redeq2}
\ee
see Eq.~\eqref{eq:DPT_probvect} for purely real eigenvalues [and similarly for eigenvalues with non-zero imaginary parts, see Eq.~\eqref{Psstime}]. Notice that since $\widetilde{\cal T}$ is a linear transformation, the brackets $\braket{\eigvectDLnk{j}}{P_0}$ do not change under $\widetilde{\cal T}$ as they are scalars. Reduced phase probability vectors can be defined in terms of the reduced eigenvectors in the degenerate subspace, see Eq.~\eqref{eq:probphasevect},
\be
\kket{\Pi_l^\lambda} = \kket{R_{0,\text{D}}^\lambda} + \sum_{j=1}^{n-1} (\phi_j)^l \kket{R_{j,\text{D}}^\lambda} \, ,
\label{redeq3}
\ee
and the reduced Doob steady state can be written in terms of these reduced phase probability vectors, $\kket{P_{\text{ss},P_0}^\lambda}  = \sum_{l=0}^{n-1} w_l \kket{\Pi_l^\lambda}$, see Eq.~\eqref{coef}. Finally, the structural relation between the Doob eigenvectors spanning the degenerate subspace, Eq.~\eqref{eq:eigvectsfromeigvect0}, is also reflected in the order-parameter space. In particular, for a statistically-relevant value of $\mu$
\be
\bbraket{\mu}{R_{j,\text{D}}^\lambda} \approx \phi_j^{-\ell_\mu} \bbraket{\mu}{R_{0,\text{D}}^\lambda} \, ,
\label{redeq4}
\ee
for $j\in[1\isep n-1]$, where $\ell_\mu=[0\isep n-1]$ 
is an indicator function which maps the different possible values of the order parameter $\mu$ with their corresponding phase index $\ell_\mu$.
It is worth mentioning that this implies that, if the steady-state distribution of $\mu$ follows a large-deviation principle, $\bbraket{\mu}{R_{0,\text{D}}^\lambda} \asymp \text{e}^{-L F(\mu)}$, with $L$ being the system size and $F(\mu)$ the associated large deviation function, then the rest of gap closing reduced eigenvectors obey the following property for the statistically-relevant values of $\mu$
\begin{equation*}
    \bbraket{\mu}{R_{j,\text{D}}^\lambda} \asymp \phi_j^{-\ell_\mu} \text{e}^{-L F(\mu)}
    .
\end{equation*}
In the following sections we shall illustrate our main results by projecting the spectral information in the order-parameter reduced Hilbert space for three paradigmatic many-body systems exhibiting continuous DPTs.

\section{Dynamical criticality in the boundary-driven WASEP}
\label{secoWASEP}

\begin{figure}
\includegraphics[width=0.9\linewidth]{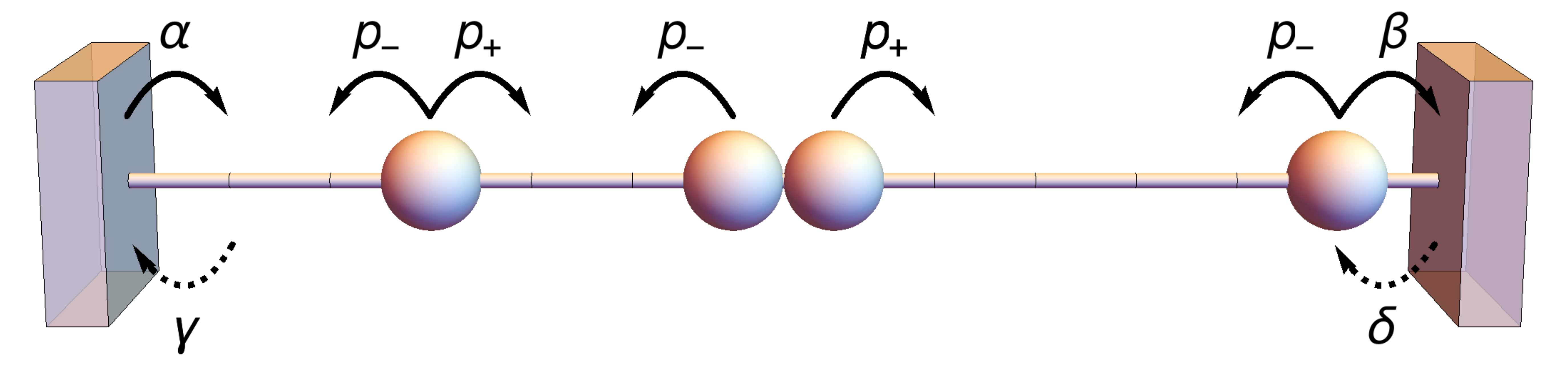}
\caption{{\bf Sketch of the boundary-driven WASEP.} $N$ particles in a lattice of $L$ sites which randomly jump to empty neighboring sites with asymmetric rates  $p_\pm$, and coupled to boundary reservoirs which inject and remove particles at different rates.
\label{fig:WASEP_def_open}}
\end{figure}

We shall first illustrate the ideas introduced above using the boundary-driven (or open) weakly asymmetric simple exclusion process (WASEP), which is an archetypical stochastic lattice gas modeling a driven diffusive system \cite{gartner87a, de-masi89a, bodineau05a}.
It consists of $N$ particles in a one-dimensional lattice of $\lattsize$ sites which can be either empty or occupied by one particle at most, so that the state $\conf$ of the system at any time is defined by the set of occupation numbers of all sites, $\conf = \{\occup_k\}_{k=1, ..., \lattsize}$, with $n_k=0,1$. Such state is encoded in a column vector $\ket{\conf} = \bigotimes^{\lattsize}_{k=1} (\occup_k, 1 - \occup_k)^{\transposesymb}$, where $\transposesymb$ denotes transposition, in a Hilbert space ${\cal H}$ of dimension $2^L$. Particles hop randomly to empty adjacent sites with asymmetric rates $\jumprateRL = \frac{1}{2} \text{e}^{\pm \field / \lattsize}$ to the right and left respectively, due to the action of an external field $\field$ applied to the particles of the system, see Fig.~\ref{fig:WASEP_def_open}. The ends of the lattice are connected to particle reservoirs which inject and remove particles with rates $\rateinL$, $\rateoutL$ respectively in the leftmost site, and rates $\rateinR$, $\rateoutR$ in the rightmost one. These rates are related to the densities of the boundary particle reservoirs as $\densL = \rateinL/(\rateinL + \rateoutL)$ and $\densR = \rateinR/(\rateinR + \rateoutR)$ \cite{derrida07a}. Overall, the combined action of the external field and the reservoirs drive the system into a nonequilibrium steady state with a net particle current.

At the macroscopic level, namely after a diffusive scaling of the spatiotemporal coordinates, the WASEP is described by the following diffusion equation for the particle density field $\rho(x,t)$ \cite{spohn12a},
\begin{equation}
\pdv{\rho}{t} = - \pdv{\left[-\diffusivity(\dens)\pdv{\dens}{x} + \mobility(\dens) \field \right]}{x}\, ,
\label{eq:WASEP_hydro}
\end{equation}
where $\diffusivity(\dens) = \frac{1}{2}$ is the diffusion transport coefficient and $\mobility(\dens) =  \dens (1 - \dens)$ is the mobility.

At the microscopic level, the stochastic generator of the dynamics reads
\begin{equation}
\begin{split}
\label{eq:WASEP_gen}
\genmat = \sum_{k=1}^{\lattsize-1} \Big[\;
          &\jumprateR \big( \creationop_{k+1} \destructop_{k} 
        - \occupop_{k  } (\identityop - \occupop_{k+1} )\big)
    \\ 
        + &\jumprateL (\creationop_{k  } \destructop_{k+1}
        - \occupop_{k+1} (\identityop - \occupop_{k  } )\big)\Big]
    \\
    + &\rateinL [ \creationop_{1} - (\identityop - \occupop_{1}) ]
        + \rateoutL [ \destructop_{1} - \occupop_{1} ]
    \\
    + &\rateinR [ \creationop_{\lattsize} - (\identityop - \occupop_{\lattsize}) ]
        + \rateoutR [ \destructop_{\lattsize} - \occupop_{\lattsize} ]
    ,
\end{split}
\end{equation}
where $\creationop_{k}$, $\destructop_{k}$ are respectively the creation and annihilation operators given by $\hat{\sigma}^{\pm}_k= (\hat{\sigma}^x_k \pm i \hat{\sigma}^y_k)/2  $, with $\hat{\sigma}^{x,y}_k$ the standard $x,y$-Pauli matrices, while $\occupop_{k}=\hat{\sigma}^+_k \hat{\sigma}^-_k$ and $\identityop_k$ are the occupation and identity operators acting on site $k$.
The first row in the r.h.s of the above equation corresponds to transitions where a particle on site $k$ jumps to the right with rate $p_+$, whereas the second one corresponds to the jumps from site $k+1$ to the left with rate $p_-$. The last two rows correspond to the injection and removal of particles at the left and right boundaries.

Interestingly, if the boundary rates are such that $\rateinL =  \rateoutR$ and $\rateoutL = \rateinR$, implying that $\densR = 1 - \densL$, the dynamics becomes invariant under a particle-hole (PH) transformation \cite{perez-espigares18a}, $\symmop_{\textrm{PH}}$, which thus commutes with the generator of the dynamics, $[\symmop_{\textrm{PH}},  \genmat] = 0$. This transformation simply amounts to changing the occupation of each site, $\occup_k \to 1 - \occup_k$, and inverting the spatial order, $k \to \lattsize -k +1$, and it is represented by the microscopic operator
\begin{equation}
\begin{split}
\symmop_{\textrm{PH}} = \prod_{k'=1}^{\lattsize} \hat{\sigma}^x_{k'} \prod_{k=1}^{\floor{\lattsize/2}} \Big[\creationop_{k}\destructop_{\lattsize-k+1} + \creationop_{\lattsize-k+1}\destructop_{k} \\
+ \frac{1}{2}(\hat{\sigma}^z_k \hat{\sigma}^z_{\lattsize-k+1} + \identityop)\Big] \, ,
\end{split}
\end{equation}
where $\floor{\cdot}$ is the floor function and $\hat{\sigma}^z_k = \identityop - 2\occupop_{k}$ is the $z$-Pauli matrix.
The operator in brackets exchanges the occupancy of sites $k$ and $\lattsize-k+1$. 
In particular, the first two terms act on particle-hole pairs, while the last one affects pairs with the same occupancy.
Notice that $\symmop_{\textrm{PH}}$ is a $\Zsymm{2}$ symmetry, since $(\symmop_{\textrm{PH}})^2 = \identityop$. Macroscopically the PH transformation means to change $\rho\to 1-\rho$ and $x\to 1-x$, which leaves invariant Eq. (\refeq{eq:WASEP_hydro}) due to the symmetry of the mobility $\sigma(\rho)$ around $\rho=1/2$ and the constant diffusion coefficient \cite{baek17a,baek18a,perez-espigares18a}.

A key observable in this model is the time-integrated and space-averaged current $\intcurrent$, and the corresponding time-intensive observable $\current=Q/\tau$. The current $Q$ is defined as the number of jumps to the right minus those to the left per bond (in the bulk) during a trajectory of duration $\tau$. For any given trajectory $\omega_\tau$, this observable remains invariant under the PH transformation, $Q(\mathcal{S}_{\textrm{PH}}\omega_\tau) = Q(\omega_\tau)$, since the change in the occupation and the inversion of the spatial order gives rise to a double change in the flux sign, leaving the total current invariant. In this way, the symmetry under the transformation $\symmop_{\textrm{PH}}$ will be inherited by the Doob driven process associated with the fluctuations of the current, see \S\ref{ssecSym}. Indeed, the current statistics and the corresponding driven process can be obtained by biasing or \emph{tilting} the original generator according to Eq.~\eqref{eq:genmatT},
\begin{equation}
\begin{split}
\genmatT = \sum_{k=1}^{\lattsize-1} \Big[\; &\jumprateR \big( \text{e}^{\currenttilt/(\lattsize - 1)}\creationop_{k+1} \destructop_{k} - \occupop_{k  } (\identityop_{k+1} - \occupop_{k+1} )\big) \\ 
+ &\jumprateL (\text{e}^{-\currenttilt/(\lattsize - 1)} \creationop_{k  } \destructop_{k+1} - \occupop_{k+1} (\identityop_{k} - \occupop_{k  } )\big)\Big] \\
+ &\rateinL [ \creationop_{1} - (\identityop_{1} - \occupop_{1}) ] + \rateoutL [ \destructop_{1} - \occupop_{1} ] \\
+ &\rateinR [ \creationop_{\lattsize} - (\identityop_{\lattsize} - \occupop_{\lattsize}) ] + \rateoutR [ \destructop_{\lattsize} - \occupop_{\lattsize} ] \, ,
\end{split}
\label{eq:WASEPopen_tilted_gen}
\end{equation}
where we have used that the contribution to the time-integrated current $\intcurrent$ of a particle jump in the transition $C\to C'$ is $\eta_{C,C'}=\pm 1/(\lattsize - 1)$, depending on the direction of the jump. 

\begin{figure}
\includegraphics[width=1.0\linewidth]{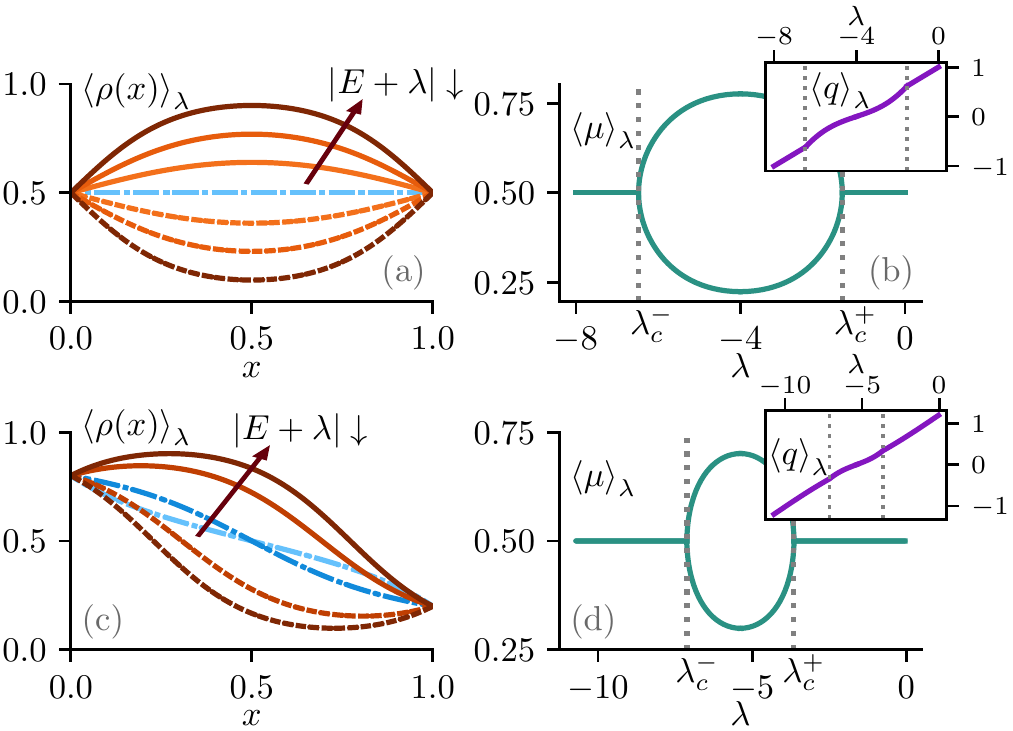}
\caption{{\bf Particle-hole dynamical symmetry breaking in the open WASEP.} Macroscopic Fluctuation Theory results for the DPT in current fluctuations observed in the boundary-driven WASEP for $\field = 4>E_c$ \cite{perez-espigares18a}. (a) Density profiles of the system for $\rho_L=\rho_R=0.5$ and different values of the biasing field $\currenttilt$. The flat, PH-symmetry-preserving profile (blue, dash-dotted line) is the optimal solution for moderate current fluctuations, corresponding to values above $\crit{\currenttilt}^+ = -1.52$, while the PH-symmetry-breaking density profiles (reddish, full and dashed lines) correspond to $\currenttilt = -1.6, -1.75, -2, -2.5$ and $\lambda=-4 =-E$. (b) Mean lattice occupation $\langle \mu\rangle_\lambda$ in terms of the biasing field $\lambda$ for $\rho_L=\rho_R=0.5$. The observed bifurcations at $\lambda_c^\pm$ signal the symmetry-breaking DPT. The inset shows the average current as a function of $\lambda$, which becomes nonlinear in the symmetry-breaking regime. Panels (c) and (d) are equivalent to (a) and (b), respectively, but for $\rho_L=0.8$ and $\rho_R=1-\rho_L=0.2$.}
\label{fig:DPT_open}
\end{figure}

Remarkably, when $\densR = 1 - \densL$ and the stochastic dynamics is hence invariant under $\symmop_{\textrm{PH}}$, the boundary-driven WASEP displays, for a sufficiently strong external field $E$, a second-order DPT for fluctuations of the particle current below a critical threshold. Such DPT, illustrated in Fig.~\ref{fig:DPT_open}, was predicted in \cite{baek17a,baek18a} and further explored in \cite{perez-espigares18a} from a macroscopic perspective. In order to sustain a long-time current fluctuation, the system adapts its density field so as to maximize the probability of such event, according to the MFT action functional \cite{bertini15a,baek17a,baek18a,perez-espigares18a}. For moderate current fluctuations, this optimal density profile might change, but retains the PH symmetry of the original action. However, for 
current fluctuations below a critical threshold in absolute value, the PH symmetry of the original action breaks down and two different but equally probable optimal density fields appear, both connected via the symmetry operator [see full and dashed reddish lines in Fig.~\ref{fig:DPT_open}(a) and Fig.~\ref{fig:DPT_open}(c)]. The emergence of these two action minimizers can be understood by noticing that, when $\densR = 1 - \densL$, either crowding the lattice with particles or depleting the particle population define two equally-optimal strategies to hinder particle motion, thus reducing the total current through the system \cite{baek17a,baek18a,perez-espigares18a}. More precisely, the DPT appears for an external field $|E|>E_c=\pi$ and current fluctuations such that $\abs{\current} \leq \currentcrit = \sqrt{\field^2 - \pi^2}/4$, which correspond to biasing fields in the range $\currenttiltcritmin < \currenttilt < \currenttiltcritmax$, with $\currenttiltcritminmax = -\field \pm \sqrt{\field^2 - \pi^2}$, see insets to Fig.~\ref{fig:DPT_open}(b) and Fig.~\ref{fig:DPT_open}(d). To characterize the phase transition, a suitable choice of the order parameter is the mean occupation of the lattice, defined as $\mass = \lattsize^{-1} \sum_{k=1}^{\lattsize} \occup_k$, with $\mu\in [0,1]$ since $n_k=0,1$ due to the exclusion rule. The behavior of this order parameter as a function of the biasing field is displayed in Fig.~\ref{fig:DPT_open}(b) and Fig.~\ref{fig:DPT_open}(d). 

\begin{figure}
\includegraphics[width=1\linewidth]{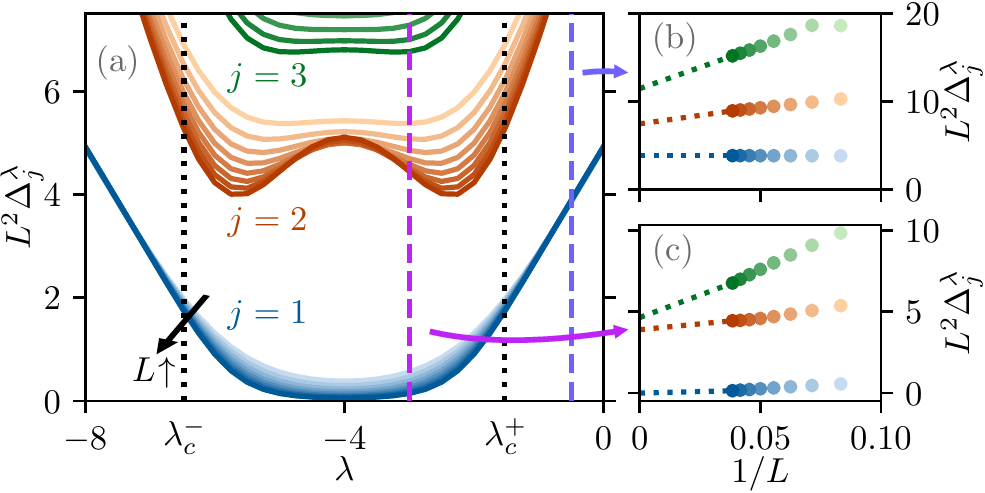}
\caption{ {\bf DPT and quasi-degeneracy in the open WASEP.} (a) Scaled spectral gaps, $L^2\Delta_j^{\lambda}$, with $j=1,2,3$, as a function of the biasing field $\currenttilt$ for different lattice sizes ($L= 12, 14, 16, 18, 20, 22, 24, 26$) in the open WASEP.  Each set of lines is associated with a different value of $j$, with darker colors corresponding to larger system sizes. The small panels show the scaled spectral gaps as a function of the inverse lattice size for (b) $\lambda=-0.5>\lambda_c^+$ and (c) $\lambda=-3.5<\lambda_c^+$. The spectral gap $L^2\Delta_1^{\lambda}$ vanishes as $L$ increases for $\lambda_c^-<\lambda<\lambda_c^+$, but remains finite outside this region.}
\label{fig:eigvalues_open}
\end{figure}
   
We now proceed to explore the spectral fingerprints of the phase transition for current fluctuations in the open WASEP, a $\mathbb{Z}_2$ symmetry-breaking DPT. In the following, we set $E=4$ and $\alpha=\beta=\gamma=\delta=0.5$, corresponding to equal densities $\rho_L=\rho_R=0.5$, though all our results apply also to arbitrary strong drivings as far as $\rho_L=1-\rho_R$ and $|E|>\pi$. As discussed in previous sections, the hallmark of any symmetry-breaking DPT is the emergence of a degenerate subspace for the leading eigenvectors of the Doob driven process. Our first goal is hence to analyze the scaled spectral gaps $L^2 \Delta_j^{\lambda}$ associated with the first eigenvalues of the Doob generator $\genmatD$ for the boundary-driven WASEP. 
Note that the $L^2$ scaling in the spectral gaps is required because the system dynamics is diffusive \cite{derrida07a,perez-espigares18a}. These spectral gaps, obtained from the numerical diagonalization of the tilted generator in Eq.~\eqref{eq:WASEPopen_tilted_gen}, are displayed in Fig.~\ref{fig:eigvalues_open}(a) as a function of $\lambda$ for different system sizes. Recall that by definition $\Delta_0^{\lambda}(\lattsize)=0$ $\forall \lambda$ since $\genmatD$ is a probability-conserving stochastic generator. Moreover, we observe that $L^2 \Delta_{2,3}^{\lambda}(L)>0$ for all $\lambda$ and $L$, so their associated eigenvectors do not contribute to the Doob stationary subspace. On the other hand, $L^2 \Delta_1^{\lambda}(L)$ exhibits a more intricate behavior.
In particular, for $\lambda_c^-< \lambda < \lambda_c^+$ this spectral gap vanishes as $L$ increases, 
signaling that the DPT has already kicked in, while outside of this range $L^2 \Delta_1^{\lambda}(L)$ converges to a non-zero value as $L$ increases.
Note however that the change in the spectral gap behavior across $\lambda_c^{\pm}$ is not so apparent due to the moderate system sizes at reach with numerical diagonalization. In any case, these two markedly different behaviors are more clearly appreciated in Figs.~\ref{fig:eigvalues_open}(b)-\ref{fig:eigvalues_open}(c), which display $L^2\Delta_j^{\lambda}$ for $j=1,2,3$ as a function of $1/L$ for $\lambda=-0.5>\lambda_c^+$ (top panel) and $\lambda=-3.5<\lambda_c^+$ (bottom panel). In particular, a clear decay of $L^2\Delta_1^{\lambda}(L)$ to zero as $1/L\to 0$ is apparent in Fig~\ref{fig:eigvalues_open}(c), while $L^2 \Delta_{2,3}^{\lambda}(L)$ remain non-zero in this limit. In this way, outside the critical region, i.e. for $\lambda > \lambda_c^+$ or $\lambda < \lambda_c^-$, the Doob steady state is unique and given by the first Doob eigenvector, $\probTvectst = \eigvectDR{0}$, as shown in Eq.~\eqref{ssnopt}. This Doob steady state preserves the PH symmetry of the original dynamics. On the other hand, for $\lambda_c^-< \lambda < \lambda_c^+$, the spectral gap vanishes in the asymptotic $L\to\infty$ limit. As a consequence, the second eigenvector of $\genmatD$, $\eigvectDR{1}$, enters the degenerate subspace so that the Doob steady state is now doubly degenerated in the thermodynamic limit, and given by $\probTvectst = \eigvectDR{0} + \eigvectDR{1} \braket{\eigvectDLnk{1}}{\probvecttimenk{0}}$, see Eq.~\eqref{eq:DPT_probvect}, thus breaking spontaneously the PH symmetry of the original dynamics. Note that this is exact in the macroscopic limit, while for finite sizes it is just an approximation valid on timescales $1/\Delta_2^\lambda(L) \ll t \ll 1/\Delta_1^\lambda(L)$, the long time limit being always $\probTvectst \approx  \eigvectDR{0}$ for finite system sizes.

\begin{figure}
\includegraphics[width=1\linewidth]{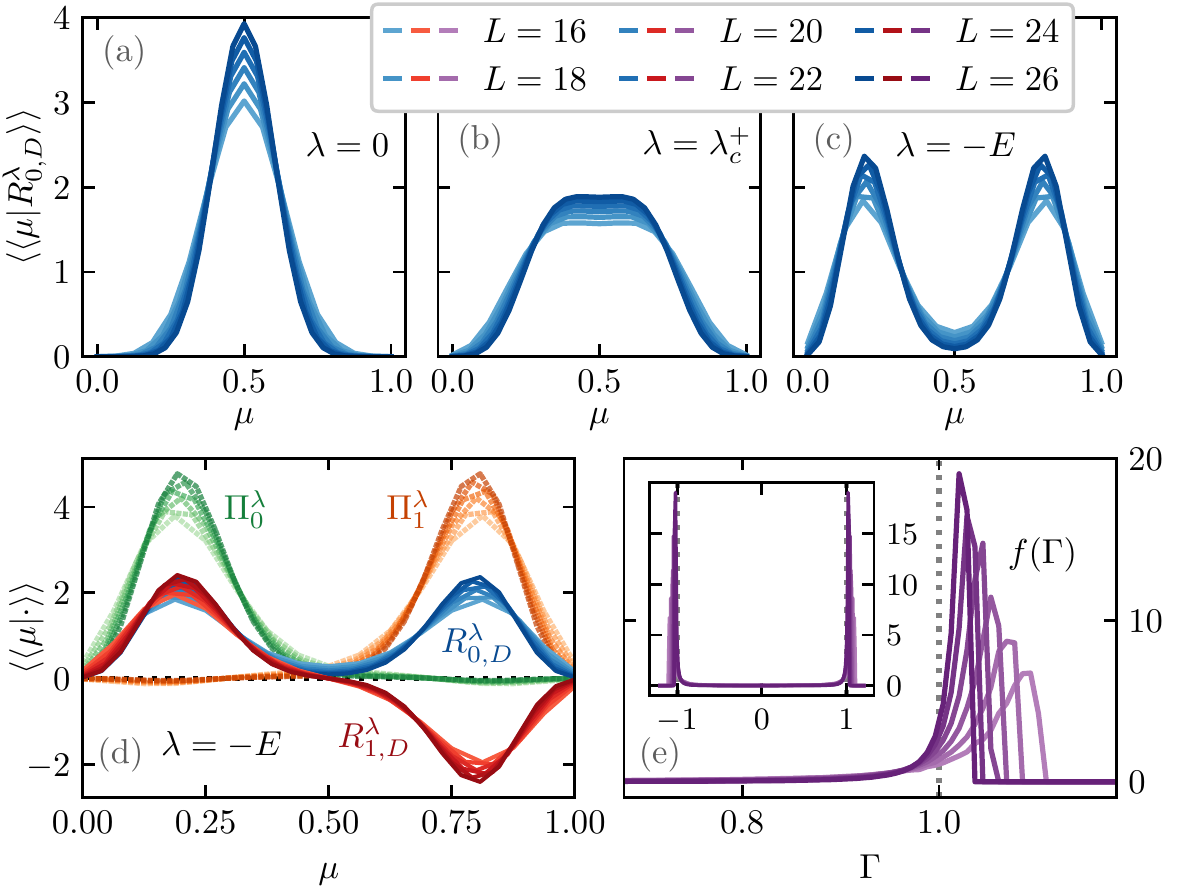}
\caption{{\bf Structure of the degenerate subspace in the open WASEP.}
The top panels show the structure of the leading reduced Doob eigenvector $\bbraket{\mu}{R_{0,\text{D}}^\lambda}$ for different values of $\currenttilt$ across the DPT and varying $L$. From left to right: (a) $\currenttilt = -0.5$ (symmetry-preserving phase, before the DPT), (b) $\currenttilt = \lambda_c^+=-1.52$ (critical), (c) $\currenttilt = -4$ (symmetry-broken phase, after the DPT). (d) Comparative structure of the first and second reduced Doob eigenvectors in the degenerate subspace,  $\bbraket{\mu}{R_{j,\text{D}}^\lambda}$ with $j=0, 1$, in the symmetry-broken phase ($\currenttilt = -4$, full lines). The structure of the resulting reduced phase probability vectors $\bbraket{\mu}{\Pi_l^\lambda}$, $l=0,1$, is also shown (dotted lines). (e) Histogram for $\Gamma=\braket{C}{\eigvectDRnk{1}}/\braket{C}{\eigvectDRnk{0}}$ obtained for a large set of configurations $\ket{C}$ sampled from the Doob steady-state distribution for $\lambda = -4$.}
\label{fig:eigvect_open}
\end{figure}
       
In order to analyze the structure of eigenvectors contributing to the Doob steady state as predicted in Section \S\ref{secDPTsymm}, we now turn to the order parameter space and the reduced vectors defined in Eq.~\eqref{Pconserv3}, using as order parameter the mean occupation of the lattice $\mass = \lattsize^{-1} \sum_{k=1}^{\lattsize} \occup_k$. This is a good order parameter as defined by its behavior under the symmetry operator $\symmop_{\textrm{PH}}$, see discussion in \S\ref{ssecOparam}. In this way we extract the relevant macroscopic information contained in the leading Doob eigenvectors, which is displayed in Fig.~\ref{fig:eigvect_open} for different values of the biasing field $\lambda$. In particular, Figs.~\ref{fig:eigvect_open}(a)-\ref{fig:eigvect_open}(c) show the order parameter structure of the leading reduced Doob eigenvector $\kket{R_{0,\text{D}}^\lambda}$ before the DPT [$\lambda=0$, Fig.~\ref{fig:eigvect_open}(a)], at the critical point [$\lambda=\lambda_c^+$, Fig.~\ref{fig:eigvect_open}(b)], and after the DPT [$\lambda=-E$, Fig.~\ref{fig:eigvect_open}(c)]. These panels fully confirm the predictions of section \S\ref{secDPTsymm}. In particular, before the DPT happens there is only a single phase contributing to the Doob steady state, $\kket{R_{0,\text{D}}^\lambda} = \kket{\Pi_0^\lambda}$, which preserves the $\mathbb{Z}_2$ symmetry of the original dynamics. This is reflected in the unimodality of the distribution $\bbraket{\mu}{R_{0,\text{D}}^\lambda}$ shown in Fig.~\ref{fig:eigvect_open}(a) for different system sizes. Indeed, in this phase $\bbraket{\mu}{R_{0,\text{D}}^\lambda}$ is nothing but the steady state probability distribution of the order parameter $\mu$, see Eq.~\eqref{Pconserv}. Upon approaching the critical point $\lambda=\lambda_c^+$, the distribution $\bbraket{\mu}{R_{0,\text{D}}^\lambda}$ is still unimodal but becomes flat around the peak (i.e. with zero second derivative), see Fig.~\ref{fig:eigvect_open}(b), a feature very much reminiscent of second-order, $\mathbb{Z}_2$ symmetry-breaking phase transitions \cite{binney92a}. In fact, once $\currenttilt$ enters the symmetry-broken region, the distribution $\bbraket{\mu}{R_{0,\text{D}}^\lambda}$ becomes bimodal as shown in Fig.~\ref{fig:eigvect_open}(c), where two different but symmetric peaks around $\mu \approx 0.25$ and $\mu \approx 0.75$, respectively, can be distinguished. The leading reduced Doob eigenvector is still invariant under the symmetry operation, i.e. $\bbraket{\mu}{R_{0,\text{D}}^\lambda}=\bbra{\mu}\hat{S}_\mu\kket{R_{0,\text{D}}^\lambda}$ (recall that $\mass \to 1 -\mass$ under such transformation), but the degenerate subspace also includes now (in the $L\to\infty$ limit) the second reduced Doob eigenvector $\kket{R_{1,\text{D}}^\lambda}$, whose order parameter structure is compared to that of $\kket{R_{0,\text{D}}^\lambda}$ in Fig.~\ref{fig:eigvect_open}(d). Clearly, while $\kket{R_{0,\text{D}}^\lambda}$ is invariant under $\hat{S}_\mu$ as stated, i.e. it has a symmetry eigenvalue $\phi_0=1$, $\kket{R_{1,\text{D}}^\lambda}$ is on the other hand antisymmetric, $\hat{S}_\mu\kket{R_{1,\text{D}}^\lambda}=-\kket{R_{1,\text{D}}^\lambda}$, i.e. $\phi_{1} = \text{e}^{i\pi}=-1$. This result can be also confirmed numerically for the unreduced eigenvectors $\ket{R_{0,\text{D}}^\lambda}$ and $\ket{R_{1,\text{D}}^\lambda}$ in the space of configurations. The reduced phase probability vectors can be written according to Eq.~\eqref{redeq3} as $\kket{\Pi_l^\lambda} = \kket{R_{0,\text{D}}^\lambda} + (-1)^l \kket{R_{1,\text{D}}^\lambda}$, with $l=0,1$, and they simply correspond to the degenerate reduced Doob steady states in each of the symmetry branches, see Fig.~\ref{fig:eigvect_open}(d). Such distributions correspond to each of the symmetry-broken profiles previously shown in Fig.~\ref{fig:DPT_open}(a) for $\lambda=-E$. In general, the reduced Doob steady state in the $L\to\infty$ limit will be a weighted superposition of these two degenerate branches, 
\be
\kket{P_{\text{ss},P_0}^\lambda}  = w_0 \kket{\Pi_0^\lambda} + w_1 \kket{\Pi_1^\lambda} \, , \nonumber
\ee
see Eq.~\eqref{coef} and \S\ref{ssecOparam}, with weights $w_l=(1 + (-1)^l \braket{\eigvectDLnk{1}}{\probvecttimenk{0}})/2$ depending on the initial state distribution $\ket{P_0}$. This illustrates as well the phase selection mechanism via initial state preparation discussed in \S\ref{ssecPhase}: choosing $\ket{\probvecttimenk{0}}$ such that $\braket{\eigvectDLnk{1}}{\probvecttimenk{0}}=(-1)^l$ leads to a \emph{pure} steady state $\probTvectst = \probDphasevect{l}$.

To end this section, we want to test the detailed structure imposed by symmetry on the Doob degenerate subspace. In particular we showed in \S\ref{ssecSpec} that, in the symmetry-broken regime, the components $\braket{C}{\eigvectDRnk{j}}$ of the subleading Doob eigenvectors $j\in[1,n-1]$ associated with statistically-relevant configurations $\ket{C}$ are almost equal to those of the leading eigenvector, $\braket{C}{\eigvectDRnk{0}}$, except for a complex phase, see Eq.~\eqref{eq:eigvectsfromeigvect0}. This will be true provided that $\ket{C}$ belongs to a given symmetry class $\ell_C$. For the open WASEP, we have that
\be
\braket{C}{\eigvectDRnk{1}} \approx (-1)^{-\ell_C} \braket{C}{\eigvectDRnk{0}} \, , \nonumber
\label{eq:eigvectsfromeigvect0v2}
\ee
where $\ell_C=0,1$ depending whether configuration $\ket{C}$ belongs to the high-$\mu$ or low-$\mu$ symmetry sector, respectively. To test this prediction, we sample a large number of statistically-relevant configurations in the Doob steady state, and study the histogram for the quotient $\Gamma(\conf) = \braket{\conf}{\eigvectDRnk{1}} / \braket{C}{\eigvectDRnk{0}}$, see Fig.~\ref{fig:eigvect_open}(e). As expected from the previous equation, the frequencies $f(\Gamma)$ peak sharply around $(\phi_1)^0=1$ and $\phi_1=-1$, see also inset to Fig.~\ref{fig:eigvect_open}(e), and concentrate around these values as $\lattsize$ increases. This confirms that the structure of the subleading Doob eigenvector $\ket{R_{1,\text{D}}}$ is enslaved to that of $\ket{R_{0,\text{D}}}$ depending on the symmetry basin of each configuration. Moreover, this observation also supports \emph{a posteriori} 
that statistically-relevant configurations can be partitioned into disjoint symmetry classes. In the reduced order-parameter space, the relation between eigenvectors in the degenerate subspace implies that $\bbraket{\mu}{R_{1,\text{D}}^\lambda} \approx (-1)^{-\ell_\mu} \bbraket{\mu}{R_{0,\text{D}}^\lambda}$, where $\ell_\mu=\Theta(\mu-0.5)$ is an indicator function identifying each phase in $\mu$-line, with $\Theta(x)$ the Heaviside step function. In this way, the magnitude and shape of the peaks of $\bbraket{\mu}{R_{1,\text{D}}^\lambda}$ are directly related to those of $\bbraket{\mu}{R_{0,\text{D}}^\lambda}$, despite their antisymmetric (resp. symmetric) behavior under $\hat{S}_\mu$, as corroborated in Fig.~\ref{fig:eigvect_open}(d).

We end by noting that, despite the moderate lattice sizes at reach with numerical diagonalization, the results presented above for the boundary-driven WASEP show an outstanding agreement with the macroscopic predictions of \S\ref{secDPTsymm}.

\section{Energy fluctuations in spin systems: the $r$-state Potts model}
\label{secPotts}

\begin{figure}
\includegraphics[width=1\linewidth]{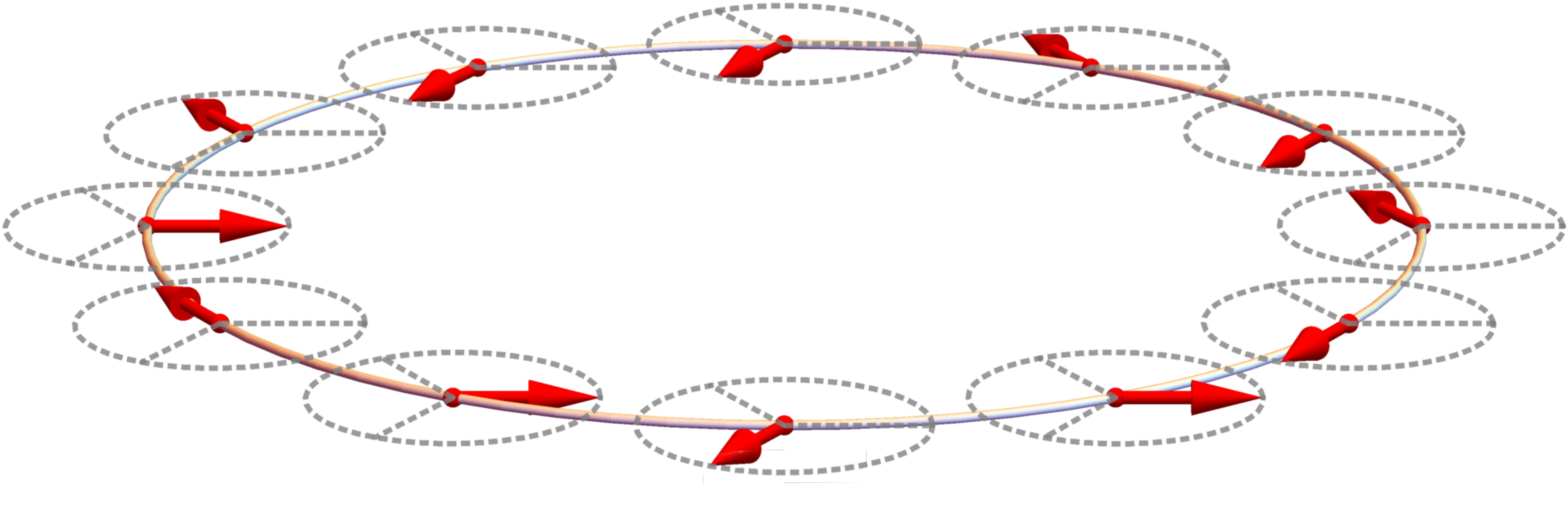}
\caption{{\bf Sketch of the 3-state Potts model.} Each lattice site contains a spin variable with 3 possible in-plane orientations, and neighboring spins interact depending on their relative orientations as described by the Hamiltonian~\eqref{HPott}.}
\label{fig:sketch_potts}
\end{figure}

The next example is the one-dimensional Potts model of ferromagnets \cite{potts52a}, a generalization of the Ising model. The system consists of a 1D periodic lattice with $\lattsize$ spins $\{s_k\}_{k=1,...,L}$, which can be in any of $\nstatesP$ different states ${s}_k \in \{0, 1, ..., \nstatesP-1\}$ distributed in a unit circle with angles $\spinangle_k = 2\pi s_k /\nstatesP$, as sketched in Fig.~\ref{fig:sketch_potts} for the particular case $r=3$. Nearest-neighbor spins interact according to the Hamiltonian
\begin{equation}
\hamilt = -J \sum_{k=1}^{\lattsize} \cos(\spinangle_{k+1} - \spinangle_k) \, ,
\label{HPott}
\end{equation}    
with $J>0$ a coupling constant favouring the parallel orientation of neighboring spins. Configurations $\{C\}=\{s_k\}_{k=1,...,L}$ can be represented as vectors in a Hilbert space ${\cal H}$ of dimension $r^L$,    
\be
\ket{C}=\bigotimes_{k=1}^L \left( \delta_{s_k,0},\delta_{s_k,1},...,\delta_{s_k,r-1} \right)^T\, , \nonumber
\ee
such that $s_k=0$, $s_k=1$, $\ldots$ , $s_k=r-1$ correspond to $\ket{0}_k=(1,0,\ldots,0)$, $\ket{1}_k=(0,1,\ldots,0)$, $\ldots$, $\ket{r-1}_k=(0,0,\ldots,1)$, respectively. Spins evolve stochastically in time according to the single spin-flip Glauber dynamics at inverse temperature $\beta$ \cite{glauber63a}. The stochastic generator $\genmat$ for this model can be hence obtained from the Hamiltonian as $\bra{C'}\genmat\ket{C}=\transitionrate{\conf}{\conf^{\prime}} = 1/\left(1 + \text{e}^{\invtemp \Delta E_{\conf', \conf}}\right)$, where $\Delta E_{\conf', \conf}$ is the energy change in the transition $C\to C'$, which involves a spin rotation. The explicit operator form for $\genmat$ can be then easily obtained from these considerations, but is somewhat cumbersome (see e.g. Appendix B of \cite{marcantoni20a} for an explicit expression of $\genmat$ in the case $\nstatesP=3$). 

Interestingly, the Hamiltonian~\eqref{HPott} is invariant under any global rotation of all the spins for angles multiple of $2\pi/r$. For convenience we define an elementary rotation of angle $2\pi/r$, which transforms $\ket{s}_k$ into $\ket{s+1}_k$ for every spin $k$, with the operator
\begin{equation}
\symmopR = \bigotimes_{k=1}^L \sum_{s=0}^{\nstatesP-1} \left(\ket{s+1}\bra{s}\right)_k \, ,
\label{symmopR}
\end{equation}
where we identify $\ket{\nstatesP}_k = \ket{0}_k$. Note that 
\be
(\symmopR)^r=\hat{S}_{2\pi}=\identityop \, .
\label{symmopR2}
\ee
Since the Glauber dynamics inherits the symmetries of the Hamiltonian, the generator is also invariant under the action of the rotation operator $\symmopR$, i.e.~$[\genmat, \symmopR]=0$. This hence implies that $\genmat$ has a $\Zsymm{r}$ symmetry in the language of \S\ref{ssecSym}, see Eq.~\eqref{symmopR2}, and makes the $r$-state Potts model a suitable candidate for illustrating our results beyond the $\Zsymm{2}$ symmetry-breaking phenomenon of the previous section, provided that this model exhibits a DPT in its fluctuation behavior.

It is well known that the 1D Potts model does not present any standard phase transition for finite values of $\invtemp$ \cite{baxter16a}. However, we shall show below that it does exhibit a paramagnetic-ferromagnetic DPT for $r=3$ and $r=4$ when trajectories are conditioned to sustain a fluctuation of the time-averaged energy per spin well below its typical value. Indeed, in order to sustain such a low energy fluctuation, the $r$-spin system eventually develops ferromagnetic order so as to maximize the probability of such event, aligning a macroscopic fraction of spins in the same (arbitrary) direction and thus breaking spontaneously the underlying $\Zsymm{\nstatesP}$ symmetry. This symmetry breaking process in energy fluctuations leads to $\nstatesP$ different ferromagnetic phases, each one corresponding to one of the $\nstatesP$ possible spin orientations.
This DPT is well captured by the average magnetization per spin $\meanmag = L^{-1}\sum_{k=1}^{\lattsize} \text{e}^{i\spinangle_k}$, a complex number which plays the role of the order parameter in this case.  
Note that a similar DPT has been reported for the 1D Ising model \cite{jack2010}, which can be seen as a  $\nstatesP = 2$ Potts model.

Notice that, apart from the higher order $\Zsymm{\nstatesP}$ symmetry-breaking process involved in this DPT, a crucial difference with the one observed in the open WASEP is that here the observable whose fluctuations we are interested in (i.e. the energy) is \emph{configuration-dependent}, as opposed to the particle current in WASEP, which depends on state transitions [see Eq.~\eqref{eq:current_def} and related discussion]. Note also that, as in \cite{jack2010}, temperature does not play a crucial role in this DPT. In particular, a change in the temperature just amounts to a modification of the critical bias $\lambda_c$ at which the DPT occurs, which becomes more negative as the temperature increases. Since the aim of this work is not to characterize in detail the DPT but to analyze the spectral fingerprints of the symmetry-breaking process, we consider $\invtemp = 1$ in what follows without loss of generality.

\begin{figure}
\includegraphics[width=1\linewidth]{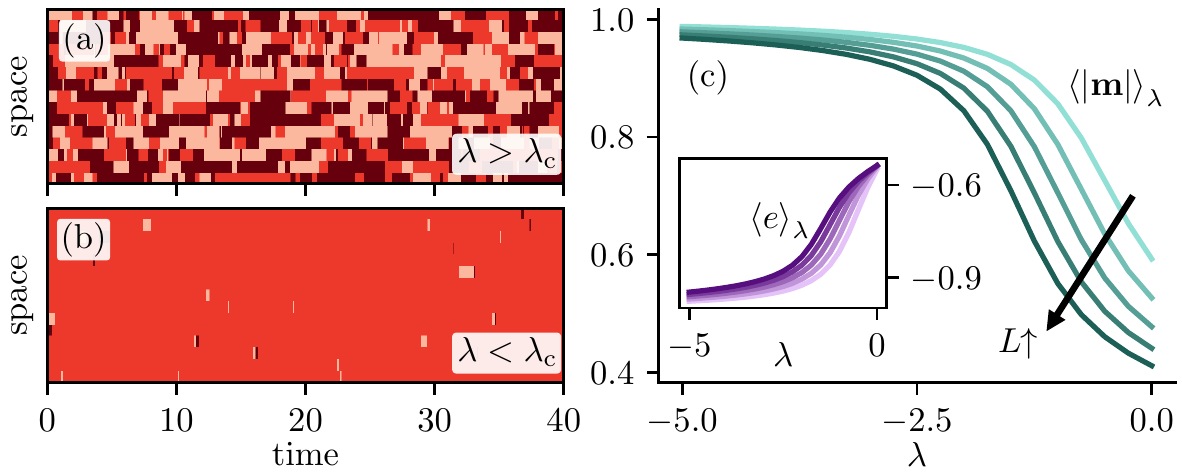}
\caption{{\bf $\mathbb{Z}_3$ dynamical symmetry breaking in the 3-state Potts model.} Left panels: Spatio-temporal raster plots of typical trajectories of the spin system (a) before the DPT ($\lambda>\lambda_c\approx -1$), and (b) once the DPT kicks in ($\lambda<\lambda_c$) for $\lattsize=16$ and $\beta=1$. Each color corresponds to one of the three posible spin states. (c) Magnitude of magnetization  as a function of the biasing field $\energytilt$ for increasing $L$ in the Doob stationary regime. The inset shows the average energy per spin vs $\lambda$. Each color represents a different $\lattsize = 8, 10, 12, 14, 16$, with darker colors corresponding to larger sizes.}
\label{fig:eigvals_Pottsq3}
\end{figure}

The statistics of the biased trajectories can be obtained from the tilted generator, see Eq.~\eqref{eq:genmatT}, which now reads 
\begin{equation}
\genmatT = \genmat + \lambda \sum_{\conf}e(C) \ketbra{\conf} \, ,
\label{genmatT}
\end{equation}
where $e(C)=\hamilt(\conf)/\lattsize$ is the energy per spin in configuration $\conf$, see Eq.~\eqref{HPott}. We start by analyzing the 3-state model, and summarize the results of the 4-state model at the end of this section. The main features of the Potts DPT are illustrated in Fig.~\ref{fig:eigvals_Pottsq3}. In particular, Figs.~\ref{fig:eigvals_Pottsq3}(a)-\ref{fig:eigvals_Pottsq3}(b) show two typical trajectories for different values of the biasing field $\lambda$. These trajectories are obtained using the Doob stochastic generator for each $\lambda$, see Eq.~\eqref{genmatD}. Interestingly, typical trajectories for moderate energy fluctuations [Fig.~\ref{fig:eigvals_Pottsq3}(a), $\lambda>\lambda_c\approx -1$] are disordered, paramagnetic while, for energy fluctuations well below the average, trajectories exhibit clear ferromagnetic order, breaking spontaneously the $\mathbb{Z}_3$ symmetry [Fig.~\ref{fig:eigvals_Pottsq3}(b), $\lambda<\lambda_c$]. The emergence of this ferromagnetic dynamical phase is well captured by the magnetization order parameter. Fig.~\ref{fig:eigvals_Pottsq3}(c) shows the average magnitude of the magnetization $\avgT{|\meanmag|}$ as a function of $\lambda$, while the inset shows the behavior of the average energy per spin, $\avgT{e}$. These observables are calculated from the Doob stationary distribution. As expected, the energy decreases as $\energytilt$ becomes more negative, while the magnitude of the magnetization order parameter increases sharply, the more the larger $L$ is. The presence of a DPT around $\energytilt =\lambda_c \approx -1$ is apparent, although the system sizes at reach via numerical diagonalization (recall that the generator is a $3^L\times 3^L$ matrix) do not allow for a more precise determination of the critical threshold.

\begin{figure}
\includegraphics[width=1\linewidth]{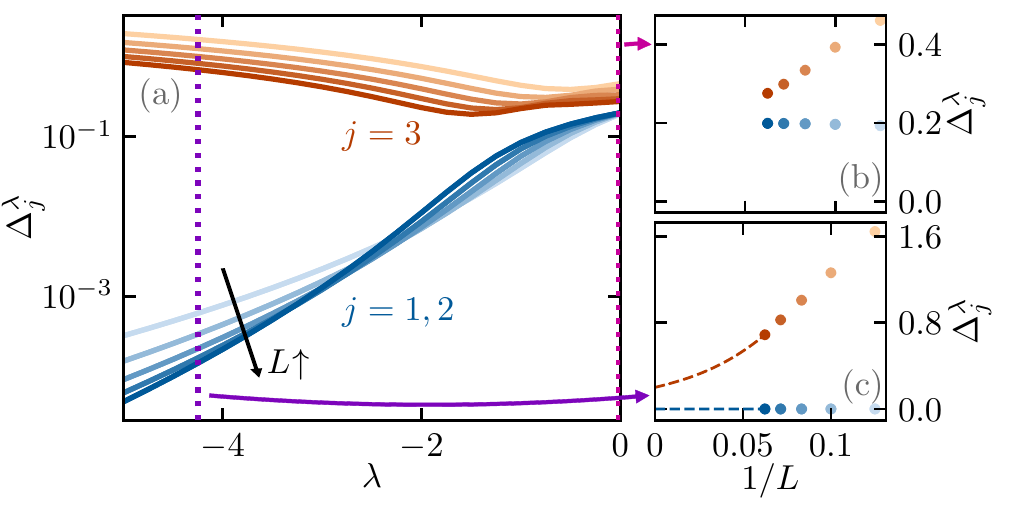}
\caption{{\bf DPT and quasi-degeneracy in the 3-state Potts model.} (a) Evolution of the three leading spectral gaps $\Delta_j^\lambda$, $j=1,2,3$, with the biasing field $\energytilt$ for different lattice sizes $\lattsize$. Note that $\Delta_1^{\lambda}=\Delta_2^{\lambda}$ $\forall \lambda,L$. The right panels show the spectral gaps as a function of the inverse lattice size for (b) $\lambda=0>\lambda_c$ and (c) $\lambda=-4.25<\lambda_c$. For $\lambda>\lambda_c$ the system remains gapped $\forall L$ so the Doob steady state is unique and no symmetry breaking happens. On the other hand, for $\lambda<\lambda_c$ the (equal) spectral gaps $\Delta_{1,2}^{\lambda}$ vanish as $L\to \infty$, leading to a $\mathbb{Z}_3$ dynamical symmetry breaking. Blue symbols correspond to eigenvectors $j=1,2$ and orange symbols to $j=3$, while the dashed lines display the expected behavior. The lattices sizes used are $\lattsize = 8, 10, 12, 14, 16$ (ordered by increasing color intensity).}
\label{fig:gap_Pottsq3}
\end{figure}

As explained before, the alignment of a macroscopic fraction of spins along a preferential (but arbitrary) direction breaks spontaneously the underlying $\mathbb{Z}_3$ symmetry. We hence expect this DPT to be accompanied by the appearance of a degenerate subspace spanned by the three leading Doob eigenvectors, and the corresponding decay of the second and third spectral gaps $\Delta_j^\lambda$ in the thermodynamic limit ($j=1, 2$, recall that $\Delta_0^\lambda=0$ $\forall \lambda$). This is confirmed in Fig.~\ref{fig:gap_Pottsq3}, which explores the spectral signatures of the Potts DPT. In particular, Fig.~\ref{fig:gap_Pottsq3}(a) shows the evolution with $\lambda$ of the three leading spectral gaps $\Delta_j^\lambda$, $j=1,2,3$, for different system sizes. As expected, while the system remains gapped for $\lambda>\lambda_c$ $\forall L$ [see Fig.~\ref{fig:gap_Pottsq3}(b)], once the DPT kicks in ($\lambda<\lambda_c$) the spectral gaps $\Delta_{1,2}^{\lambda}$ vanish as $L\to \infty$, as confirmed in Fig.~\ref{fig:gap_Pottsq3}(c).
On the other hand, $\Delta_3^{\lambda}$ is expected to remain non-zero, although this is not evident in Fig.~\ref{fig:gap_Pottsq3}(c) due to the limited system sizes at our reach.
Note that  $\Delta_1^{\lambda}=\Delta_2^{\lambda}$ $\forall \lambda$ since $\Delta_j^\lambda = -\Re(\eigvalD{j})$ and $\eigvalD{1}=(\eigvalD{2})^*$ (and similarly for eigenvectors, $\eigvectDR{1} = \eigvectDR{2}^{*}$), as complex eigenvalues and eigenvectors of real matrices such as $\genmatD$ and $\hat{S}_{\frac{2\pi}{3}}$ come in complex-conjugate pairs.
In this case only the eigenvectors are complex, the second and third eigenvalues of $\genmatD$ are real and therefore equal, $\eigvalD{1}=\eigvalD{2}$.
 In fact the corresponding eigenvalues of the symmetry operator $\hat{S}_{\frac{2\pi}{3}}$ are $\symmeigval{0}=1$, $\symmeigval{1}=\text{e}^{i2\pi/3}$, and $\symmeigval{2}=\text{e}^{-i2\pi/3}$.
 Therefore, for $\lambda>\lambda_c$, where the spectrum is gapped, the resulting Doob steady state will be unique as given by the leading Doob eigenvector, $\probTvectst = \eigvectDR{0}$, which remains invariant under $\hat{S}_{\frac{2\pi}{3}}$. 
On the other hand, for $\lambda<\lambda_c$ the two subleading spectral gaps $\Delta_{1,2}^{\lambda}$ vanish as $L\to\infty$, so the Doob stationary subspace is three-fold degenerated in the thermodynamic limit, and the Doob steady state depends on the projection of the initial state along the eigen-directions of the degenerate subspace,
\begin{equation}
\begin{split}
\probTvectst = \eigvectDR{0} &+ \eigvectDR{1} \braket{\eigvectDLnk{1}}{\probvecttimenk{0}} \\
&+ \eigvectDR{2} \braket{\eigvectDLnk{2}}{\probvecttimenk{0}} \, .
\end{split}
\nonumber
\end{equation}
Since we also have that $\eigvectDL{2} = \eigvectDL{1}^{*}$, the above expression can be rewritten as 
\be
\probTvectst = \eigvectDR{0} + 2\Re\left[\eigvectDR{1} \braket{\eigvectDLnk{1}}{\probvecttimenk{0}}\right] \, ,
\label{probPotts}
\ee    
that is, the Doob steady state in the thermodynamic limit is completely specified by the magnitude and complex argument of $\braket{\eigvectDLnk{1}}{\probvecttimenk{0}}$. This steady state for $\lambda<\lambda_c$ breaks the $\mathbb{Z}_3$ symmetry of the spin dynamics since $\hat{S}_{\frac{2\pi}{3}} \probTvectst \ne \probTvectst$. 

\begin{figure}
\includegraphics[width=1\linewidth]{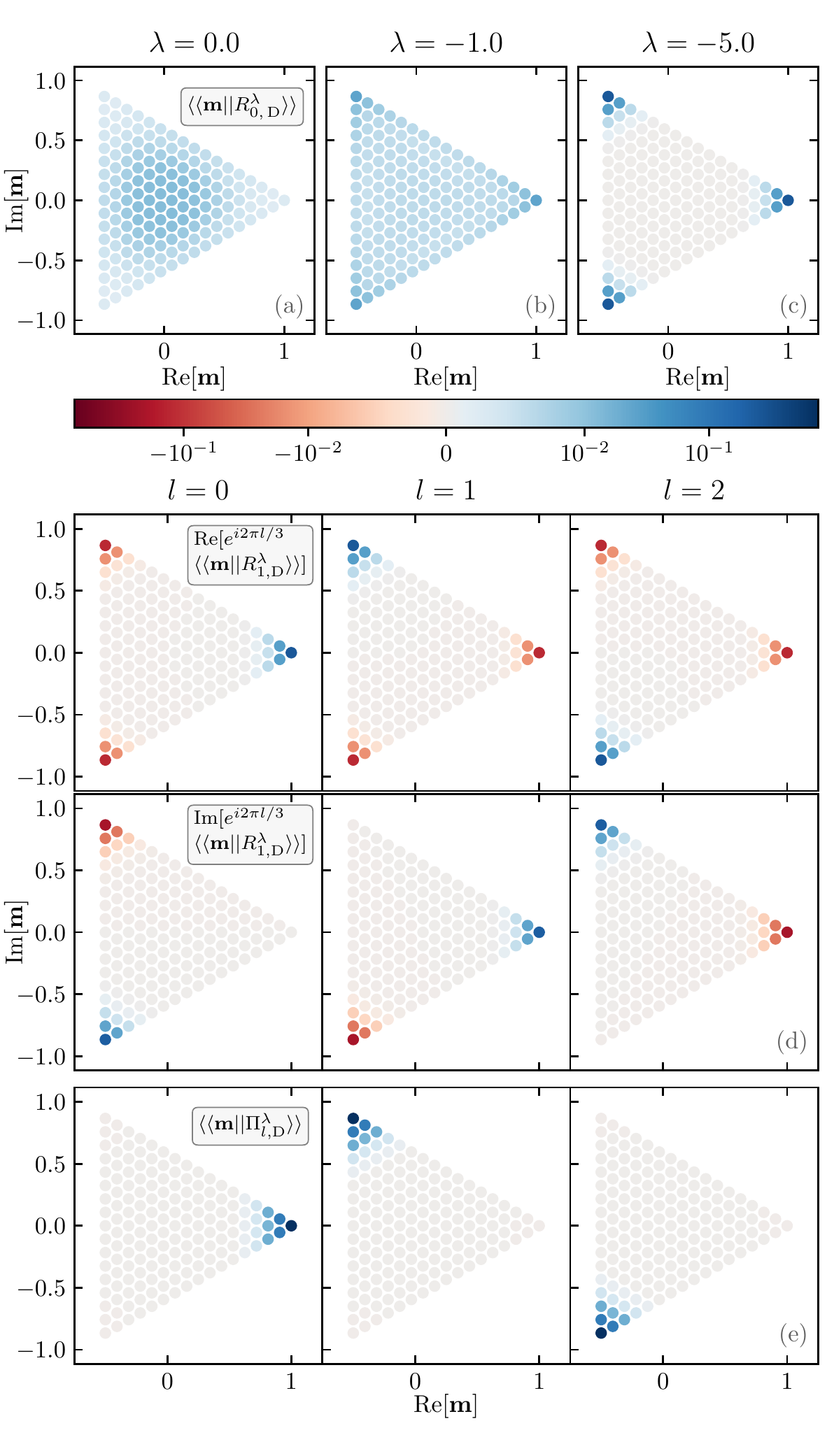}
\caption{{\bf Quasi-degenerate reduced eigenvectors in the 3-state Potts model.} (a)-(c) Structure of $\bbraket{\meanmag}{R_{0,\text{D}}^\lambda}$ in the complex $\meanmag$-plane for different values of $\currenttilt$ across the DPT and $L=16$. From left to right: (a) $\currenttilt = 0$ (symmetry-preserving phase, before the DPT), (b) $\currenttilt = -1 \approx \lambda_c$, (c) $\currenttilt = -5$ (symmetry-broken phase, after the DPT). 
(d) Structure of $\bbraket{\meanmag}{R_{1,\text{D}}^\lambda}$. The top panels display $\Re[\text{e}^{i 2\pi l/3}\bbraket{\meanmag}{R_{1,\text{D}}^\lambda}]$, while the mid panels show $\Im[\text{e}^{i 2\pi l/3}\bbraket{\meanmag}{R_{1,\text{D}}^\lambda}]$ for $l=0,1,2$. This enables to illustrate the phase selection mechanism of Eq.~\eqref{redeq3}. (e) Structure of the resulting reduced phase probability vectors $\bbraket{\meanmag}{\Pi_l^\lambda}$.}
\label{fig:Potts3_eigvect}
\end{figure}

Again, as in the example for the boundary-driven WASEP, we now turn to the reduced magnetization Hilbert space to analyze the structure of the eigenvectors spanning the Doob stationary subspace. In particular, Fig.~\ref{fig:Potts3_eigvect} shows the structure of the leading reduced Doob eigenvector $\kket{R_{0,\text{D}}^\lambda}$ in terms of the (complex) magnetization order parameter $\meanmag = L^{-1}\sum_{k=1}^{\lattsize} \text{e}^{i\spinangle_k}$ before the DPT [$\lambda=0>\lambda_c$, Fig.~\ref{fig:Potts3_eigvect}(a)], around the critical point [$\lambda=-1\approx \lambda_c$, Fig.~\ref{fig:Potts3_eigvect}(b)], and once the DPT is triggered [$\lambda=-5< \lambda_c$, Fig.~\ref{fig:Potts3_eigvect}(c)]. We recall that
\be
\bbraket{\meanmag}{R_{0,\text{D}}^\lambda} = \sum_{\substack{\ket{C}\in{\cal H}: \\ \meanmag(C)=\meanmag}} \braket{C}{R_{0,\text{D}}^\lambda} \, , \nonumber
\ee
see Eq.~\eqref{Pconserv}, and note that $\ket{R_{0,\text{D}}^\lambda}$ is always real, and so is the projection $\bbraket{\meanmag}{R_{0,\text{D}}^\lambda}$ which can be hence considered as a probability distribution in the complex $\meanmag$-plane. For $\lambda>\lambda_c$, before the DPT happens, $\bbraket{\meanmag}{R_{0,\text{D}}^\lambda}$ is peaked around $|\meanmag|=0$, see Fig.~\ref{fig:Potts3_eigvect}(a). In this case the spectrum is gapped and there exists a unique, symmetry-preserving reduced Doob steady state $\kket{P_{\text{ss},P_0}^\lambda} = \kket{R_{0,\text{D}}^\lambda}$. When $\lambda\approx\lambda_c$ the distribution $\bbraket{\meanmag}{R_{0,\text{D}}^\lambda}$ flattens and spreads out, see Fig.~\ref{fig:Potts3_eigvect}(b), hinting at the emerging DPT which becomes apparent once $\lambda<\lambda_c$, when $\bbraket{\meanmag}{R_{0,\text{D}}^\lambda}$ develops well-defined peaks in the complex $\meanmag$-plane around regions with $|\meanmag|\approx 1$ and complex phases $\varphi=0,$ $2\pi/3$ and $4\pi/3$, see Fig.~\ref{fig:Potts3_eigvect}(c). In all cases $\kket{R_{0,\text{D}}^\lambda}$ is invariant under the reduced symmetry transformation, $\bbra{\meanmag}\hat{S}_\meanmag\kket{R_{0,\text{D}}^\lambda}=\bbraket{\meanmag}{R_{0,\text{D}}^\lambda}$, where $\hat{S}_\meanmag$ amounts now to a rotation of angle $2\pi/3$ in the complex $\meanmag$-plane that keeps constant $|\meanmag|$. However, for $\lambda<\lambda_c$ the Doob stationary subspace is three-fold degenerate in the $L\to\infty$ limit, and includes now the complex-conjugate pair of eigenvectors $\kket{R_{1,\text{D}}^\lambda}$ and $\kket{R_{2,\text{D}}^\lambda}$, which transform under the reduced symmetry operator as $\hat{S}_\meanmag\kket{R_{j,\text{D}}^\lambda} = \phi_j \kket{R_{j,\text{D}}^\lambda}$ with $\phi_j=\text{e}^{i2\pi j/3}$ and $j=1,2$. The reduced phase probability vectors now follow from Eq.~\eqref{redeq3},
\be
\begin{split}
\kket{\Pi_l^\lambda} &= \kket{R_{0,\text{D}}^\lambda} +  \text{e}^{i 2\pi l/3}\kket{R_{1,\text{D}}^\lambda} + \text{e}^{-i2\pi l/3}\kket{R_{2,\text{D}}^\lambda} \\
& = \kket{R_{0,\text{D}}^\lambda} +  2\Re[\text{e}^{i 2\pi l/3}\kket{R_{1,\text{D}}^\lambda}] \, , \nonumber
\end{split}
\ee
with $l=0,1,2$, and define the 3 degenerate Doob steady states, one for each symmetry branch, once the DPT appears. The order parameter structure of these reduced phase probability vectors, as well as that of the real and imaginary parts of the reduced eigenvector $\kket{R_{2,\text{D}}^\lambda}=\kket{R_{1,\text{D}}^\lambda}^*$, are shown in Figs.~\ref{fig:Potts3_eigvect}(d)-\ref{fig:Potts3_eigvect}(e). In particular we display $\Re[\text{e}^{i 2\pi l/3}\bbraket{\meanmag}{R_{1,\text{D}}^\lambda}]$ and $\Im[\text{e}^{i 2\pi l/3}\bbraket{\meanmag}{R_{1,\text{D}}^\lambda}]$, instead of $\Re[\bbraket{\meanmag}{R_{1,\text{D}}^\lambda}]$ and $\Im[\bbraket{\meanmag}{R_{1,\text{D}}^\lambda}]$, respectively, to illustrate the phase selection mechanism of Eq.~\eqref{eq:probphasevect} while conveying the full complex structure of this eigenvector. For instance, the phase vector $\kket{\Pi_0^\lambda}$ can be selected by just adding $2\Re[\kket{R_{1,\text{D}}^\lambda}]$ to $\kket{R_{0,\text{D}}^\lambda}$, see the $l=0$ in Figs.~\ref{fig:Potts3_eigvect}(d)-\ref{fig:Potts3_eigvect}(e), while for the other two $\kket{\Pi_l^\lambda}$ a complex phase is required to \emph{rotate} $\kket{R_{1,\text{D}}^\lambda}$ and cancel out the undesired peaks in $\kket{R_{0,\text{D}}^\lambda}$. A generic steady state will correspond to a weighted superposition of these phase probability vectors, $\ket{\probTvectst} = \sum_{l=0}^{2} w_l \ket{\probDphasevect{l}}$, with the statistical weights depending on the initial state, see Eq.~\eqref{coef}.

\begin{figure}
\includegraphics[width=1\linewidth]{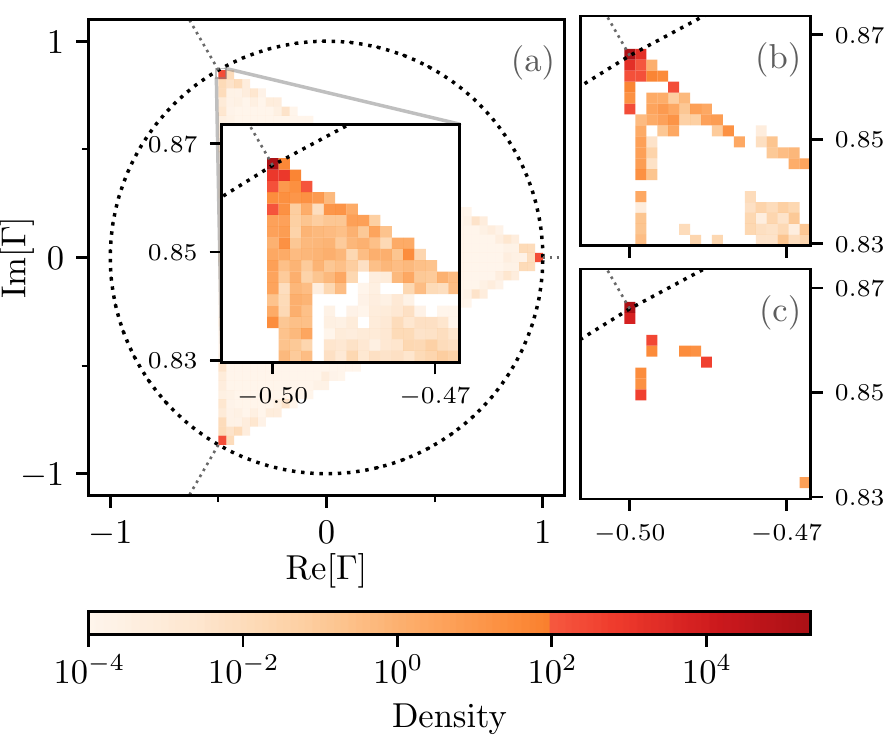}
\caption{{\bf Structure of the degenerate subspace in the 3-state Potts model.}
(a) Density plot for $\Gamma=\braket{C}{\eigvectDRnk{1}}/\braket{C}{\eigvectDRnk{0}}$ in the complex $\Gamma$-plane obtained for a large set of configurations $\ket{C}$ sampled from the Doob steady-state distribution for $L=16$ and $\lambda = -5$. The inset is a zoom on one of the compact regions around the complex unit circle where points converge. The small panels to the right show the same density plot, as obtained for different system sizes (b) $L=12$ and (c) $L=8$. 
A different color is used for the highest values of the density to highlight the sharp concentration around the cube roots of unity.
}
\label{fig:Potts3_deg}
\end{figure}

The intimate structural relation among the leading eigenvectors defining the degenerate subspace in the symmetry-broken regime ($\lambda<\lambda_c$) can be studied now by plotting $\Gamma(\conf) = \braket{\conf}{\eigvectDRnk{1}} / \braket{C}{\eigvectDRnk{0}}$ in the complex plane for a large sample of statistically-relevant configurations $\ket{C}$. As predicted in Eq.~\eqref{eq:eigvectsfromeigvect0}, this quotient should converge as $L$ increases to
\be
\Gamma(C) \approx (\text{e}^{i 2\pi/3})^{-\ell_C} \, ,
\label{cregions}
\ee
with $\ell_C=0,1,2$ identifying the symmetry sector to which configuration $\ket{C}$ belongs in. Fig.~\ref{fig:Potts3_deg} shows the density histogram in the complex $\Gamma$-plane obtained in this way for $\lambda=-5<\lambda_c$ and a large sample of important configurations. As predicted, all points concentrate sharply around three compact regions around the complex unit circle at phases $\varphi=0,~2\pi/3$ and $4\pi/3$, see Eq.~\eqref{cregions} and the inset in Fig.~\ref{fig:Potts3_deg}.
Notice that, even though a log scale is used to appreciate the global structure, practically all density is contained in a very small region around the cube roots of unity.
Moreover, the convergence to the predicted values as $L$ increases is illustrated in Figs.~\ref{fig:Potts3_deg}(b)-\ref{fig:Potts3_deg}(c). Equivalently, in the reduced order parameter space we expect that
\be
\bbraket{\meanmag}{R_{j,\text{D}}^\lambda} \approx (\text{e}^{i 2\pi j/3})^{-\ell_\meanmag} \bbraket{\meanmag}{R_{0,\text{D}}^\lambda} \, ,\nonumber
\ee
for $j=1,2$, where $\ell_\meanmag=0,1,2$ is a characteristic function identifying each phase in the $\meanmag$-plane. This relation implies that the size and shape (and not only the positions) of the different peaks in $\bbraket{\meanmag}{R_{j,\text{D}}^\lambda}$, $j=0,1,2$, in the complex plane are the same, see Eq.~\eqref{redeq4}, a general relation also confirmed in the boundary-driven WASEP.

\begin{figure}
\includegraphics[width=1\linewidth]{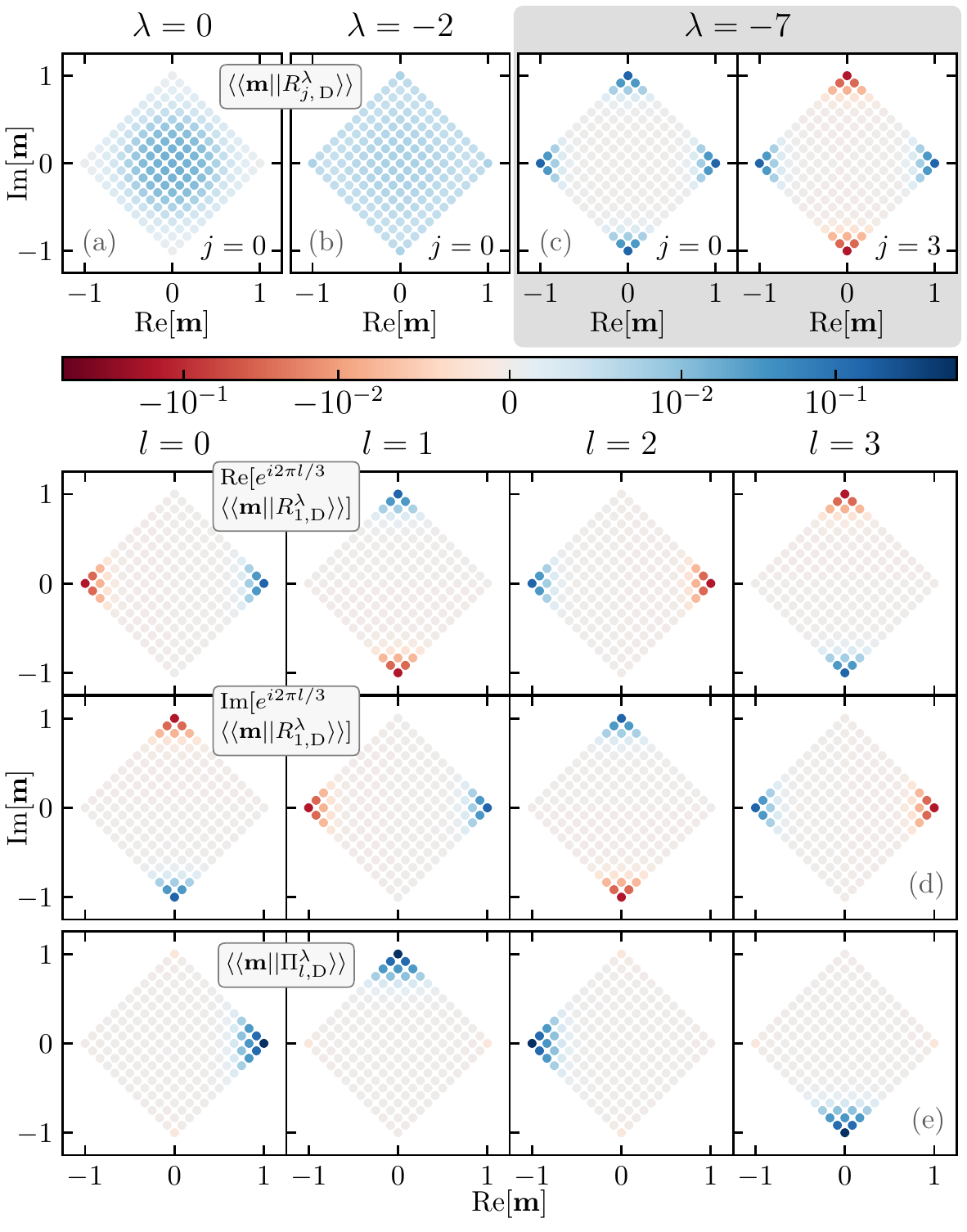}
\caption{{\bf Quasi-degenerate reduced eigenvectors in the 4-state Potts model.} 
Panels (a)-(c) show the structure of $\bbraket{\meanmag}{R_{0,\text{D}}^\lambda}$ in the complex $\meanmag$-plane for different values of $\currenttilt$ across the DPT and $L=12$. From left to right: (a) $\currenttilt = 0$ (symmetry-preserving phase, before the DPT), (b) $\currenttilt = -2 \approx \lambda_c$, (c) $\currenttilt = -7$ (symmetry-broken phase, after the DPT). The panel to the right of (c)
shows $\bbraket{\meanmag}{R_{3,\text{D}}^\lambda}$, which is also real. (d) Real and imaginary structure of $\bbraket{\meanmag}{R_{1,\text{D}}^\lambda}$. The top panels display $\Re[\text{e}^{i 2\pi l/4}\bbraket{\meanmag}{R_{1,\text{D}}^\lambda}]$, while the mid panels show $\Im[\text{e}^{i 2\pi l/4}\bbraket{\meanmag}{R_{1,\text{D}}^\lambda}]$ for $l=0,1,2,3$. This illustrates the phase selection mechanism of Eq.~\eqref{redeq3}. Panels (e) show the structure of the resulting reduced phase probability vectors $\bbraket{\meanmag}{\Pi_l^\lambda}$. 
}
\label{fig:Potts4_eigvect}
\end{figure}

\begin{figure}
\includegraphics[width=1\linewidth]{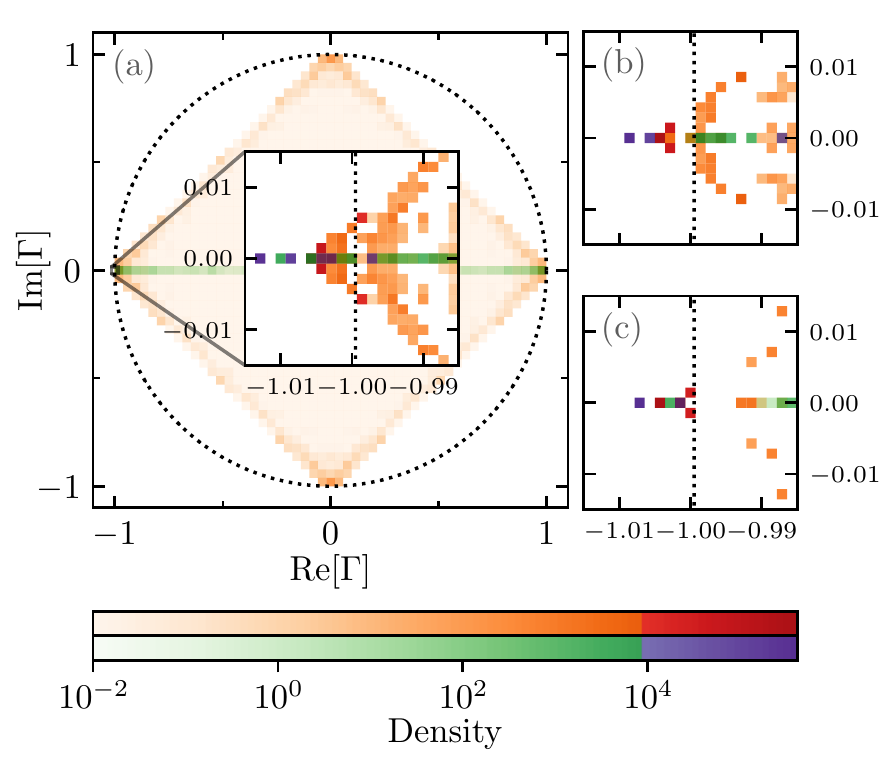}
\caption{{\bf Structure of the degenerate subspace for the 4-state Potts model.} 
(a) Density plot for $\Gamma_j=\braket{C}{\eigvectDRnk{j}}/\braket{C}{\eigvectDRnk{0}}$ in the complex $\Gamma$-plane for $j=1,3$ obtained for a large set of configurations $\ket{C}$ sampled from the Doob steady-state distribution with $\lattsize = 12$ and $\lambda = -7$. The inset zooms on one of the regions, given by Eq.~\eqref{eq:eigvectsfromeigvect0} where the density peaks. The small panels to the right show the same density plot, as obtained for different system sizes: (b) $L = 10$  and (c) $L = 8$. A comparison of the different panels allows us to appreciate the convergence to the predicted values as $L$ increases, even though in this case the difference is subtle due to the similar lattice sizes. Notice the log scale in the colorbar; this shows that almost all the density is contained in a small region around $\pm 1$ and $\pm i$, as predicted.
}
\label{fig:Potts4_eigvect_complete}
\end{figure}

Equivalent ideas hold valid for the ferromagnetic dynamical phase found in the 4-state Potts model. In this case, when the system is conditioned to sustain a large time-averaged energy fluctuation well below its average, a similar DPT to a dynamical ferromagnetic phase appears, breaking spontaneously the $\Zsymm{4}$ discrete rotational symmetry of this model. Hence we expect a degenerate Doob stationary subspace spanned by the first four leading eigenvectors, with eigenvalues under the symmetry operator given by $\symmeigval{0} = 1$, $\symmeigval{1} = \text{e}^{i 2\pi/4}$ and $\symmeigval{2} = \text{e}^{-i 2\pi/4}$ and $\symmeigval{3} = -1$. The generic Doob steady state can be then written as

\begin{equation}
\begin{split}
\probTvectst = \eigvectDR{0} + 2\Re[\eigvectDR{1} \braket{\eigvectDLnk{1}}{\probvecttimenk{0}}]  \\
+ \eigvectDR{3} \braket{\eigvectDLnk{3}}{\probvecttimenk{0}} \, , \nonumber
\end{split}
\end{equation}
where $\eigvectDR{3}$ is purely real and we have used that $\eigvectDR{2} = \eigvectDR{1}^{*}$. Fig.~\ref{fig:Potts4_eigvect} summarizes the spectral signatures of the DPT in the reduced order parameter space for the 4-state Potts model. In particular, Figs.~\ref{fig:Potts4_eigvect}(a)-\ref{fig:Potts4_eigvect}(c) show $\bbraket{\meanmag}{R_{0,\text{D}}^\lambda}$ for different values of $\lambda$ across the DPT. This distribution, which exhibits now four-fold symmetry, goes from unimodal around $|\meanmag|=0$ for $\lambda> \lambda_c$ to multimodal, with 4 clear peaks, for $\lambda< \lambda_c$, as expected. Figure~\ref{fig:Potts4_eigvect}(c) also includes $\bbraket{\meanmag}{R_{3,\text{D}}^\lambda}$ for this $\lambda$, which is purely real. Figures~\ref{fig:Potts4_eigvect}(d) capture the real and imaginary structure of $\bbraket{\meanmag}{R_{1,\text{D}}^\lambda}$ for $\lambda< \lambda_c$, in a manner equivalent to Fig.~\ref{fig:Potts3_eigvect}(e) (recall that $\bbraket{\meanmag}{R_{2,\text{D}}^\lambda}$ is simply its complex conjugate). Interestingly, the presence of a fourth eigenvector in the degenerate subspace for $\lambda<\lambda_c$ make for a richer phase selection mechanism. Indeed, the reduced phase probability vectors are now
\begin{equation}
\kket{\Pi_l^\lambda} = \kket{R_{0,\text{D}}^\lambda} +  2\Re[\text{e}^{i \pi l/2}\kket{R_{1,\text{D}}^\lambda}] + (-1)^l  \kket{R_{3,\text{D}}^\lambda} \, . \nonumber
\end{equation}
Their order parameter structure $\bbraket{\meanmag}{\Pi_l^\lambda}$ is displayed in Figs.~\ref{fig:Potts4_eigvect}(e). The $j=3$ eigenvector transfers probability from the configurations with magnetizations either in the horizontal or vertical orientation to the other one, see Fig.~\ref{fig:Potts4_eigvect}(c), while the combination of the second and third eigenvectors transfers probability between the two directions as dictated by the complex argument of $\braket{\eigvectDLnk{1}}{\probvecttimenk{0}}$, see Eq.~\eqref{coef}. Finally, Fig.~\ref{fig:Potts4_eigvect_complete} confirms the tight symmetry-induced structure in the degenerate subspace for $\lambda<\lambda_c$ by plotting $\Gamma_j(\conf) = \braket{\conf}{\eigvectDRnk{j}} / \braket{C}{\eigvectDRnk{0}}$ in the complex plane for $j=1,3$ and a large sample of statistically-relevant configurations $\ket{C}$. 
As expected, the point density associated with each eigenvector peak around $(\symmeigval{j})^{-\ell_C}$, which in this case correspond to $\pm 1$, $\pm i$ 
for $\eigvectDR{1}$, and $\pm 1$ for $\eigvectDR{3}$. Also, the density concentrates more and more around these points as $L$ increases. 

Summing up, we have shown how symmetry severely constraints the spectral structure associated with a DPT characterizing the energy fluctuations of a large class of spin systems.

\begin{figure}
\includegraphics[width=0.9\linewidth]{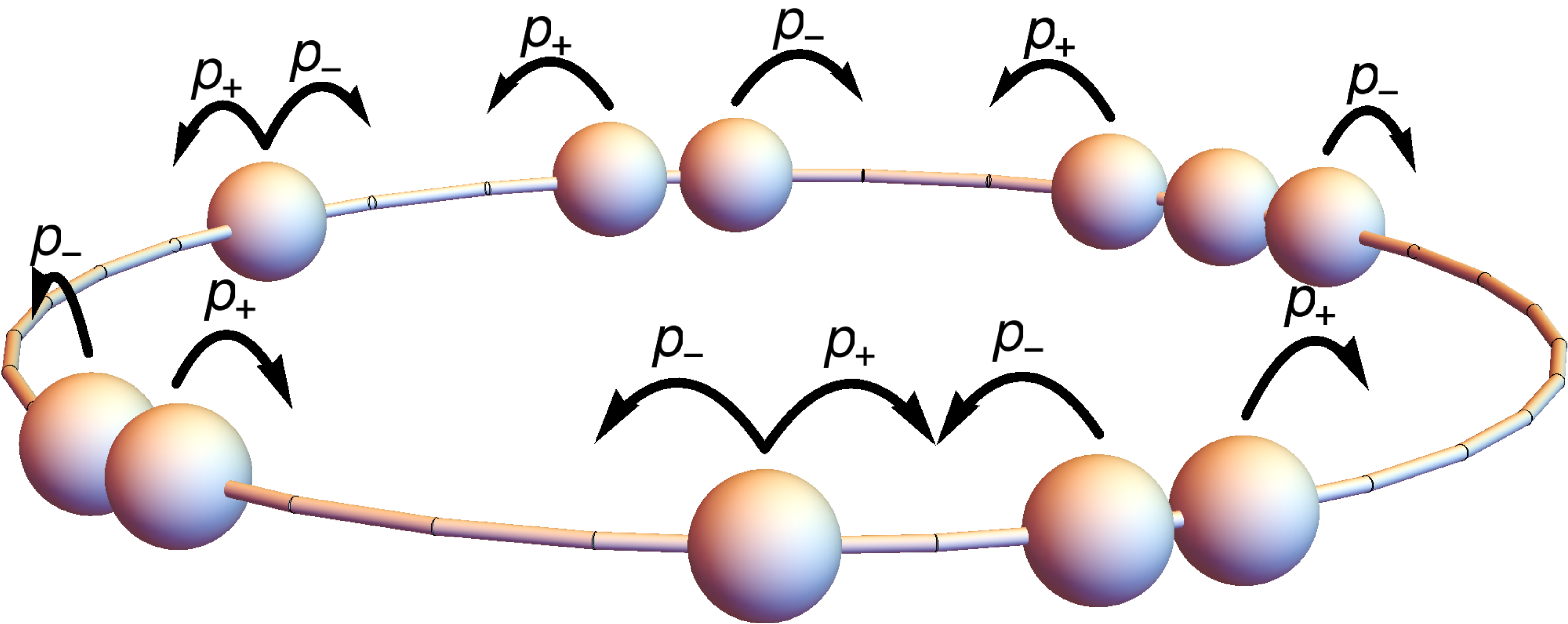}
\caption{{\bf Sketch of the WASEP with periodic boundary conditions.} 
The stochastic particle jumps occur now in a periodic lattice, so the total number of particles is conserved during the evolution.}
\label{fig:WASEP_def_closed}
\end{figure}

\section{A spectral perspective on time crystals: the closed WASEP}
\label{seccWASEP}

For the last example we go back to the WASEP model, using now periodic (or closed) boundary conditions, as illustrated in Fig.~\ref{fig:WASEP_def_closed}. Despite the absence of boundary driving, the steady state of the closed WASEP sustains a net particle current due to the external field. In this system, unlike the boundary-driven case, the total number of particles $\nprtcls$ remains constant during the evolution, so that the mean density $\dens_0 = \nprtcls/\lattsize$ becomes an additional control parameter. This means in particular that the PH symmetry present in the boundary-driven WASEP (when the reservoir densities obey $\rho_\text{R}=1-\rho_\text{L}$) is lost except when the global density is $\dens_0 = 0.5$, since the PH transformation changes the density as $\dens_0 \to 1 - \dens_0$. Instead, the closed WASEP is invariant under the translation operator, $\symmop_{T}$, which moves all the particles one site to the right, $k \to k+1$. Such operator reads,
\begin{equation}
\symmop_{\textrm{T}} = \prod_{k=1}^{\lattsize-1} \Big[\creationop_{k}\destructop_{k+1} + \creationop_{k+1}\destructop_{k} + \frac{1}{2}(\hat{\sigma}^z_k \hat{\sigma}^z_{k+1} + \identityop) \Big] \, , \nonumber
\end{equation}
where we identify site $L+1$ with site 1. Thus we have $[\genmat,\symmop_{\textrm{T}}]=0$ for the stochastic generator in this model. Note that $(\symmop_{\textrm{T}})^L=\identityop$, and hence the closed WASEP will exhibit a $\mathbb{Z}_L$ symmetry. As we will discuss below, this case is subtly different from the previous examples, as the order of the $\mathbb{Z}_L$ symmetry increases with the lattice size, approaching a continuous symmetry in the thermodynamic limit. Still, our results remain valid in this case.
The current large deviation statistics is encoded in the spectral properties of the tilted generator, which now reads
\begin{equation}
\begin{split}
\genmatT = \sum_{k=1}^{\lattsize} \Big[\; &\jumprateR \big( \text{e}^{+\currenttilt/\lattsize}\creationop_{k+1} \destructop_{k} - \occupop_{k  } (\identityop_{k+1} - \occupop_{k+1} )\big) \\ 
+ &\jumprateL (\text{e}^{-\currenttilt/\lattsize} \creationop_{k  } \destructop_{k+1} - \occupop_{k+1} (\identityop_{k} - \occupop_{k  } )\big)\Big] \, , \nonumber
\end{split}
\label{eq:WASEPclosed_tilted_gen}
\end{equation}
to be compared with the boundary-driven case, Eq.~\eqref{eq:WASEPopen_tilted_gen}. The original generator $\genmat$ can be recovered by setting $\currenttilt = 0$ above, and it can be easily checked that $[\genmat,\symmop_{\textrm{T}}]=0$.

\begin{figure}
\includegraphics[width=0.9\linewidth]{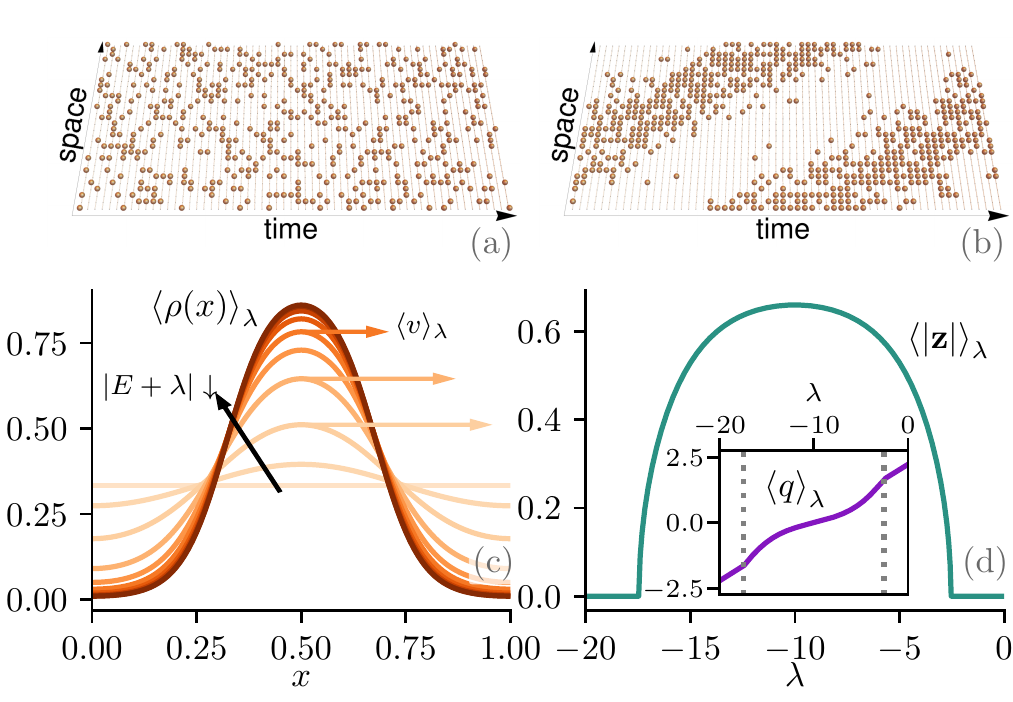}
\caption{{\bf $\mathbb{Z}_L$ symmetry-breaking DPT in the closed WASEP.} Top panels: Typical spacetime trajectories of the closed WASEP for current fluctuations above (a) and below (b) the critical current. Note the periodic boundary conditions, and the emergence of a jammed matter wave below the critical current. (c) Density profile of the rotating condensate for different values of $\currenttilt$. (d) Average magnitude of the packing order parameter as a function of $\lambda$. The inset shows the average current vs $\lambda$, which becomes nonlinear in the symmetry-breaking regime. In all panels the global density is $\rho_0=1/3$ and $E=10>E_c$.}
\label{fig:DPT_WASEPclosed}
\end{figure}

Interestingly, the closed WASEP also presents a symmetry-breaking DPT when the system is biased towards currents well below its typical value and in the presence of a strong enough field \cite{bodineau05a,perez-espigares13a,hurtado-gutierrez20a}. In this case the optimal strategy to sustain a low current fluctuation cannot be depleting or crowding the lattice with particles to hamper the flow, as in the boundary-driven case, since now the total number of particles is constant. Instead, when this DPT kicks in, the particles pack together creating a jammed, rotating condensate which hinders particle motion to facilitate such a low current fluctuation, see Figs.~\ref{fig:DPT_WASEPclosed}(a)-\ref{fig:DPT_WASEPclosed}(b). This condensate breaks spontaneously the translation symmetry $\symmop_{T}$ and, whenever $\rho_0\ne 1/2$, travels at constant velocity along the lattice, breaking also time-translation symmetry \cite{bodineau05a,perez-espigares13a}. These features are the fingerprint of the recently discovered time-crystal phase of matter \cite{wilczek12a,shapere12a,zakrzewski12a,moessner17a,richerme17a,yao18a,sacha18a,hurtado-gutierrez20a}. Specifically, this DPT appears for external fields $\abs{\field} > E_c=\pi/\sqrt{\dens_0(1-\dens_0)}$ and for currents $\abs{\current} \le \crit{\current}=\dens_0(1-\dens_0)\sqrt{E^2-E_c^2}$, which correspond to biasing fields $\currenttiltcritmin < \currenttilt < \currenttiltcritmax$, with $\currenttiltcritminmax = -\field \pm\sqrt{\field^2 - \crit{\field}^2}$ \cite{bodineau05a,perez-espigares13a}, see the inset to Fig.~\ref{fig:DPT_WASEPclosed}(b). For $\lambda$ outside this regime, the typical density field sustaining the fluctuation is just flat, structureless [Fig.~\ref{fig:DPT_WASEPclosed}.(a)], while within the critical region a matter density wave [Fig.~\ref{fig:DPT_WASEPclosed}(b)] with a highly nonlinear profile develops. Figure \ref{fig:DPT_WASEPclosed}(c) shows the density profiles of the resulting jammed condensate for different values of $\currenttilt$ in the macroscopic limit.

\begin{figure}
\includegraphics[width=1\linewidth]{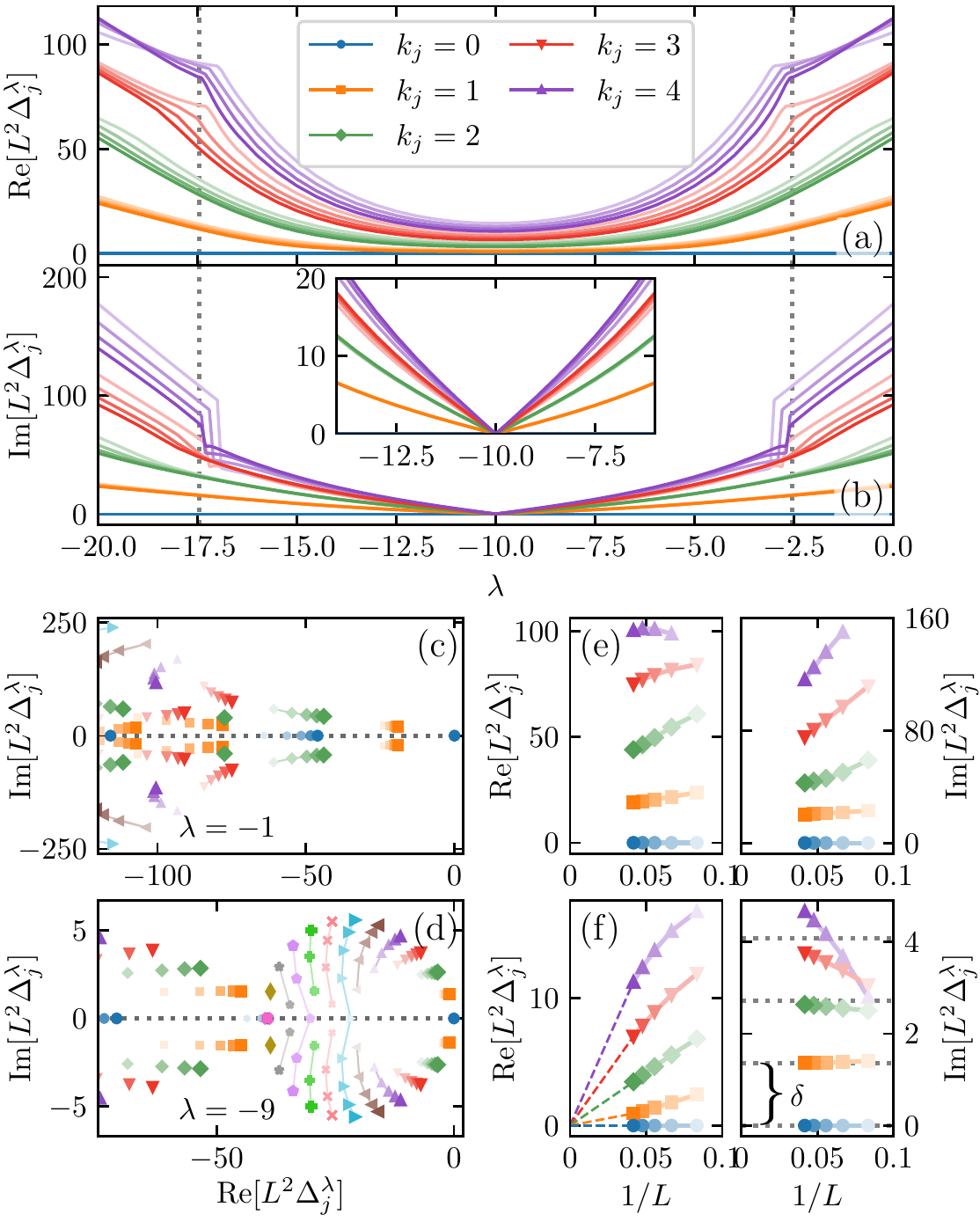}
\caption{{\bf Spectral signatures of a continuous time crystal DPT.} Diffusively-scaled eigenvalues of the Doob stochastic generator $\genmatD$ for the closed WASEP with $\rho_0=1/3$ and $E=10>E_c$ and lattice sizes $L=9, 12, 15, 18, 21, 24$.
The different colors and marker types denote the symmetry eigenvalue $\symmeigval{j} = e^{i 2\pi k_j/\lattsize}$ corresponding to each $\Delta^{\lambda}_j$.
Top panels show the evolution with $\lambda$ of the real (a) and imaginary (b) parts of the first few leading eigenvalues of $\genmatD$ for increasing values of $L$, denoted by increasing color intensity. 
More specifically, they show the largest $\Delta^{\lambda}_j$ corresponding to each $k_j$.
Bottom panels show the spectrum in the complex plane in (c) the homogeneous phase for $\lambda=-1$, and (d) the condensate phase for $\lambda=-9$.
The size of each marker indicates the lattice size (bigger marker correspond to bigger $L$), showing their evolution as $L$ increases.
The colors and markers beyond the ones in the legend correspond to $k_j=[5\isep 12]$, which appear in order in panel (d).
Panels (e)-(f) show the finite-size scaling analysis for the real and imaginary parts of the leading eigenvalues in the homogeneous (e) and condensate (f) phases. The real parts converge to zero as a power law of $1/L$ in the condensate phase, while the imaginary parts exhibit a clear band structure with constant frequency spacing $\delta$, proportional to the condensate velocity.
}
\label{fig:eigvalues_closed}
\end{figure}

A suitable way to characterize this DPT consist in measuring the packing of the particles in the 1D ring.
For a configuration $C=\{n_k\}_{k=1,\ldots,L}$, with $n_k=0,1$ the occupation number of site $k$, 
the \emph{packing order parameter} $\cm$ is defined as 
\be
\cm = \frac{1}{N} \sum_{k=1}^{\lattsize} \occup_k \, \text{e}^{i 2\pi k/\lattsize} = |\cm| \text{e}^{i \varphi} \, .
\label{coher}
\ee
This measures the position of the center of mass of the system in the two-dimensional plane. The magnitude $|\cm|$ of this packing parameter is close to zero for any homogeneous distribution of particles in the ring, but increases significantly for condensed configurations, while its complex phase $\varphi$ signals the angular position of the condensate's center of mass. In this way, we expect $|\cm|$ to increase from zero  when the condensate first appears at the DPT. This is confirmed in Fig.~\ref{fig:DPT_WASEPclosed}(b), which shows the evolution of $\langle |\cm|\rangle_\lambda$ as a function of the biasing field. 

As in the previous cases, the DPT in the closed WASEP is accompanied by the emergence of a degenerate Doob stationary subspace spanned by multiple Doob eigenvectors with vanishing spectral gaps in the thermodynamic limit. However, in stark contrast with previous examples, in this case the number of degenerating eigenvectors is not fixed but increases linearly with the system size. This can be observed in Fig.~\ref{fig:eigvalues_closed}, which shows the spectrum of the Doob stochastic generator $\genmatD$ for the closed WASEP with $L=24$, $\rho_0=1/3$ and $E=10>E_c$. In particular, Figs.~\ref{fig:eigvalues_closed}(a)-\ref{fig:eigvalues_closed}(b) show the evolution with $\lambda$ of the real (a) and imaginary (b) parts of the first few leading eigenvalues of $\genmatD$ for different system sizes $L$. A clear change of behavior is apparent at $\lambda_c^\pm$. Indeed, the whole structure of the spectrum in the complex plane changes radically as we move across $\lambda_c^\pm$, see Figs.~\ref{fig:eigvalues_closed}(c)-\ref{fig:eigvalues_closed}(d), with a gapped phase for $\lambda>\lambda_c^+$ or $\lambda<\lambda_c^-$ [see Fig.~\ref{fig:eigvalues_closed}(e)], and an emerging gapless phase for $\lambda_c^-<\lambda<\lambda_c^+$ characterized by a vanishing spectral gap $\Delta_j^\lambda(L)$ of a macroscopic fraction of eigenvalues $j\in[1,{\cal O}(L)]$ as $L\to\infty$, which decay linearly as $1/L$ and with hierarchical structure in $j$, see Fig.~\ref{fig:eigvalues_closed}(f). Moreover, the imaginary parts of the gap-closing eigenvalues in this regime are non-zero (except for the leading one) and exhibit an emerging band structure with a constant frequency spacing $\delta$, see dashed horizontal lines in Fig.~\ref{fig:eigvalues_closed}(f). We will show below that this band structure in the imaginary axis can be directly linked with the velocity $v$ of the moving condensate.

\begin{figure}
\includegraphics[width=1\linewidth]{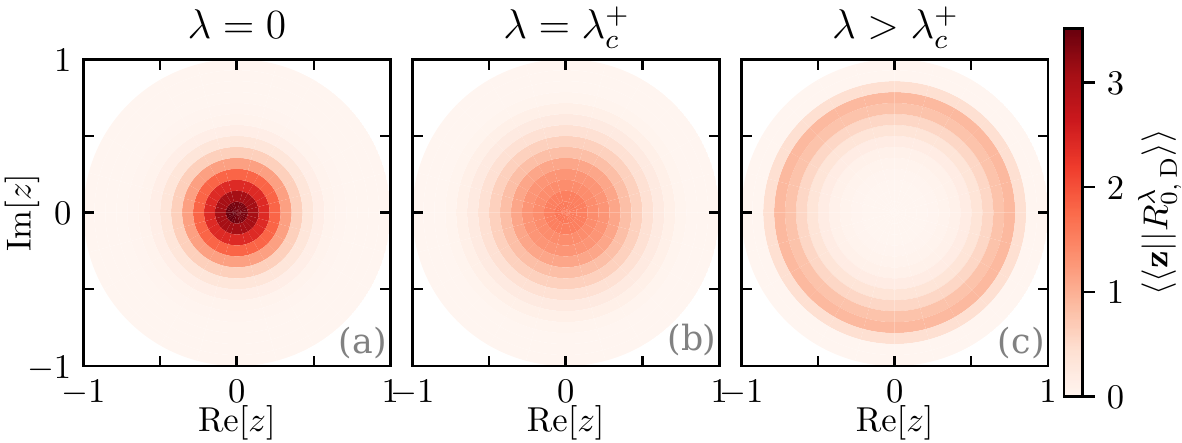}
\caption{{\bf The leading eigenvector across the DPT in the closed WASEP.} Structure of $\bbraket{\cm}{R_{0,\text{D}}^\lambda}$ in the complex $\cm$-plane for different values of $\currenttilt$ across the DPT for $\rho_0=1/3$, $E=10>E_c$ and $L=24$. From left to right: (a) $\currenttilt = 0$ (symmetry-preserving phase, before the DPT), (b) $\currenttilt = -2.5 \approx \lambda_c^+$, (c) $\currenttilt = -9$ (symmetry-broken phase, after the DPT). Note the transition from unimodal$\bbraket{\cm}{R_{0,\text{D}}^\lambda}$ peaked around $|\cm|\approx 0$ for $\lambda>\lambda_c^+$ to the inverted Mexican-hat structure with a steep ridge around $|\cm|\approx 0.7$ for $\lambda_c^-<\lambda<\lambda_c^+$.}
\label{fig:eigvect0_closed}
\end{figure}

In view of these spectral properties, we naturally expect a unique Doob steady state in the regime where the spectrum is gapped, i.e. $\lambda>\lambda_c^+$ or $\lambda<\lambda_c^-$, given by the leading Doob eigenvector $\probTvectst = \eigvectDR{0}$. This steady state remains invariant under $\hat{S}_{\textrm{T}}$. On the other hand, in the gapless regime $\lambda_c^-<\lambda<\lambda_c^+$ the Doob \emph{stationary} subspace will be ${\cal O}(L)$-degenerate, and the resulting Doob \emph{steady state} will be in fact time-dependent and approximately equal to
\be
\probTvectst(t) \approx \eigvectDR{0} + \sum_{j=1}^{L-1} \text{e}^{+i t \Im(\theta_{j,\text{D}}^\lambda)}\eigvectDR{j} \braket{\eigvectDLnk{j}}{\probvecttimenk{0}}\, ,
\label{PssTC}
\ee
see Eq.~\eqref{Psstime} in \S\ref{ssecDeg} and the associated discussion. It is important to notice that this is an approximation, the more accurate the larger $L$ is.
For any finite $L$ the leading spectral gaps won't be completely closed, in fact they decay as $1/L$ in a hierarchical manner, see Fig.~\ref{fig:eigvalues_closed}(d) and Fig.~\ref{fig:eigvalues_closed}(f), and the Doob stationary subspace will be \emph{quasi}-degenerate \cite{gaveau98a,gaveau06a,minganti18a}. As $L$ increases, the steady state is better approximated by Eq.~\eqref{PssTC}, i.e. as more and more eigenvectors enter the quasi-degenerate subspace. As expected, the Doob \emph{steady state} in this quasi-degenerate regime breaks spontaneously the translation symmetry, 
so $\hat{S}_{\textrm{T}}\probTvectst(t) \ne \probTvectst(t)$, see below.

Assuming now $L$ to be odd for simplicity (all results can be trivially generalized to even $L$), and recalling that the complex eigenvalues and eigenvectors of $\genmatD$ come in complex-conjugate pairs, so $\Im(\theta_{2k+1,\text{D}}^\lambda)=-\Im(\theta_{2k,\text{D}}^\lambda)$, the band structure with constant spacing $\delta$ observed in the imaginary parts of the gap-closing eigenvalues implies that
\be
\Im(\theta_{j,\text{D}}^\lambda) =
\begin{cases}
+ j \delta/2 \, , & j=2,4,\ldots,L-1  \\
-(j+1) \delta/2 \, , & j=1,3,\ldots,L-2
\end{cases} \nonumber
\ee
and hence asymptotically
\be
\probTvectst (t) \approx \eigvectDR{0} + 2 \sum_{j=2\atop j\, \text{even}}^{L-1} \Re\left[ \text{e}^{+i t \frac{j\delta}{2}}\eigvectDR{j} \braket{\eigvectDLnk{j}}{\probvecttimenk{0}} \right]\, . \nonumber
\label{PssTC2}
\ee
Similarly, the symmetry eigenvalues $\phi_j$ of the different Doob eigenvectors, such that $\hat{S}_\text{T} \eigvectDR{j}=\phi_j\eigvectDR{j}$, come in complex-conjugate pairs and obey
\be
\phi_j =
\begin{cases}
\text{e}^{+i \pi j/L} \, , & j=2,4,\ldots,L-1  \\
\text{e}^{-i \pi (j+1)/L} \, , & j=1,3,\ldots,L-2
\end{cases} \nonumber
\ee
such that $\phi_{2k+1}=\phi_{2k}^*$. In this way, we can easily see that
\be
\begin{split}
\hat{S}_\text{T} \probTvectst (t) & \approx  \eigvectDR{0} + 2 \sum_{j=2\atop j\, \text{even}}^{L-1} \Re\Big[ \text{e}^{+i (t+\frac{2\pi}{L\delta}) \frac{j\delta}{2}}\eigvectDR{j} \\
& \times \braket{\eigvectDLnk{j}}{\probvecttimenk{0}} \Big] =  \probTvectst(t+\frac{2\pi}{L\delta}) \, . \nonumber
\end{split}
\label{PssTC3}
\ee
This shows that, in the quasi-degenerate phase $\lambda_c^-<\lambda<\lambda_c^+$, (i) the symmetry is spontaneously broken, $\hat{S}_\text{T} \probTvectst (t)\ne \probTvectst (t)$, but (ii) spatial translation and time evolution are two sides of the same coin in this regime. In particular, we have shown that a spatial translation of a unit lattice site is equivalent to a temporal evolution of time $2\pi/L\delta$, i.e. $\hat{S}_\text{T} \probTvectst (t) =\probTvectst (t+\frac{2\pi}{L\delta})$, leading to a time-periodic motion of period $2\pi/\delta$ or equivalently a density wave of velocity $v=L\delta/2\pi$.

For the phase probability vectors in the symmetry-broken regime, Eq.~\eqref{eq:probphasevect} implies that
\be
\ket{\Pi_l^\lambda} = \eigvectDR{0} + 2 \sum_{j=2\atop j\, \text{even}}^{L-1} \Re\Big[\text{e}^{+i \frac{\pi l}{L}j}\eigvectDR{j} \Big] \, ,
\label{PssTC4}
\ee
such that $\hat{S}_\text{T}\ket{\Pi_l^\lambda}=\ket{\Pi_{l+1}^\lambda}$. The dominant configurations in these phase probability vectors correspond to different static particle condensates, localized around the $L$ different lattice sites. Note that these \emph{localized} $\ket{\Pi_l^\lambda}$ are built as linear superpositions of the different \emph{delocalized} eigenvectors $\eigvectDR{j}$ shifted appropriately according to their symmetry eigenvalues, $(\phi_j)^l$. Figure {\ref{polar}(f), which will be discussed later, sketches this condensate localization mechanism in the reduced order parameter space. We can write the time-dependent \emph{stationary} Doob state in terms of the static phase probability vectors as
\be
\probTvectst (t) \approx \sum_{l=0}^{L-1} w_l(t) \ket{\Pi_l^\lambda} \, ,
\label{PssTC5}
\ee
where the different phase weights $w_l(t)$ are now time-dependent, see Eq.~\eqref{coeft},
\be
w_l (t)= \frac{1}{L} + \frac{2}{L}\sum_{j=2\atop j\, \text{even}}^{L-1}  \Re\Big[ \text{e}^{+i (\frac{t\delta}{2}-\frac{\pi l}{L})j} \braket{\eigvectDLnk{j}}{\probvecttimenk{0}} \Big] \, .
\label{PssTC6}
\ee
The periodicity of the resulting symmetry-broken state is reflected in the fact that $w_l(t+2\pi/\delta)=w_l(t)$ $\forall l\in[0\isep L-1]$.

\begin{figure}
\includegraphics[width=1\linewidth]{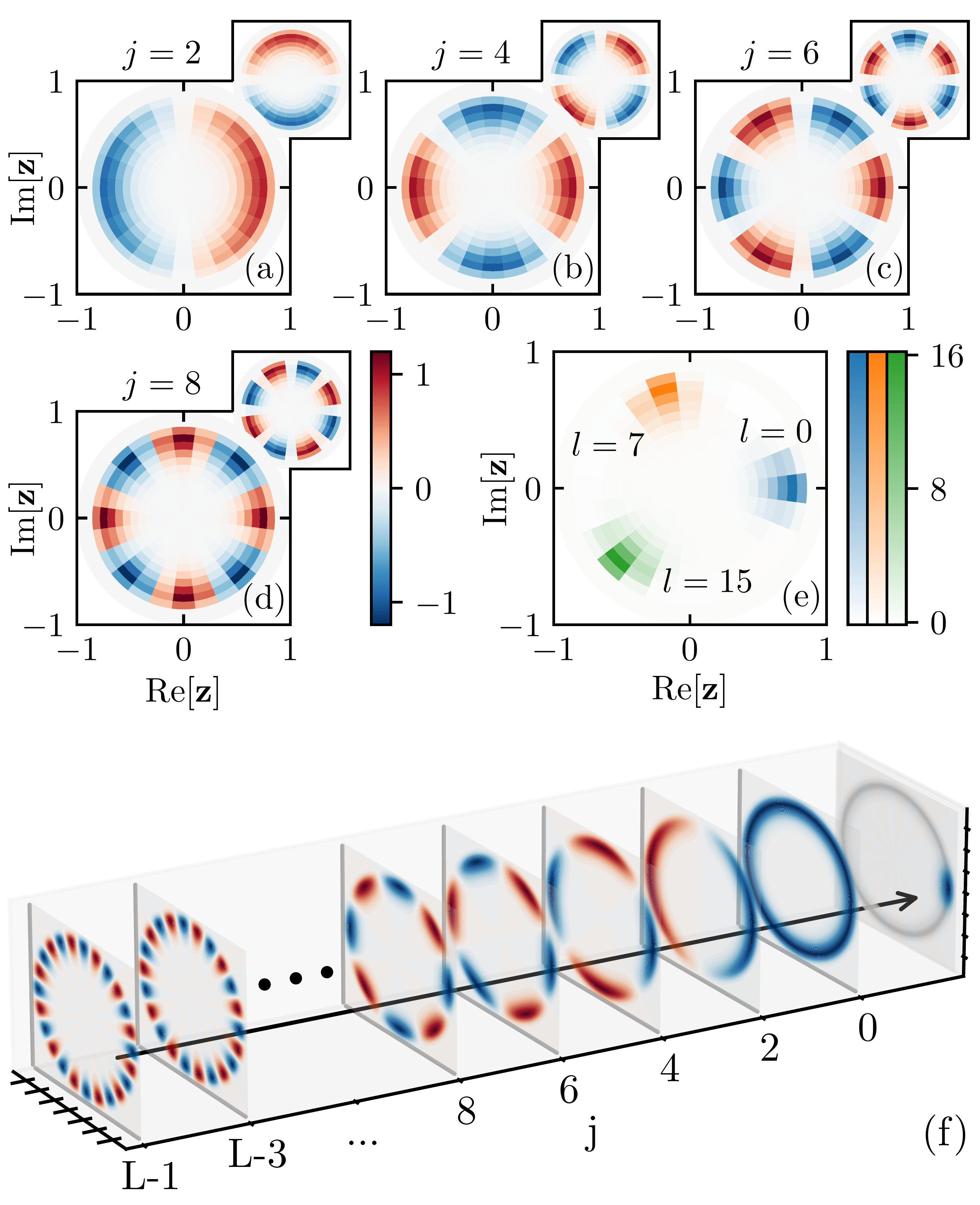}
\caption{{\bf Quasi-degenerate reduced eigenvectors in the closed WASEP and condensate localization.} (a)-(e) First few reduced eigenvectors of the Doob degenerate subspace in the closed WASEP. In particular, the real and imaginary parts of $\bbraket{\cm}{R_{j,\text{D}}^\lambda}$ are displayed for $\lambda_c^-<\lambda=-9<\lambda_c^+$ and different (even) values of $j=2,4,6,8$. Recall that the complex eigenvalues and eigenvectors of $\genmatD$ come in complex-conjugate pairs, so $\bbraket{\cm}{R_{2k+1,\text{D}}^\lambda}={\bbraket{\cm}{R_{2k,\text{D}}^\lambda}}^*$. The main panels show the real parts, while the insets display the imaginary parts for each $j$. Note the $j$-fold symmetry of reduced eigenvectors, and that non-negligible structure appears in all cases in the region $|\cm|\approx 0.7$, as expected in the symmetry-broken dynamical phase. (f) Sample of the resulting reduced phase probability vectors $\bbraket{\cm}{\Pi_l^\lambda}$ for $l=0,6,17$. (g) Sketch of the spectral localization mechanism that gives rise to a compact condensate. Each slice shows $\Re[\phi_j^l\bbraket{\cm}{R_{j,\text{D}}^\lambda}]$ for the corresponding $j$.}
\label{polar}
\end{figure}

The spectral structure of the Doob stationary subspace is better explored in the reduced Hilbert space associated with the packing order parameter $\cm$ introduced in Eq.~\eqref{coher}. Figure \ref{fig:eigvect0_closed} shows the structure of the leading reduced eigenvector $\bbraket{\cm}{R_{0,\text{D}}^\lambda}$ in the complex $\cm$-plane for varying $\lambda$ across the DPT. As observed in previous examples, before the DPT occurs (i.e. for $\lambda>\lambda_c^+$ or $\lambda<\lambda_c^-$, when the spectrum of $\genmatD$ is gapped) the real distribution $\bbraket{\cm}{R_{0,\text{D}}^\lambda}$ is unimodal and peaked around $|\cm|\approx 0$, indicating the absence of order in this symmetry-preserving phase. As $\lambda$ approaches the critical point $\bbraket{\cm}{R_{0,\text{D}}^\lambda}$ flattens and spreads over the unit complex circle, see Fig.~\ref{fig:eigvect0_closed}(b) for $\lambda\approx\lambda_c^+$, while deep inside the critical regime $\lambda_c^-<\lambda<\lambda_c^+$ the distribution $\bbraket{\cm}{R_{0,\text{D}}^\lambda}$ develops an inverted Mexican-hat shape, see Fig.~\ref{fig:eigvect0_closed}(c), with a steep ridge around $|\cm|\approx 0.7$ but homogeneous angular distribution. This means that the typical configurations contributing to $\ket{R_{0,\text{D}}}$ correspond to symmetry-broken condensate configurations ($|\cm|\ne 0$), localized but with a homogeneous angular distribution for their center of mass. Indeed, the resulting reduced eigenvector is invariant under the reduced symmetry operator, $\hat{S}_\cm \kket{R_{0,\text{D}}^\lambda}=\kket{R_{0,\text{D}}^\lambda}$, where $\hat{S}_{\cm}$ is now just a rotation of $2\pi/L$ radians in the complex $\cm$-plane. As in the previous examples, the subleading eigenvectors spanning the (quasi-)degenerate subspace cooperate to break the symmetry, in this case by localizing the condensate at a particular point in the lattice. Figure \ref{polar} shows the $\cm$-structure of the real and imaginary parts of the first few subleading reduced eigenvectors in the closed WASEP, $\bbraket{\cm}{R_{j,\text{D}}^\lambda}$ for $j=2,4,6,8$. Interestingly, the $j$th-order ($j$ even) reduced eigenvector exhibits a clear $(j/2)$-fold angular symmetry in the $\cm$-plane [i.e. invariance under rotations of angles $4\pi/j=2\pi/(j/2)$], with non-negligible structure around $|\cm|\approx 0.7\ne 0$ for the particular case $\lambda=-9$. 
All (quasi-)degenerate eigenvectors hence exhibit some degree of angular symmetry but their superposition, weighted by their symmetry eigenvalues $(\phi_j)^l$,
cooperates to produce a compact condensate localized at site $l$ and captured by the reduced phase probability vector $\kket{\Pi_l^\lambda}$, see also Eq.~\eqref{PssTC4}.
A sample of the resulting reduced phase probability vectors is shown in Fig.~\ref{polar}(e), which as expected are localized around different angular positions along the ring. Figure \ref{polar}(f) shows a sketch of the spectral localization mechanism that gives rise to a compact localized condensate from the superposition of multiple delocalized reduced eigenvectors in the degenerate subspace. As described above, the time dependence introduced by the imaginary parts of the gap-closing eigenvalues, together with their imaginary band structure, lead to the motion of the condensate at constant velocity.

\begin{figure}
\includegraphics[width=1\linewidth]{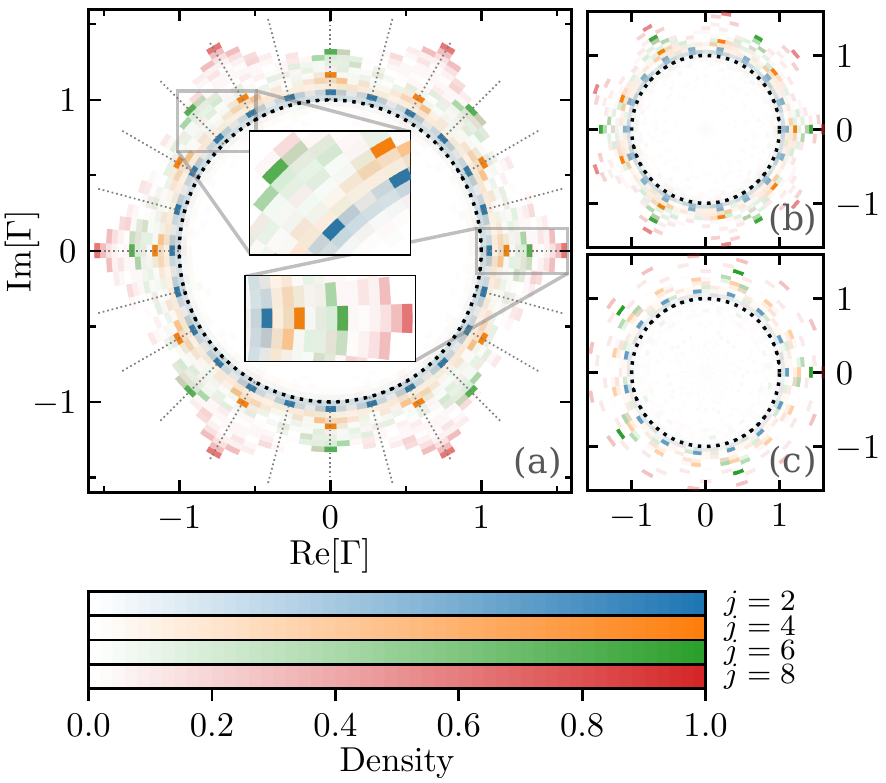}
\caption{{\bf Structure of the degenerate subspace in the closed WASEP.} Density plot of $\Gamma_j=\braket{C}{\eigvectDRnk{j}}/\braket{C}{\eigvectDRnk{0}}$ in the complex $\Gamma$-plane, with $j=2,4,6,8$, obtained for a large set of configurations $\ket{C}$ sampled from the Doob \emph{stationary} distribution in the symmetry-broken regime for $\lambda=-9$, $E=10>E_c$ and (a) $L=24$, (b) $L=18$ and (c) $L=15$. Different colors correspond to different values of index $j$. The insets show zooms on compact regions around the complex unit circle to better appreciate the emerging structure.}
\label{fig:eigvect_WASEPclosed_complete}
\end{figure}

According to Eq.~\eqref{eq:eigvectsfromeigvect0}, in the symmetry-broken regime we should expect a tight relation between the eigenvectors spanning the degenerate Doob subspace. In particular, we expect that
\be
\braket{C}{\eigvectDRnk{j}} \approx \text{e}^{-i \frac{\pi j}{L}\ell_C} \braket{C}{\eigvectDRnk{0}} 
\label{PssTC7}
\ee
for the statistically-relevant configurations $\ket{C}$ in the Doob \emph{stationary} state belonging to the basin of attraction of phase $\ell_C\in[0,L-1]$, see the associated discussion in \S\ref{ssecSpec}. To investigate this relation, we now plot in Fig.~\ref{fig:eigvect_WASEPclosed_complete} a density map in the complex plane for the quotients $\Gamma_j(C)=\braket{C}{\eigvectDRnk{j}}/\braket{C}{\eigvectDRnk{0}}$ for $j=2,4,6,8$ obtained from a large sample of configurations drawn from the Doob \emph{stationary} distribution $\probTvectst (t)$. As expected from Eq.~\eqref{PssTC7}, we observe a condensation of points around the complex unit circle, with high-density regions at nodal angles multiple of $\varphi_j=\pi j /L$. For instance, for $j=2$ we expect to observe sharp peaks in the density plot of $\Gamma_2(C)$ in the complex unit circle at angles $2\pi k/L$, $k\in[0\isep L-1]$, as confirmed in Fig.~\ref{fig:eigvect_WASEPclosed_complete}(a) for $L=24$.
The convergence to the complex unit circle improves as the system size $L$ increases, see Figs.~\ref{fig:eigvect_WASEPclosed_complete}(b) and \ref{fig:eigvect_WASEPclosed_complete}(c), and for fixed $L$ this convergence is better for smaller spectral index $j$, i.e. for the eigenvectors $\eigvectDRnk{j}$ whose (finite-size) spectral gap $\Delta_j^\lambda(L)$ is closer to zero, see Fig.~\ref{fig:eigvalues_closed}(f).
Note also that, when the (even) spectral index $j$ is conmensurate with $2L$, we expect $2L/j$ nodal accumulation points in the density plot for $\Gamma_j(C)$, e.g. 
for $j=6$ and $L=24$ we expect 8($=2\times 24/6$) nodal points, 
as seen in Fig.~\ref{fig:eigvect_WASEPclosed_complete}(a). 
For a (even) spectral index $j$ inconmensurate with $2L$ one should observe just $L$ nodal points, as shown in Fig.~\ref{fig:eigvect_WASEPclosed_complete}(b) for $j=4$ and $L=15$. Overall, this analysis confirms the tight structural relation imposed by the $\mathbb{Z}_L$ symmetry on the eigenvectors spanning the Doob \emph{stationary} subspace, confirming along the way that statistically-relevant configurations in the Doob \emph{stationary} state for the closed WASEP can be classified into different symmetry classes.

In summary, we have shown in this section that the general results derived in \S\ref{secDPTsymm} on how symmetry imposes a specific spectral structure across a DPT are also valid when it asymptotically breaks a continuous symmetry, as it is the case in the closed WASEP model and its DPT to a time-crystal phase for low enough current fluctuations. The chances are that this picture remains valid in more general, continuous symmetry-breaking DPTs.

\section{Discussion}
\label{secdisc}

In this paper we have unveiled the spectral signatures of symmetry-breaking DPTs. Such DPTs appear in the fluctuating behavior of many-body systems as non-analyticities in the large deviation functions describing the fluctuations of time-averaged observables, and are accompanied by singular changes in the trajectories responsible for such rare events. The main tools used in this work include the quantum Hamiltonian formalism for the master equation, describing the dynamics of stochastic many-body systems, together with large deviation theory whereby the symmetry of the microscopic dynamics has been fully exploited. A cornerstone in our analysis has been the Doob transform to build a driven stochastic process that makes typical a rare fluctuation of the original dynamics. Crucially, the steady state of the resulting Doob dynamics contains all the information of the most likely path leading to such rare fluctuation of the original process.

In this way, the spectral hallmark of a symmetry-breaking DPT is the emergence of a degeneracy in the stationary subspace of Doob eigenvectors. The degenerate eigenvectors exhibit different behavior under the symmetry transformation, and we show how symmetry and degeneracy cooperate to yield different, coexisting steady states once the DPT has kicked in. Such steady states are characterized by physical phase probability vectors, connected via the symmetry transformation, that we explicitly build from the gapless Doob eigenvectors in the degenerate subspace. Moreover, a generic steady state can be then written as a weighted sum of these phase probability vectors, with the different weights controlled by the initial state. This mechanism explains how the system breaks the symmetry by singling out a particular dynamical phase out of the multiple possible phases present in the first Doob eigenvector. By conjecturing that statistically-relevant configurations in the symmetry-broken regime can be partitioned into different symmetry classes, we further derive an expression for the components of the subleading Doob eigenvectors in the degenerate subspace in terms of the leading eigenvector and the symmetry eigenvalues, showcasing the stringent spectral structure imposed by symmetry on DPTs. Finally, we introduce a reduced Hilbert space based on a suitable order parameter for the DPT, with appropriate transformation properties under the symmetry operator. All the spectral signatures of the DPT are reflected in this reduced order-parameter space, which hence allows for the empirical verification of our results while providing a natural classification scheme for configurations in terms of their symmetry properties.

We have illustrated our general results by analyzing three distinct DPTs in several paradigmatic many-body systems. These include the one-dimensional boundary-driven WASEP, which exhibits a particle-hole symmetry-breaking DPT for current fluctuations, the $r$-state Potts model for spin dynamics (with $r=3,4$), which displays discrete rotational symmetry-breaking DPTs for energy fluctuations, and the closed WASEP which presents a continuous (in the $L\to\infty$ limit) symmetry-breaking DPT to a time-crystal phase characterized by a rotating condensate or density wave. Our results on the spectral fingerprints of symmetry-breaking DPTs are fully confirmed in these intriguing examples, offering a fresh view on spontaneous symmetry breaking phenomena at the fluctuating level. This is particularly interesting for the case of the time-crystal DPT in the closed WASEP, where the validity of our results suggests an extension to the limit of continuous symmetry breaking phenomena.

The spectral symmetry-breaking mechanism described in this work is completely general for $\mathbb{Z}_n$-invariant systems, so we expect these results to hold valid also in standard (steady-state) critical phenomena, where the dimensional reduction introduced by the projection on the order-parameter, reduced Hilbert space can offer new perspectives on well-known phase transitions \cite{gaveau98a,gaveau06a,binney92a}.

It would be also interesting to extend the current analysis to more complex DPTs. For instance, it would be desirable to investigate the spectral signatures of DPTs in realistic high-dimensional driven diffusive systems, as e.g. the DPTs discovered in the current vector statistics of the 2D closed WASEP \cite{tizon-escamilla17a}. In this case, the complex interplay among the external field, lattice anisotropy, and vector currents in 2D leads to a rich phase diagram, with different symmetry-broken dynamical phases separated by lines of first- and second-order DPTs, and competing time-crystal phases. The spectral fingerprints of this complex competition between DPTs would further illuminate future developments. It would be also interesting to explore the spectral signatures of possible DPTs in driven dissipative systems \cite{prados11a,prados12a,hurtado13a}, or for diffusive systems characterized by multiple local conservation laws, as e.g. the recently introduced kinetic exclusion process \cite{gutierrez-ariza19a}. Finally, though the interplay between symmetry and DPTs in open quantum systems has been investigated in recent years \cite{manzano14a,manzano18a}, the range of possibilities offered by the order-parameter reduced Hilbert space calls for further investigation. 

\begin{acknowledgments}
    The research leading to these results has received funding from the fellowship FPU17/02191 and from the Projects of I+D+i Ref. PID2020-113681GB-I00, Ref. PID2021-128970OA-I00 and Ref. FIS2017-84256-P, Ref. A-FQM-175-UGR18, Ref. P20\_00173 and Ref. A-FQM-644-UGR20 financed by the Spanish Ministerio de Ciencia, Innovación y Universidades and European Regional Development Fund, Junta de Andalucía-Consejería de Economía y Conocimiento 2014-2020. We are also grateful for the the computing resources and related technical support provided by PROTEUS, the supercomputing center of Institute Carlos I in Granada, Spain.
\end{acknowledgments}

\bibliography{referencias-BibDesk-OK-sept21}{}

\appendix

\section{Symmetry and the Doob generator}
\label{appA}

In this Appendix we show that, whenever the original stochastic generator $\genmat$ is invariant under a unitary symmetry operator $\symmop$, i.e. $[\genmat,\symmop]=0$, both the tilted ($\genmatT$) and the Doob ($\genmatD$) generators are also invariant under $\symmop$, provided that the time-integrated observable $\trajobservable$ associated with these large-deviation generators exhibits the same symmetry, i.e. $A({\cal{S}}\omega_{\tau})=A(\omega_{\tau})$ for any trajectory $\traj$, where ${\cal S}$ is the map in trajectory space induced by the symmetry operator $\hat{S}$ at the configurational level, see \S\ref{ssecSym}.

As explained in \S\ref{sec:model_tools}, the time-additive observables $A(\omega_{\tau})$ whose large deviation statistics we are interested in might depend on the state of the process and its transitions over time. For jump processes as the ones considered here, such trajectory-dependent observables can be written in general as 
\begin{equation}
A (\omega_{\tau}) = \sum_{i=0}^{m } (t_{i+1} - t_i)g(C_i) + \sum_{i=0}^{m - 1} \eta_{C_i,C_{i+1}}\, , \nonumber
\label{eq:current_defA}
\end{equation}
 see Eq.~\eqref{eq:current_def} in \S\ref{sec:model_tools}. The first sum above corresponds to the time integral of configuration-dependent observables, $g(C_i)$, while the second sum stands for observables that increase by $\eta_{C_i,C_{i+1}}$ in the transitions from $C_i$ to $C_{i+1}$. In the first sum we have defined $t_0=0$ and $t_{m+1}=\tau$. Demanding $A(\omega_{\tau})$ to remain invariant under the symmetry transformation for any trajectory implies that both the configuration-dependent $g(C)$ and the transition-dependent $\eta_{C,C'}$ functions are invariant under such transformation, so $g(C)=g(C_S)$ and $\eta_{C,C'}=\eta_{C_S,C'_S}$, with the definitions $\ket{C_S}=\hat{S}\ket{C}$ and $\ket{C'_S}=\hat{S}\ket{C'}$. From this and the definition of $\genmatT$ in Eq.~\eqref{eq:genmatT} we can see that if the original generator $\genmat$ commutes with $\symmop$, so that $W_{C\to C'}=\bra{C'}\genmat\ket{C}=\bra{C'}\symmop^{-1}\genmat\symmop\ket{C}=W_{C_S\to C'_S}$ $\forall \ket{C},\ket{C'}\in {\cal H}$, then the tilted generator $\genmatT$ will also conmute with $\hat{S}$. In particular,
\begin{equation}
\begin{split}
\symmop \genmatT \symmop^{-1} = 
\sum_{\conf, \conf^{\prime} \neq \conf} \text{e}^{\lambda \eta_{C,C'}}\transitionrate{\conf}{\conf^{\prime}} \symmop\ket{\conf^{\prime}} \bra{\conf}\symmop^{-1} \\ 
- \sum_{\conf} \escaperate{\conf} \symmop\ketbra{\conf}\symmop^{-1} + \lambda \sum_{\conf} g(C)\symmop\ketbra{\conf}\symmop^{-1} = \\ 
\sum_{\conf_S, \conf^{\prime}_S \neq \conf_S} \text{e}^{\lambda \eta_{C_S,C'_S}}\transitionrate{\conf_S}{\conf^{\prime}_S} \ket{\conf^{\prime}_S} \bra{\conf_S} \\ 
- \sum_{\conf_S} \escaperate{\conf_S} \ketbra{\conf_S} + \lambda \sum_{\conf_S} g(C_S)\ketbra{\conf_S} = \genmatT \, .
\end{split}
\nonumber
\label{eq:genmatTA}
\end{equation}
Therefore we have that $[\genmatT,\hat{S}]=0$, provided the above conditions on observable $A$ hold.

The associated Doob stochastic generator is defined as $\genmatD = \eigvectTLop{0} \genmatT (\eigvectTLop{0})^{-1} - \eigvalT{0} \identityop$, where $\eigvectTLop{0}$ is a diagonal matrix with elements $(\eigvectTLop{0})_{ij} = (\eigvectTL{0})_i \delta_{ij}$, with $\eigvectTL{0}$ the leading left eigenvector of $\genmatT$, see Eq.~\eqref{genmatD}. In order to prove that $[\genmatD,\hat{S}]=0$, we hence have to show first that $\eigvectTL{0}$ is invariant under $\symmop$, i.e. $\eigvectTL{0}\symmop = \eigvectTL{0}$. Since $\genmatT$ entries are non-negative, by virtue of Perron-Frobenius theorem the eigenvector $\eigvectTL{0}$ associated with the largest eigenvalue $\eigvalT{0}$ must be non-degenerate and their components real and positive.
Moreover, because we have shown that $\genmatT$ commutes with $\symmop$, $\eigvectTL{0}$ must also be an eigenvector of $\symmop$, $\eigvectTL{0} \symmop = \symmeigval{0} \eigvectTL{0}$.
Finally, because all components of both $\eigvectTL{0}$ and $\symmop$ are real and positive, we must have the same for $\symmeigval{0} \eigvectTL{0}$.
The only eigenvalue of $\symmop$ that satisfies this is $\symmeigval{0} = 1$, and therefore $\eigvectTL{0} \symmop = \eigvectTL{0}$. In this way, the operator $\eigvectTLop{0}$ used in the Doob transform commutes with $\symmop$,
\begin{equation}
\begin{split}
\symmop \eigvectTLop{0} \symmop^{-1} &= \sum_{\conf} \symmop\ket{C} \braket{\eigvectTLnk{0}}{\conf} \bra{\conf}\symmop^{-1} \\
&= \sum_{C_S} \ket{C_S} \braket{\eigvectTLnk{0}}{C_S} \bra{C_S} = \eigvectTLop{0} \, , \nonumber
\end{split}
\end{equation}
where we have used in the second equality that $\braket{\eigvectTLnk{0}}{\conf}=\bra{\eigvectTLnk{0}}\symmop\ket{C}$, implying that $[\genmatD,\hat{S}]=0$, which was to be proved.

\section{Phase probability vectors in terms of right eigenvectors}
\label{appB}
{
\newcommand{\phaseweight}[1]{w_{#1,\probvecttimenk{0}}}

In this Appendix we compute explicitly the coefficients relating the phase probability vectors $\ket{\Pi_l^\lambda}$ introduced in \S\ref{ssecPhase} with the Doob right eigenvectors $\eigvectDR{j}$ spanning the degenerate stationary subspace for a $\mathbb{Z}_n$ symmetry-breaking DPT. By definition, any phase probability vector can be always written as a linear combination 
\be
\probDphasevect{l} = \sum_{j=0}^{n-1} C_{l,j} \eigvectDR{j}\, , 
\label{ppvlinA}
\ee
with complex coefficients $C_{l,j}\in \mathbb{C}$. Phase probability vectors must be also normalized, $\braket{-}{\Pi_l^\lambda}=1$ $\forall l\in[0\isep n-1]$, and consecutive vectors must be related by the action of the symmetry operator, $\probDphasevect{l+1} = \symmop \probDphasevect{l}$, which implies that $\probDphasevect{l} = \symmop^l \probDphasevect{0}$ and therefore $C_{l,j} = C_{0,j} (\phi_j)^l$, see Eq.~\eqref{ppvlinA}, with $\phi_j$ the eigenvalues of the symmetry operator, $\symmop\eigvectDR{j}=\phi_j \eigvectDR{j}$. 

In order to obtain the coefficients $C_{0,j}$, we impose now that the Doob stationary distribution can be written as a statistical mixture (or convex sum)
of the different phase probability vectors
\begin{equation}
\probTvectst = \sum_{l=0}^{n-1} w_l \probDphasevect{l} = \sum_{j=0}^{n-1} \sum_{l=0}^{n-1} w_l C_{0,j} (\phi_j)^l \eigvectDR{j} \, , \nonumber
\end{equation}
with $\sum_{l=0}^{n-1} w_l = 1$ and $0\le w_l\le 1$, where we have used Eq.~\eqref{ppvlinA} in the second equality. Comparing this expression with the spectral decomposition of the Doob steady state, $\probTvectst = \sum^{n-1}_{j=0}  \eigvectDR{j} \braket{\eigvectDLnk{j}}{\probvecttimenk{0}}$, we find
\be
\braket{\eigvectDLnk{j}}{\probvecttimenk{0}} = \sum_{l=0}^{n-1} w_l C_{0,j} (\symmeigval{j})^l  \, .
\label{leftA}
\ee
Taking now the modulus on both sides of the equation, using the triangular inequality 
and noticing that eigenvalues $\phi_j$ lie in the complex unit circle so $|\phi_j|=1$, we obtain $\abs{\braket{\eigvectDLnk{j}}{\probvecttimenk{0}}} \le \abs{C_{0,j}}$. This inequality is saturated whenever the initial vector $\probvecttime{0}$ is chosen so that the Doob stationary vector coincides with one phase, i.e., $w_l = \delta_{l,l'}$ for some $l'$, see Eq.~\eqref{leftA}, so that $\probTvectst=\probDphasevect{l'}$ for this particular initial $\ket{\probvecttimenk{0}}$. In this way, we have found that in general
\be
\abs{C_{0,j}} = \max_{\ket{P_0}} \abs{\braket{\eigvectDLnk{j}}{\probvecttimenk{0}}} \, . \nonumber
\ee
We can now write
\be
\max_{\ket{P_0}} \left| \braket{\eigvectDLnk{j}}{\probvecttimenk{0}}\right|=\max_{\ket{P_0}} \left| \sum_C \braket{\eigvectDLnk{j}}{C}\braket{C}{\probvecttimenk{0}}\right|  \, ,\nonumber
\ee
and since we have chosen to normalize the left eigenvectors such that $\max_{C} \abs{\braket{\eigvectDLnk{j}}{\conf}} = 1$, see \S\ref{sec:model_tools}, and noting that $\braket{C}{\probvecttimenk{0}}\le 1$ $\forall\ket{C}$, it is clear that the maximum over $\ket{P_0}$ is reached when $\ket{P_0}=\ket{C^*}$, the configuration where $\abs{\braket{\eigvectDLnk{j}}{\conf^*}}$ takes it maximum value 1. Therefore we find that $\abs{C_{0,j}} = 1$. Note also that the normalization condition $\max_{C} \abs{\braket{\eigvectDLnk{j}}{\conf}} = 1$ specifies left eigenvectors up to an arbitrary complex phase, which can be now chosen so that $C_{0,j} = 1$ $\forall j < n$. This hence implies that the coefficients in the expansion~\eqref{ppvlinA} are $C_{l, j} = (\symmeigval{j})^l$, and we can obtain the final form of the probability vector of the phases in terms of the degenerate right eigenvectors,
\begin{equation}
\probDphasevect{l} = \sum_{j=0}^{n-1} (\symmeigval{j})^l \eigvectDR{j} \, , \nonumber
\end{equation}
see Eq.~\eqref{eq:probphasevect} in the main text.

This structure in the phase vectors $\probDphasevect{l}$ has implications on the eigenvalues of the symmetry operator.
In particular, the fact that the different phases must be linearly independent implies that the first $n$ eigenvalues $\symmeigval{j}$ must be different. 
If there were two eigenvalues such that $\symmeigval{j'} = \symmeigval{j''}$, then all the vectors $\probDphasevecttime{l}$ would live in the hyperplane given by the constraint $(\eigvectDL{j'} - \eigvectDL{j''})\ket{v} = 0$, in contradiction to our initial assumption. Therefore, for the symmetry-breaking DPT to occur, the first $n$ eigenvalues $\symmeigval{j}$ must be different, which in turn implies that they must correspond to all the $n$-th roots of unity.
}

\end{document}